\documentclass[a4paper,11pt]{article}
\pdfoutput=1

\usepackage{jheppub}
\usepackage[T1]{fontenc}
\usepackage{amssymb,amsmath}
\usepackage{braket,bbm,bm}
\usepackage{mathtools}
\usepackage[usenames,dvipsnames,svgnames,table]{xcolor}
\usepackage[utf8]{inputenc}
\usepackage{soul, color}
\usepackage{subcaption}
\usepackage{lmodern}
\usepackage{footnote}
\usepackage[normalem]{ulem}
\usepackage{glossaries-extra}
\setabbreviationstyle[acronym]{long-short}
\glssetcategoryattribute{acronym}{nohyperfirst}{true}

\usepackage{slashed}
\usepackage{multirow}
\usepackage{here}
\usepackage{hyperref}
\usepackage{verbatim}
\usepackage[justification=justified,singlelinecheck=false]{caption}
\usepackage{cleveref}
\usepackage{listings}
\usepackage{fancyvrb}
\usepackage{dsfont}
\usepackage{nicefrac,xfrac}
\usepackage{verbatim}
\usepackage{bbm}

\Crefname{equation}{eq.}{eqs.}
\Crefname{section}{section}{sections}
\Crefname{figure}{figure}{figures}
\Crefname{appendix}{appendix}{appendices}

\newcommand{\cD}[0]{\mathcal D}

\newcommand{\cK}[0]{\mathcal K}

\newcommand{\cM}[0]{\mathcal M}
\newcommand{\cO}[0]{\mathcal O}

\newcommand{\cY}[0]{\mathcal Y}

\newcommand{\wt}[0]{\widetilde}
\newcommand{\wh}[0]{\widehat}

\newcommand{\df}[0]{\mathrm{df}}

\newcommand{\Kdf}[0]{{\cK_{\df,3}}}

\newcommand{\PV}[0]{{\mathrm{PV}}}

\newcommand{\kdf}{\mathcal{K}_{\text{df},3} }

\newcommand{\bcX}{\boldsymbol{\mathcal X}}
\newcommand{\bcY}[0]{\boldsymbol{\mathcal Y}}

\newcommand{\XR}[3]{\boldsymbol{\mathcal X}_{[pab]}^{(#1#2#3)}}
\newcommand{\YL}[3]{\boldsymbol{\mathcal Y}^{[kab] \dagger}_{(#1#2#3)}}
\newcommand{\YR}[3]{\boldsymbol{\mathcal Y}^{[kab]}_{(#1#2#3)}}

\newcommand{\CR}[0]{\boldsymbol{\mathcal C}}

\newcommand{\YLp}[4]{\boldsymbol{\mathcal Y}^{[kab] \dagger}_{(#1#2#3),#4}}
\newcommand{\YRp}[4]{\boldsymbol{\mathcal Y}^{[kab]}_{(#1#2#3),#4}}

\newcommand{\ThreeBody}[0]{%
Detmold:2008gh,
Beane:2007qr,
Briceno:2012rv,
Polejaeva:2012ut,
Hansen:2014eka,
Hansen:2015zga,
Hammer:2017uqm,
Konig:2017krd,
Hammer:2017kms,
Mai:2017bge,
Briceno:2018mlh,
Briceno:2018aml,
Blanton:2019igq,
Pang:2019dfe,
Jackura:2019bmu,
Briceno:2019muc,
Romero-Lopez:2019qrt,
Hansen:2020zhy,
Blanton:2020gha,
Pang:2020pkl,
Romero-Lopez:2020rdq,
Blanton:2020gmf,
Blanton:2020jnm,
Muller:2020vtt,
Blanton:2021mih,
Muller:2021uur,
Blanton:2021eyf,
Briceno:2024txg,
Xiao:2024dyw,
Feng:2024wyg,
Hansen:2024ffk,
Jackura:2022gib,
Draper:2024qeh,
Alotaibi:2025pxz}

\usepackage{mfirstuc} 
\newcommand{\addReviewer}[2]{
  \expandafter\newcommand\csname #1\endcsname[1]{{\bf \color{#2} \capitalisewords{#1}:\,##1}}
  \expandafter\newcommand\csname #1cor\endcsname[2]{{\color{#2} \capitalisewords{#1}:\,\st{##1}{\bf ##2}}}
  \expandafter\newcommand\csname #1color\endcsname{#2}
}

\definecolor{cardinal}{rgb}{0.77, 0.12, 0.23}
\definecolor{teal}{rgb}{0.0, 0.5, 0.5}

\addReviewer{sebastian}{cardinal}
\addReviewer{fernando}{orange}
\addReviewer{steve}{Maroon}
\addReviewer{MTH}{teal}

\newcommand{\HSQCa}[0]{Hansen:2014eka}
\newcommand{\HSQCb}[0]{Hansen:2015zga}

\newcommand{\dwave}[0]{Blanton:2019igq}

\newcommand{\isospin}[0]{Hansen:2020zhy}

\newcommand{\tetraquark}[0]{Hansen:2024ffk}
\newcommand{\DRS}[0]{Dawid:2024dgy}
\newcommand{\threeneutron}[0]{Draper:2023xvu}
\newcommand{\multichannel}[0]{Draper:2024qeh}
\newcommand{\BSQC}[0]{Blanton:2020jnm}

\newcommand{\BSnondegen}[0]{Blanton:2020gmf}
\newcommand{\BStwoplusone}[0]{Blanton:2021mih}

\newcommand{\ThreeBodyNumerics}[0]{%
Detmold:2008fn,
Beane:2007es,
Detmold:2011kw,
Mai:2018djl,
Horz:2019rrn,
Blanton:2019vdk,
Mai:2019fba,
Culver:2019vvu,
Fischer:2020jzp,
Hansen:2020otl,
NPLQCD:2020ozd,
Alexandru:2020xqf,
Brett:2021wyd,
Blanton:2021llb,
Mai:2021nul,
Garofalo:2022pux,
Draper:2023boj,
Yan:2024gwp,
Dawid:2025doq,
Dawid:2025zxc}

\newcommand{\IntegralEquations}[0]{%
Jackura:2020bsk,
Hansen:2020otl,
Mai:2021nul,
Dawid:2021fxd,
Dawid:2023jrj,
Jackura:2023qtp,
Briceno:2024ehy,
Jackura:2025wbw}

\newcommand{\LQCDBaryons}[0]{Meng:2003gm,Torok:2009dg,Lang:2012db,Alexandrou:2013ata,Detmold:2015qwf,Lang:2016hnn,Andersen:2017una,BaryonScatteringBaSc:2023zvt}

\newcommand{\SignalToNoise}[0]{Parisi:1983ae}
\newcommand{\TwoHadronFormalAndNumerical}[0]{Briceno:2014oea}

\newcommand{\TwoPlusThreeBound}[0]{Dawid:2024dgy,Hansen:2024ffk}

\newcommand{\LQCDRoper}{Burkert:2017djo}

\title{\boldmath Finite-volume formalism for $N\pi\pi$ at maximal isospin}

\author[a]{Maxwell T. Hansen}
\author[b]{, Fernando Romero-L\'opez}
\author[c]{, and Stephen R. Sharpe}

\affiliation[a]{School of Physics and Astronomy, University of Edinburgh, Edinburgh EH9 3JZ, UK}
\affiliation[b]{Albert Einstein Center, Institute for Theoretical Physics, University of Bern, 3012 Bern, Switzerland}
\affiliation[c]{Physics Department, University of Washington, Seattle, WA 98195-1560, USA}

\emailAdd{maxwell.hansen@ed.ac.uk}
\emailAdd{fernando.romero-lopez@unibe.ch}
\emailAdd{srsharpe@uw.edu}

\abstract{We extend the relativistic field theoretic finite-volume formalism to $N \pi \pi$ scattering states at maximal isospin, $I=5/2$. As in previous work using the relativistic field theory approach, we work to all orders in a generic low-energy effective theory, and determine the quantization condition that relates finite-volume energies to intermediate K matrices, and the integral equations connecting the latter to the physical scattering amplitudes. We discuss the parametrization of the K matrices, and explain in detail the new features that arise in implementing the quantization condition due to the spin of the nucleon in combination with the use of non-degenerate particles. As a concrete example, we provide a sample numerical application including the $\Delta$ resonance in the $N\pi$ subchannel. The extension to the $I=3/2$ and $1/2$ channels is more involved, due to mixing with $N\pi$ states, and we do not provide a complete formalism for these cases. We explain why $N\pi$ states cannot be included by treating the nucleon as a pole in $p$-wave $N\pi$ scattering, an approach that has been successful in studying $D D^*$ scattering using the three-particle $DD\pi$ formalism. We additionally provide results for all isospins under the assumption of no two-to-three mixing, thereby laying the groundwork for a follow-up paper in which all $N\pi\pi \leftrightarrow N\pi$ systems are fully treated. Finally, we study the singularities in $N\pi\pi$ amplitudes arising from $N\pi\pi\pi$ intermediate states, and find that our subthreshold cutoff functions must be modified to avoid such singularities.}
\allowdisplaybreaks

\begin{document}
\maketitle
\clearpage
\flushbottom


\section{Introduction}
\label{sec:intro}

Combined experimental efforts at many facilities have produced a rich spectrum of baryonic excitations~\cite{ParticleDataGroup:2024cfk}. However, a first-principles quantum chromodynamics (QCD) description of these excitations beyond the lowest $\Delta(1232)$ resonance is still lacking. A striking example is the $N(1440)$ $J^P=\tfrac12^+$ Roper resonance, which is not understood in the quark model. There has been much work discussing its status and composition, as reviewed in ref.~\cite{\LQCDRoper}.

Lattice QCD (LQCD) offers a promising option for providing first-principles predictions for baryon resonances, and indeed there has been substantial progress in recent years~\cite{\LQCDBaryons}. However, there are several challenges to a LQCD approach. On the numerical side, one must deal with signal-to-noise degradation for baryonic states~\cite{\SignalToNoise} and the need to accurately determine multiple energy levels in a dense spectrum. On the theoretical side, one must develop finite-volume formalism that applies to baryons that, generically, have multiple decay channels involving two, three, or more particles. For example, the Roper resonance decays primarily into $N\pi$ and $N\pi\pi$ channels (where $N$ denotes a nucleon). For the $N\pi$ system, the formalism for determining the properties of resonances that decay to multiple two-particle channels, including particles with spin, is well established~\cite{\TwoHadronFormalAndNumerical}. For the $N\pi\pi$ system, however, the presently existing formalism is insufficient---it concerns either spinless particles~\cite{\ThreeBody}, or three identical spin-$1/2$ particles~\cite{\threeneutron,Bubna:2023oxo}, rather than systems of mixed spins. Furthermore, the only works addressing systems involving both two- and three-particle channels do so in the context of spinless particles~\cite{Briceno:2017tce,Severt:2022jtg}.

In this work, we take the first step towards a general formalism for studying baryon resonances. In particular, in isosymmetric QCD, we consider the $N\pi\pi$ system with general isospin, but ignore mixing with $N\pi$ states. The resulting formalism is only complete for the case of maximal $N\pi\pi$ isospin, i.e. $I=5/2$, for then mixing with the $N\pi$ channel is forbidden. Although the $I=5/2$ channel is nonresonant, and thus less interesting than those with $I=3/2$ and $1/2$, it contains the nontrivial dynamics of mixing with $\Delta \pi$, and will likely be a stepping stone for lattice studies of baryon resonances.

Our results successfully deal with two of the three challenges that formalism for the Roper resonance must overcome: incorporating the nucleon spin into the formalism, for which we adapt the approach for three neutrons~\cite{\threeneutron}, and determining the appropriate flavor structure for the three choices of isospin, which has proven to be nontrivial in previous applications~\cite{\isospin,\tetraquark,Draper:2024qeh}. We present both steps of the three-particle formalism for ${I=5/2}$, within the relativistic field theoretic (RFT) framework. First, in \Cref{sec:QC}, we give the finite-volume quantization condition that relates the spectrum to two- and three-particle K matrices. Second, in \Cref{sec:inteqs}, we present the infinite-volume integral equations that related these K matrices to two- and three-particle scattering amplitudes. The derivation of these results is given in \Cref{app:deriv}.

We do not address here the third, and most difficult challenge for a Roper formalism---incorporating mixing between $N\pi$ and $N\pi\pi$ channels. Our initial plan was to use an approach in which the nucleon is viewed as a bound state in the $p$-wave $N\pi$ subchannel. This approach has been successfully applied in various contexts~\cite{\TwoPlusThreeBound}, including the $DD\pi \leftrightarrow D^* D$ system~\cite{\tetraquark,\DRS}, and allows one to study ``$2+3$'' systems using only the three-particle formalism. Unfortunately, we find that it is not applicable in the present case, and we explain this failure in \Cref{sec:dimer}. Thus a full Roper formalism will need to extend the 2+3 formalism of ref.~\cite{Briceno:2017tce} to include spin and flavor.

In the course of this work, we realized that, for the $N\pi\pi$ system, the standard RFT parameter space, in which the invariant mass of one pair of particles is taken below threshold by some distance, allows, in some cases, the invariant mass of another pair to go above its inelastic threshold. As discussed in \Cref{app:singularities}, this can lead to singularities in $3\to3$ kernels, the avoidance of which requires alteration of the standard RFT cutoff functions.

Finally, we provide, in \Cref{sec:num}, a first implementation of the finite-volume step in this formalism. We illustrate its utility by determining the spectrum for parameters corresponding to heavier than physical quarks ($M_\pi \approx 300\;$MeV), such that the $\Delta$ baryon is stable. These parameters should be accessible to lattice calculations in the near future.

We conclude and provide an outlook in \Cref{sec:concl}. Technical issues related to the implementation are presented in two appendices, \Cref{app:intD} determining needed integrals over Wigner rotation matrices, and \Cref{app:K0decomp} providing an explicit implementation of the leading term in the three-particle K matrix.


\section{Three-particle finite-volume quantization condition}
\label{sec:QC}

In this section, we describe the formalism needed to extract infinite-volume K matrices from the finite volume $N\pi\pi$ spectrum for maximal isospin. Here we state only the main results; an outline of the derivation, along with further technical details, is presented in \Cref{app:deriv}.

As noted in the introduction, the two major challenges in the generalizing previous work to the $N\pi\pi$ system are the presence of spin, and the flavor structure. Concerning spin, the key question, discussed at length in ref.~\cite{\threeneutron} in the study of three neutrons, is in which frame to define the spin components: the finite-volume or ``lab'' frame (which is where the calculations are done) or the center of mass frames (CMFs) of the two-particle subchannels. The relationship of spin degrees of freedom between these two frames is nontrivial due to Wigner rotations. For three neutrons, we adopted a hybrid choice using different frames for the spins of different particles. Here, having only a single spin, the situation is simpler, and we adopt the choice of using the lab frame for the nucleon spin. Why this simplifies the result will become fully clear as we present the results, but the essential point is that only the contribution of two-particle interactions ends up involving Wigner rotations. We postpone further discussion of this issue until \Cref{sec:K2} below, and turn now to the flavor structure of the formalism.

The flavor structure is considerably simplified by the fact that we are considering only $I=5/2$. This isospin appears only once in the decomposition of $N\pi\pi$ states,
\begin{equation}
\frac{1}{2} \otimes 1 \otimes 1 = \left(\frac{1}{2} \oplus \frac{3}{2} \right) \otimes 1 = \frac{1}{2} \oplus \frac{3}{2} \oplus \frac{1}{2} \oplus \frac{3}{2} \oplus \frac{5}{2}\,.
\label{eq:isospin_states}
\end{equation}
As recalled in \Cref{app:deriv}, the formalism divides the three particles into a spectator and a pair. For $I=5/2$, the spectator can be either a nucleon or a pion, leading to a two-dimensional flavor basis,
\begin{align}
&\left\{\ket{(\pi\pi)_{2}N}, \ket{(N \pi)_{3/2}\pi}\right\} \,.
\label{eq:I5basis}
\end{align}
Here, the first two entries in each triplet form the pair, with isospin denoted by the subscript. The first element of each pair is the primary member, the significance of which is discussed in \Cref{app:deriv}.

Our first main result, a relation between the finite-volume spectrum and intermediate K matrices, takes the standard RFT form~\cite{\HSQCa}. Up to exponentially suppressed finite-volume effects, the quantization condition
\begin{align}
\det_{i \bm p \ell m m_s} \left( 1+ \widehat{\mathcal K}_{\rm df,3}(E^\star) \widehat{F}_3(E, \bm P, L) \right) &= 0\,,
\label{eq:QC3}
\end{align}
determines the interacting finite-volume energies $E_n(L)$.

Solutions are valid provide that the corresponding CMF energy, given in terms of the finite-volume energy $E$ by
\begin{equation}
E^\star = \sqrt{E^2 - \bm P^2} \,,
\end{equation}
lies within the range
\begin{equation}
\sqrt{M_N^2 + 2 M_\pi^2} + M_\pi < E^\star < M_N + 3 M_\pi \,,
\end{equation}
i.e., above nonanalyticities induced by left-hand cuts, and below the inelastic threshold. We emphasize that $\widehat {\mathcal K}_{\rm df, 3}$ depends on the CMF energy $E^\star$ whereas $\widehat F_3$ depends separately on the finite-volume-frame energy $E$, the total three-momentum in that frame $\bm P$ (an integer three-vector multiple of $2 \pi/L$), and the box length $L$. Suppressing these coordinates going forward, the two key objects are defined as
\begin{equation}
\widehat{F}_3 \equiv \frac{\widehat{F}}{3}+\widehat{F} \frac{1}{1-\widehat{\mathcal{M}}_{2, L} \widehat{G}} \widehat{\mathcal{M}}_{2, L} \widehat{F}, \qquad \widehat{\mathcal{M}}_{2, L} \equiv \frac{1}{[\widehat{\mathcal{K}}_{2,L}]^{-1}-\widehat{F}}\,.
\label{eq:F3}
\end{equation}
Each quantity with a caret is a matrix in the tensor-product space labeled by several indices: the spectator-flavor index $i$, running over the two choices in \Cref{eq:I5basis}; the spectator momentum $\bm p$, running over allowed finite-volume momenta; the spin $\ell$ and $z$-component $m$ of the pair; and, finally, the $z$ component of the spin of the nucleon, taking values $m_s=\pm 1/2$. The set of allowed $\bm p$ is truncated by a cutoff function that is built into the formalism: see \Cref{eq:cutoff} and surrounding discussion. The sum over $\ell$ is truncated by hand in a manner to be discussed below (see, in particular, \Cref{sec:K2}). The net result is that, in practical applications, the matrices are finite dimensional, and numerical implementation is straightforward.

The derivation sketched in \Cref{app:deriv} yields the following results for the component matrices of the quantization condition.
The $F$ matrix is
\begin{align}
\widehat F & =
\begin{pmatrix}
F_N & 0 \\
0 & F_\pi
\end{pmatrix}\,,
\label{eq:FI5}
\end{align}
where $F_N$ and $F_\pi$ are defined in \Cref{eq:FNt,eq:Fpit}, respectively.
Compared to results in \Cref{app:deriv}, we have dropped projectors on to even or odd values of $\ell$. In particular, we leave it implicit that for the pairs $(\pi\pi)_{2,0}$ only even values of $\ell$ contribute, while for $(\pi\pi)_1$ only odd values contribute.

The $G$ matrix is given by
\begin{align}
&\widehat G = \begin{pmatrix}
0 & \sqrt{2} \mathbb{P}_{\ell} G_{N \pi} \\
\sqrt{2} G_{\pi N }\mathbb{P}_{\ell} & \mathbb{P}_\ell G_{\pi \pi} \mathbb{P}_\ell \\
\end{pmatrix}\,,
\label{eq:GI5}
\end{align}
where $G_{\pi\pi}$, $G_{\pi N}$, and $G_{N\pi}$ are given by \Cref{eq:Gpipit,eq:GpiNt,eq:GNpit}, respectively, and
\begin{equation}
[\mathbb P_\ell]_{i \bm p' \ell' m' m'_s, j \bm p \ell m m_s} = \delta_{ij} \delta_{\bm p' \bm p}
\delta_{\ell' \ell} \delta_{m' m} \delta_{m'_s m_s} (-1)^\ell\,.
\label{eq:Pell}
\end{equation}

The final ingredient is
\begin{align}
\widehat \cK_{2,L} &=
\begin{pmatrix}
\tfrac{1}{2} \cK_{2,L}{(\pi\pi, I=2)} & 0 \\
0 & \cK_{2,L}{(N\pi, I=3/2)}
\end{pmatrix}\,.
\label{eq:K2L5}
\end{align}
The explicit form of the quantities on the right-hand side involves the consideration of spin in different frames and is discussed in the following subsection.


\subsection{Contributions from two-particle scattering}
\label{sec:K2}

There are two cases to consider: $\pi\pi$ scattering with a nucleon spectator, and $N\pi$ scattering with a pion spectator. The former is straightforward, for two reasons. First, the particle with spin is the spectator, so the spin structure is trivial. Second, since the scattering does not involve spin, there is no recoupling between spin and angular momentum in the K matrix. Thus the result is a simple generalization of that discussed in previous RFT works on $2+1$ systems~\cite{\BStwoplusone,\tetraquark,\multichannel,\DRS}, and is given by
\begin{equation}
[\cK_{2,L}(\pi\pi, I)]_{\bm p'\ell' m' m'_s; \bm p\ell m m_s} =
2 \omega_N(\bm p) L^3 \delta_{\bm p' \bm p}\, \delta_{\ell' \ell}\, \delta_{m' m}\, \delta_{m'_s m_s}\, \cK_{2,\ell}^{\pi\pi,I}(q_N^\star(\bm p))\,,
\label{eq:K2Lpipi}
\end{equation}
where
\begin{align}
\left[\cK_{2,\ell}^{\pi\pi,I}(q)\right]^{-1} &= \frac1{16 \pi E_{\pi\pi}(q)}
\left\{ q \cot\delta_\ell^{\pi\pi,I}(q) + |q| [1 - H_{N}(\bm p)] \right\}\,,
\label{eq:K2ellpipi}
\\
E_{\pi\pi}(q) &= 2 \sqrt{q^2 + M_\pi^2 }\,.
\end{align}
Here $\delta_\ell^{\pi\pi,I}$ is the phase shift for $\pi\pi$ scattering with isospin $I$. In practical applications one sets the phase shift to zero for $\ell > \ell_{\rm max}$, for some choice of $\ell_{\rm max}$.

The function $H_N(\bm p)$ appearing in \Cref{eq:K2ellpipi} is a smooth cutoff function, which also appears in the matrices $F$ and $G$. Its argument is the spectator momentum, which is related to $q\equiv q_N^\star(\bm p)$ by \Cref{eq:qpstarN}. The standard RFT cutoff function has the general form
\begin{equation}
H_X(\boldsymbol p) = J \big (z_X(\boldsymbol p) \big), \qquad z_X(\boldsymbol p) = (1+\epsilon_H)\frac{\sigma_X(\boldsymbol p) - \sigma^{\rm min}_{X}}{ \sigma^{\rm th}_{X} - \sigma^{\rm min}_{X}},
\label{eq:cutoff}
\end{equation}
where $J(z)$ is given, for instance, in eq.~(2.23) of ref.~\cite{Blanton:2021eyf}. This function is defined to vanish for $z<0$, to equal unity for $z \geq 1$, and to vary smoothly in the region $0 < z < 1$. The quantities entering $z_X(\bm p)$ depend on the spectator. $\sigma_X(\bm p)$, defined in \Cref{eq:qpstarN}, is the squared invariant mass of the nonspectator pair for a spectator of flavor $X$ having momentum $\bm p$. The value of $\sigma_X$ at the pair threshold is denoted $\sigma_X^{\rm th}$, while $\sigma_X^{\rm min}$ is the subthreshold value of the squared invariant mass at which the cutoff function vanishes. For $X=N$, we have
\begin{equation}
\sigma_N^{\rm min} = 0\,, \quad
\sigma_N^{\rm th} = 4M_\pi^2\,,
\label{eq:sigmaNmin}
\end{equation}
chosen such that $z_N(\bm p)$ vanishes at the left-hand cut corresponding to two-pion exchange. Finally, $\epsilon_H$ is a small positive quantity, which ensures that the $H_X(\bm p)$ reaches unity slightly below the subchannel threshold.

\bigskip

Now we turn to the more complicated and interesting case in which the pion is the spectator. We first remove the trivial $L$ dependence following \Cref{eq:K2Ldef},
\begin{equation}
[\cK_{2,L}(N\pi,I)]_{\bm p'\ell' m' m'_s;\bm p\ell m m_s} =
2 \omega_\pi(\bm p) L^3 \delta_{\bm p' \bm p}\, [\cK_2^{N\pi,I}(q_\pi^\star(\bm p))]_{\ell' m' m'_s;\ell m m_s}\,,
\label{eq:K2LNpi}
\end{equation}
with $\cK_2^{N\pi,I}$ being an infinite-volume quantity, depending on $q_\pi^\star(\bm p)$, which we denote as simply $q$ in the following.
$N\pi$ scattering is described most naturally in terms of angular momentum variables in the pair CMF. While $\ell,m$ are defined in this frame, $m_s$ is not, and our first task is to connect spin variables in these two frames.

To do so, we recall results from ref.~\cite{\threeneutron}, itself based largely on ref.~\cite{Chung:1971ri}. The lab-frame nucleon state with momentum $\bm a$ and spin-component $m_s$ is defined by
\begin{equation}
\left|\bm a, m_s(\bm a)\right\rangle \equiv U(L(\boldsymbol{\beta}_N(\bm a)))\left|\mathbf{0}, m_s\right\rangle\,,
\end{equation}
where $U(L(\boldsymbol{\beta}_N(\bm a)))$ is a unitary representation of a Lorentz boost with velocity $\boldsymbol{\beta}_N(\bm a) = \boldsymbol{a}/\omega_N(\boldsymbol{a})$, and $\ket{\mathbf{0},m_s}$ is a nucleon state at rest with azimuthal component $m_s$ relative to the lab-frame $z$ axis. Following ref.~\cite{\threeneutron}, we have introduced the notation that the argument of $m_s$ indicates the nature of the moving state. If, as is the case here, the state is obtained by a single boost from the state at rest, then the argument of $m_s$ matches the momentum of the state. The action of rotations on the moving state is
\begin{equation}
U(R)\left|\bm a, m_s(\bm a)\right\rangle=\left|R \bm a, m'_s(R \bm a)\right\rangle \mathcal{D}_{m'_s m_s}^{(1/2)}(R)\,,
\end{equation}
i.e.~the transformation is exactly as in nonrelativistic quantum mechanics. This moving state is that corresponding to the relativistic spinors $u(\bm a,m_s)$.

The complications due to Wigner rotations arise when boosting from the lab frame to the CMF of the $N\pi$ pair. The boost velocity required to achieve this is given by
\begin{equation}
\boldsymbol{\beta}_{N \pi}(\bm p) = - \frac{\boldsymbol{P} - \boldsymbol{p} }{E-\omega_{\pi}(\boldsymbol{p})}\,,
\label{eq:boostvelocity}
\end{equation}
where $\bm p$ is the spectator momentum (i.e.~the momentum of the pion that is not in the $N \pi$ pair). The action of this boost on the lab-frame state is given by~\cite{\threeneutron}
\begin{equation}
\left|\bm a^\star, m_s(\bm a) \right\rangle \equiv
U\left(L\big(\boldsymbol{\beta}_{N\pi}(\bm p)\big)\right)\left|\bm a, m_s(\bm a)\right\rangle=\left|\bm a^\star, m'_s(\bm a^\star)\right\rangle \mathcal{D}_{m'_s m_s}^{(1/2)}(R(\theta(\bm a, \bm p), \hat {\bm n} (\bm a, \bm p)))\,,
\label{eq:boostedstate}
\end{equation}
where $\bm a^\star$ is the spatial part of $(\omega_N(\bm a),\bm a)$ after the boost, and $R(\theta(\bm a, \bm p), \hat {\bm n} (\bm a, \bm p))$ is the rotation that results from combining two boosts. The axis and angle of the rotation are given by
\begin{align}
\hat{\bm n} (\bm a, \bm p) & =\frac{\boldsymbol{\beta}_{N\pi}(\bm p) \times \boldsymbol{\beta}_N(\bm a)}{\left|\boldsymbol{\beta}_{N\pi}(\bm p) \times \boldsymbol{\beta}_N(\bm a)\right|},
\label{eq:nhat}
\\
\cos \theta(\bm a, \bm p) & = \frac{\left(1+\gamma_{N\pi}(\bm p)+\gamma_{N}(\bm a)+\gamma^{\prime}(\bm a, \bm p)\right)^2}{\left(1+\gamma_{N\pi}(\bm p)\right)\left(1+\gamma_{N}(\bm a)\right)\left(1+\gamma^{\prime}(\bm a, \bm p)\right)}-1 \,,
\label{eq:axisangleWD}
\end{align}
with the convention $\sin \theta(\bm a, \bm p) \geq 0$, and we have also used the definitions
\begin{align}
\gamma^{\prime}(\bm a, \bm p) &= \gamma_{N \pi}(\bm p) \gamma_N(\boldsymbol a)(1+\boldsymbol{\beta}_{N\pi}(\bm p)\cdot\boldsymbol{\beta}_N(\bm a)) \,, \\
\gamma_{N\pi}(\bm p) &= \frac{1}{\sqrt{1-\bm{\beta}_{N\pi}(\bm p)^2}} \,, \ \ {\rm and}\ \
\gamma_N(\bm a) = \frac{1}{\sqrt{1-\bm{\beta}_N(\bm a)^2}} \,.
\end{align}

On the left-hand side of \Cref{eq:boostedstate} we have introduced notation for a new type of moving state, that which is obtained from the state at rest by two boosts (with velocities $\boldsymbol{\beta}_N(\bm a)$ and $\boldsymbol{\beta}_{N\pi}(\bm p)$, respectively). The argument of $m_s$, here $\bm a$, does not match the momentum of the state, here $\bm a^\star$, unlike in the single-boost states on the right-hand side of \Cref{eq:boostedstate}. The choice of argument $\bm a$ on the left-hand side indicates that the spin component is associated with the lab-frame momentum, which is the index that we use in the matrices appearing in the quantization condition. On the other hand, $m'_s(\bm a^\star)$ is the spin component that transforms as in nonrelativistic quantum mechanics under rotations in the pair CMF, and thus which can be combined straightforwardly with the pair angular-momentum $\ell,m$.

In the following, to lighten the notation, we replace $m'_s(\bm a^\star)$ with $m^{\star\prime}_s$, i.e. we use a star to denote spin components that are associated with the pair CMF, while those in the lab frame will not have stars. Thus \Cref{eq:boostedstate} becomes
\begin{equation}
\left|\bm a^\star, m_s \right\rangle =
\left|\bm a^\star, m^{\star}_s\right\rangle {\bm {\mathcal{D}}}_{m^\star_s m_s}^{(1/2)}( \bm a, \bm p )\,,\qquad
{\bm {\mathcal{D}}}_{m^\star_s m_s}^{(1/2)}( \bm a, \bm p ) \equiv {{\mathcal{D}}}_{m^\star_s m_s}^{(1/2)} \Big (R \big (\theta (\bm a, \bm p),{\hat {\bm n}} (\bm a, \bm p) \big ) \Big )\,,
\label{eq:boostedstatenew}
\end{equation}
where we have used the presence of the star to drop the prime on the dummy index $m^\star_s$, and have also introduced a compact notation for the Wigner rotation matrix, now explicitly dependent on both $\bm a$ and $\bm p$.

With these results in hand, it is in principle straightforward to relate the matrix $[\cK_2^{N\pi,I}(q_\pi^\star(\bm p))]_{\ell' m' m'_s;\ell m m_s}$ in \Cref{eq:K2LNpi} to that in which the indices are those in the pair CMF, which we denote $[\cK_2^{N\pi, \star ,I}(q_\pi^\star(\bm p))]_{\ell^{\star \prime} m^{\star \prime} m^{\star \prime}_s;\ell^\star m^\star m^\star_s}$. We just need to conjugate with the appropriate Wigner D matrix obtained from \Cref{eq:boostedstatenew}. There is a complication, however, arising from the fact that the Wigner rotation depends on the direction of the momenta of the scattering pair, coupled with the fact that the dependence on $\ell$ and $m$ is determined by performing a decomposition of the angular dependence of the two-particle amplitude. For this reason we have introduced the indices $\ell^\star$ and $m^\star$, which indicate angular decomposition in the pair CMF, without the inclusion of the Wigner rotation. Thus to proceed, we need to, first, recombine $[\cK_2^{N\pi,\star,I}(q_\pi^\star(\bm p))]_{\ell^{\star \prime} m^{\star \prime} m^{\star \prime}_s;\ell^\star m^\star m^\star_s}$ with spherical harmonics in order to have a function of the angles of the primary particle as well as the pair CMF spin indices; second, apply the Wigner rotation matrix from \Cref{eq:boostedstatenew}, which acts on these spin indices, but also contains dependence on angles; and, finally, decompose the resulting angular dependence into the $\ell, m$ basis. We stress that this final decomposition is done in terms of angles in the pair CMF, even though the spin index is now defined in the lab frame.

Following these steps, we find
\begin{multline}
[\cK_{2}^{N\pi,I}(q_\pi^\star(\bm p))]_{\ell' m' m_s'; \ell m m_s} \\
=
[D(\bm p)]_{\ell' m' m_s'; \ell^{\star \prime} m^{\star \prime} m^{\star \prime}_s}
[\cK_2^{N\pi,\star,I}(q_\pi^\star(\bm p))]_{\ell^{\star \prime} m^{\star \prime} m^{\star \prime}_s;\ell^\star m^\star m^\star_s}
[\overline D(\bm p)]_{\ell^\star m^\star m^\star_s; \ell m m_s}\,,
\label{eq:Daction}
\end{multline}
or, suppressing the indices
\begin{equation}
\cK_{2}^{N\pi,I}(q_\pi^\star(\bm p)) = D(\bm p) \cdot \cK_2^{N\pi,\star,I}(q_\pi^\star(\bm p)) \cdot \overline D(\bm p)\,,
\label{eq:Daction2}
\end{equation}
where
\begin{align}
[D(\bm p)]_{\ell m m_s; \ell^\star m^\star m^\star_s} &= \int d\Omega_{a^\star}
Y_{\ell m}^*(\hat {\bm a}^\star) [\bm \cD^{(1/2)}( \bm a, \bm p)^{-1}]_{m_s m^\star_s} Y_{\ell^\star m^\star}(\hat {\bm a}^\star)\,,
\label{eq:intD}
\\
[\overline D(\bm p)]_{\ell^\star m^\star m^\star_s; \ell m m_s} &= \int d\Omega_{a^\star}
Y_{\ell^\star m^\star}^*(\hat {\bm a}^\star) [{\bm \cD}^{(1/2)}(\bm a, \bm p)]_{m^\star_s m_s} Y_{\ell m}(\hat {\bm a}^\star)\,.
\label{eq:intDbar}
\end{align}

The evaluation of the $D(\bm p)$ and $\overline D(\bm p)$ matrices is straightforward but tedious, and is described in \Cref{app:intD}. We note that, in the nonrelativistic limit, both matrices equal the identity matrix, as expected since Wigner rotations are not needed in this limit. Another feature is that, even if one truncates the sum over $\ell^\star$, these matrices will be nonzero in general for all values of $\ell$. However, we find in numerical examples that the expected threshold suppression as one increases $\ell^\star$ is carried over to that in $\ell$. In practice, we therefore use the same maximum value for $\ell$ as for $\ell^\star$.

\bigskip

In the remainder of this section we describe the form of $\cK_2^{N\pi,\star,I}(q_\pi^\star(\bm p))$, i.e. the two-particle K matrix in the pair CMF. Here the complication due to spin is that $\cK_2^{N\pi,I}(q_\pi^\star(\bm p))$ is not diagonal in the $\ell,s$ basis, due to the recoupling of $\ell$ and $s$ into the total angular momentum, $j$. The allowed values are $j=\ell \pm 1/2$ for $\ell\ge 1$, so two (adjacent) values of $\ell$ lead to each choice of $j$. However, parity conversation forbids mixing between these choices, so, in fact, the amplitudes are diagonal in both $j$ and $\ell$. Thus we find the following expression for the $N\pi$ K matrix,
\begin{align}
[\cK_2^{N\pi,\star,I}(q_\pi^\star(\bm p))]_{\ell^{\star \prime} m^{\star \prime} m^{\star \prime}_s;\ell^\star m^\star m^\star_s} &= \delta_{\ell^{\star \prime} \ell^\star} \sum_j
P_{m^{\star \prime} m^{\star \prime}_s; m^\star m^\star_s}^{j\ell^\star} \cK_{2; j\ell^\star}^{N\pi,I}(q_\pi^\star(\bm p))\,,
\\
P_{m^{\star \prime} m^{\star \prime}_s; m^\star m^\star_s}^{j\ell^\star} &=
\sum_{ \mu_j} \langle \ell^\star m^{\star \prime}, \tfrac12 m^{\star \prime}_s | j \mu_j\rangle
\langle j \mu_j | \ell^\star m^\star, \tfrac12 m^\star_s \rangle\,.
\label{eq:Pjell}
\end{align}
The restriction to allowed values of $j$ is enforced by the Clebsch-Gordon coefficients in the projectors $P^{j\ell^\star}$. If we restrict to $\ell^\star_{\rm max}=1$, then only three channels contribute, $\{j,\ell\}=\{1/2,0\}$, $\{1/2,1\}$, and $\{3/2,1\}$. The (inverses of their) K matrices are given by
\begin{align}
\frac1{\cK_{2; j, \ell}^{N\pi,I} (q_\pi^\star(\bm p))} &= \frac1{8\pi q^{2 \ell} \sqrt{\sigma_\pi(\bm p)}}
\left\{ q^{2 \ell + 1} \cot \delta_\ell^{j} + |q| q^{2 \ell}[1 - H_\pi(\bm p) ] \right\}\,,
\label{eq:K2general}
\end{align}
where we have expressed the result in terms of the combination $ q^{2 \ell + 1} \cot \delta_\ell^{j}$ as it starts at constant order in an effective-range expansion. The standard form of the cutoff function $H_\pi(\bm p)$ by given in \Cref{eq:cutoff} above, with the parameters
\begin{equation}
\sigma_\pi^{\rm th} = (M_N+M_\pi)^2 \,, \quad
\sigma_\pi^{\rm min} = M_N^2 + 2 M_\pi^2\,.
\label{eq:sigmapimin}
\end{equation}
This choice of $\sigma_\pi^{\rm min}$ ensures that $z_\pi(\bm p)$ vanishes at the left-hand cut corresponding to a $u$-channel exchange of a nucleon. We discuss the left-hand singularity structure of the $N\pi$ amplitude further in \Cref{sec:dimer}. As noted in the Introduction, there are additional singularities in $3\to3$ kernels that are not avoided using the standard RFT cutoff function. This issue turns out to apply only for $H_\pi(\bm p)$, and is discussed in \Cref{app:singularities}, along with a proposal for a generalized cutoff function that does avoid the singularities.

Up to this point, we have been discussing $\cK_{2,L}$, rather than its inverse, the latter being what enters the quantization condition. Inversion is, however, straightforward, because, for a given choice of $\ell^\star_{\rm max}$, the projectors form a complete set, and the matrices $D(\bm p)$ and $\overline{D}(\bm p)$ entering \Cref{eq:Daction} are invertible.


\subsection{Flavor structure of \texorpdfstring{$\Kdf$}{the three-particle K matrix}}
\label{sec:Kdfflavor}

The matrix $\widehat{\cK}_{\rm df,3}$ entering the quantization condition, \Cref{eq:QC3}, is a three-particle K matrix, describing the short distance three-particle interaction~\cite{\HSQCa}. By construction it is free of $s$-channel cuts, and thus smooth aside from singularities related to three-particle resonances. It has the same symmetries as the corresponding three-particle scattering amplitude $\widehat{\cM}_3^I$, as will be discussed in more detail in \Cref{sec:Kdfparam} below. It should be kept in mind, however, that it is a cutoff-dependent quantity, and thus not directly physical. It is related to the physical amplitude, $\widehat{\cM}_3$, by integral equations to be described in \Cref{sec:inteqs}.

In this section we describe the flavor structure of $\widehat{\cK}_{\rm df,3}$, which forms a $2\times 2$ flavor matrix, as seen for $\widehat{F}$ and $\widehat{G}$ above. The underlying K matrix, by contrast, involves only a single term, corresponding to the appearance of only a single $I=5/2$ entry in the decomposition \Cref{eq:isospin_states}. We denote this underlying K matrix by $\cK_{{\rm df,3;}\; m'_s,m_s}(\{\bm p \}, \{\bm k\} )$, with $\{\bm p\}=\{\bm p_1,\bm p_2, \bm p_3\}$ being the final state momenta, and $\{\bm k\}$ the corresponding triplet for the initial state. We choose the nucleon momentum to be $\bm p_N=\bm p_3$ (and similarly $\bm k_N=\bm k_3$), while the other two momenta are those of the pions. The final and initial (lab-frame) spin indices are $m'_s$ and $m_s$, respectively. Explicit forms for the K matrix will be described in \Cref{sec:Kdfparam}. Here we only need to know that the K matrix is symmetric under the interchange of the final-state pion momenta $\bm p_1 \leftrightarrow \bm p_2$, and separately under the analogous initial-state interchange, $\bm k_1 \leftrightarrow \bm k_2$. This is because the pion pair is necessarily in a symmetric $I=2$ state when the total isospin is $I=5/2$.

The elements of the $2\times 2$ flavor matrix are all related to the single underlying amplitude, implying an outer-product structure. This corresponds to the different choices of initial and final-state spectator, either nucleon or pion. Thus one has to determine the flavor factors that come with each entry in the matrix. As explained in \Cref{app:Kdfflavor,app:chtoiso}, the result is
\begin{equation}
\left[ \widehat{\cK}_{\rm df,3}\right]_{p \ell' m' m'_s; k \ell m m_s} =
\boldsymbol {\mathcal Y}^{I=5/2}_{p\ell' m'} \circ \cK_{\df,3;\;m'_s, m_s}(\{\bm p\},\{\bm k\})
\circ \boldsymbol {\mathcal Y}^{I=5/2,\dagger}_{k \ell m}
\,,
\label{eq:Kdf3I5form}
\end{equation}
where the flavor structure enters through (leaving $k \ell m$ indices implicit for now)
\begin{align}
\bcY^{I=5/2} =
\begin{pmatrix}
\sqrt{\tfrac12} \YR312
\\
\YR132
\end{pmatrix}\,,
\qquad
\bcY^{I=5/2,\dagger} &= \left(\sqrt{\tfrac12} \YL312,\ \YL132 \right)\,.
\label{eq:Y5}
\end{align}
The operators $\boldsymbol{\mathcal Y}^{[kab]}_{\boldsymbol \sigma}$ act on functions $g(\{\bm p \})$ of three on-shell momenta, and give rise to objects that have $\{p\ell m\}$ indices:\footnote{%
The $\bcY$ operators were introduced in ref.~\cite{\tetraquark}. Those used here differ in the choice of conjugation of the spherical harmonics, consistent with the convention adopted above for the other matrices entering the quantization condition. Also, the overall factor has been corrected.%
}
\begin{align}
{\boldsymbol {\mathcal Y}}_{{{\boldsymbol \sigma}},\;p\ell m}^{[kab]} \circ g
&=
\sqrt{\frac{1}{4\pi}} \int d\Omega_{a^\star} Y^*_{\ell m}(\hat {\bm a}^\star)
g(\{ \bm p \})\bigg|_{p_{\sigma(1)}\to k,\ p_{\sigma(2)}\to a, \ p_{\sigma(3)}\to b} \,.
\label{eq:YRdef}
\end{align}
Here $\boldsymbol \sigma$ is a permutation of $\{1,2,3\}$, while $k$, $a$, and $b$ label, respectively, the spectator momentum, the primary momentum of the pair, and the secondary momentum of the pair. Thus the meaning of the subscript in \Cref{eq:YRdef} is that we take $p_{\sigma_1}$ to be the spectator momentum, with $p_{\sigma_2}$ and $p_{\sigma_3}$ forming the remaining pair, with the former primary. We boost to the CMF of this pair, and decompose into spherical harmonics, defining $\hat {\bm a}^\star$ as the direction of $\bm p_{\sigma_2}$ in this frame. An analogous definition holds for the conjugate operator $\boldsymbol{\mathcal Y}^{[kab]\dagger}_{\boldsymbol \sigma}$, which acts from the right and in which the spherical harmonic is not conjugated.

In the above, the normalization of the $\bcY$s are arbitrary, since changes can be absorbed into the underlying $\Kdf$ functions. Following Ref.~\cite{\tetraquark}, the normalizations are chosen so that $\bcX^{I;x}\circ \bcY^{I;y} = 3 \delta_{xy}$, where the $\bcX$s are defined below in \Cref{sec:inteqs}. This leads to the standard form of the relation between scattering amplitude and K matrix, $\widehat{\cM}_{\rm df,3} = \widehat{\cK}_{\rm df,3} + \cO(\cK_{\rm df,3}^2)$, where $\widehat{\cM}_{\rm df,3}$ is also defined in \Cref{sec:inteqs}.


\subsection{Parametrizing \texorpdfstring{$\Kdf$}{the three-particle K matrix}}
\label{sec:Kdfparam}

In the previous section, we have described how, given an underlying form of $\cK_{\rm df,3}$ as an expression in terms of $\{\bm p\}$ and $\{\bm k\}$, we can construct the matrix $\widehat{\cK}_{\rm df,3}$ that enters the quantization condition. In this section we discuss the underlying forms that are allowed by symmetries.

In order to incorporate the nucleon spin, individual terms in $\cK_{\rm df,3}$ must have the general form
\begin{equation}
\bar{u}_{p_3} \Gamma u_{k_3} f(\{{p}\},\{{k}\}),
\end{equation}
where $\Gamma$ is one of the matrices of the Clifford algebra, and $f$ is a function of all the (on-shell) four-momenta. We stress the spin is defined is the lab frame. Several properties must be fulfilled by this expression: it must be a Lorentz scalar, invariant under parity, and $f$ must be symmetric under the exchange $\{\boldsymbol{p}\} \leftrightarrow \{\boldsymbol{k}\}$. In addition, as seen in the previous section, it must be symmetric separately under the interchanges $k_1 \leftrightarrow k_2$ and $p_1 \leftrightarrow p_2$.

We work in an expansion about the $N\pi\pi$ threshold, an approach that has been used in previous RFT works involving mesons and baryons; see, e.g., refs.~\cite{\dwave,Draper:2023xvu}. The leading two terms with the desired symmetry properties are
\begin{equation}
\cK_{\rm df,3}
\supset \cK_0 \left[\bar{u}_{p_3} u_{k_3}\right] +
{\cK_1} \left[
\bar{u}_{p_3}\gamma_\mu u_{k_3} \frac{P^\mu}{M_N+2M_\pi} - \bar{u}_{p_3} u_{k_3}
\right]\,.
\label{eq:K0SS}
\end{equation}
The subtraction in the second term is chosen so that it vanishes at threshold. A possible third term involving $p_3^\mu+k_3^\mu$ leads, using the Dirac equation, to a contribution to $\cK_0$.

Despite the simplicity of the $\cK_0$ term, its conversion into a contribution to the matrix $\cK_{\rm df,3}$ is nontrivial. The implementation is described in \Cref{app:K0decomp}.

Finally, we note that the coefficients in the expansion of $\kdf$ are {\em a priori} unknown, but they could be estimated using effective field theories, as has been done for mesons~\cite{Baeza-Ballesteros:2023ljl,Baeza-Ballesteros:2024mii}. This is left for future work.


\section{Integral equations determining \texorpdfstring{$\cM_3$}{the three-particle scattering amplitude}}
\label{sec:inteqs}

The second step in deriving the three-particle formalism is to obtain the integral equations that relate the two- and three-particle K matrices---obtained from fits to finite-volume spectra---to the three-particle amplitude $\widehat{\cM}_3$. In the RFT approach, the method is the same for all systems: one takes the ordered double limit of finite-volume amplitudes,
\begin{equation}
\widehat{\cM}_{3} = \lim_{\epsilon \to 0^+} \lim_{L\to \infty} \widehat{\cM}_{3,L} \,.
\label{eq:M3hatI}
\end{equation}
Here $\epsilon$ is the coefficient of the imaginary part introduced in the usual way in the denominators of $F_\pi$, $F_N$, $G_{\pi \pi}$, $G_{\pi N}$, $G_{N \pi}$ [quantities defined in \Cref{eq:FNt,eq:Fpit,eq:GNpit,eq:GpiNt,eq:Gpipit}]. In the $L\to\infty$ limit, the matrix expressions for $\widehat{\cM}_{3,L}$, given below, become integral equations, as explained in ref.~\cite{\HSQCb}. Methods to solve these integral equations are discussed in refs.~\cite{\IntegralEquations}.

The finite-volume amplitude $\widehat{\cM}_{3,L}$ is obtained from an unsymmetrized amplitude by a symmetrization procedure to be described below. The unsymmetrized amplitude has definite choices of final and initial spectators, in contrast to the full amplitude, for which there is no notion of a spectator. In particular, one can divide the contributions in an all-orders perturbative effective-field-theory calculation of the full amplitude into those with specific spectators. This partitioning is described in ref.~\cite{\HSQCb} in the Feynman-diagram-based approach, and in refs.~\cite{\BSQC,\BSnondegen} in the time-ordered perturbation theory (TOPT) approach. The derivations presented in these works lead to the following expression for the unsymmetrized amplitude in terms of the quantities appearing in the quantization condition:
\begin{equation}
\widehat{\cM}_{3,L}^{(u,u)} = \widehat{\cD}_L^{(u,u) }+ \widehat{\cM}_{\df,3,L}^{(u,u)}\,.
\label{eq:MhatuuI}
\end{equation}
Here $\widehat{\cD}_L^{(u,u)}$ is the ladder amplitude, containing pairwise rescattering,
\begin{align}
\widehat{\cD}_L^{(u,u)} &= - \widehat{\cM}_{2,L} \widehat G \widehat{\cM}_{2,L}
\frac1{1 + \widehat{G} \widehat{\cM}_{2,L} }\,,
\label{eq:DhatuuI}
\end{align}
while $\widehat{\cM}_{\df,3,L}^{(u,u)}$ is the contribution containing the three-particle K matrix,
\begin{align}
\widehat{\cM}_{\df,3,L}^{(u,u)} &= \left[ \frac13 - \left(\widehat{\cM}_{2,L} + \widehat{\cD}_{L}^{(u,u)}\right) \widehat F \right]
\widehat{\cK}_{\df,3} \frac1{1 + \widehat F_3 \widehat{\cK}_{\df,3} }
\left[\frac13 - \widehat F \left(\widehat{\cM}_{2,L} + \widehat{\cD}_{L}^{(u,u)}\right) \right]\,.
\label{eq:Mhatdf3LI}
\end{align}

We now turn to the symmetrization procedure. This consists of two steps: first, the quantities in the expressions above, which are matrices in the $i \bm p\ell m m_s$ basis, are combined with spherical harmonics (and summed over $\ell$) so as to create functions of momenta; and, second, the contributions from different choices of spectator are combined with appropriate flavor factors. The analysis in \Cref{app:inteqs} determines these flavor factors; here we simply report the results. We find
\begin{equation}
\cM_{3; m'_s,m_s}(\{\bm p\},\{\bm k\}) = \lim_{\epsilon\to 0^+}\lim_{L\to\infty}
\boldsymbol{\mathcal X}^{I=5/2} \circ
\widehat{\cM}_{3,L; m'_s, m_s }^{(u,u)}
\circ \boldsymbol{\mathcal X}^{I=5/2\,\dagger} \,,
\label{eq:M3I5}
\end{equation}
where we are using a notation introduced in ref.~\cite{\tetraquark}, which we now explain. We recall that $\widehat{\cM}^{(u,u)}_{3,L}$ is a $2\times 2$ matrix in flavor space, each element of which has indices $\bm p\ell m m_s$. In \Cref{eq:M3I5} we display only the $m_s$ indices, keeping $i$ and $\bm p\ell m$ implicit. The operator $\bcX^{I=5/2}$ is defined on the two-dimensional flavor space by
\begin{equation}
\bcX^{I=5/2} =
\left(\sqrt2 \XR312,\ \XR132+\XR231\right) \,,
\label{eq:XI5}
\end{equation}
and thus contracts the flavor indices of $\widehat{\cM}^{(u,u)}_{3,L}$. Each entry $\boldsymbol{\mathcal X}^{\boldsymbol \sigma}_{[kab]}$ acts on the $\bm p \ell m$ indices to give a function of momenta:\footnote{%
This differs from the definition in ref.~\cite{\tetraquark} because the spherical harmonic here is not complex conjugated. This follows similar changes to other quantities noted above. In addition, overall normalizations have been corrected.%
}
\begin{align}
\left[\boldsymbol{\mathcal X}_{[pab]}^{\boldsymbol \sigma} \circ f\right] (\{ \bm p \})
= \sqrt{4\pi}\left[\sum_{\ell m} Y_{\ell m}(\hat {\bm a}^\star) f_{p\ell m}
\right]_{\bm p\to \bm p_{\sigma_1} , \, \bm a\to \bm p_{\sigma_2}, \, \bm b \to \bm p_{\sigma_3}}
\,.
\label{eq:XRdef}
\end{align}
As in the definition of $\bcY^\dagger$ in \Cref{eq:YRdef}, $\boldsymbol \sigma$ is a permutation of $\{1,2,3\}$. The meaning of \Cref{eq:XRdef} is that the sum over $\ell m$ yields a function of the spectator momentum $\bm p$ and the direction of the primary member of the pair, $\hat {\bm a}^\star$, the latter defined in the pair's CMF. The momentum $\bm p$ is then equated to $\bm p_{\sigma_1}$, while $\bm p_{\sigma_2}$ is chosen to be the lab-frame momentum that, when boosted to the pair CMF, has the direction given by $\hat {\bm a}^\star$. Note that, for three on-shell momenta, this direction, along with the spectator momentum and the total momentum $\bm P$, completely determines $\bm p_{\sigma_2}$. The third momentum is then given by $\bm p_{\sigma_3}=\bm P - \bm p_{\sigma_1}-\bm p_{\sigma_2}$. The left-acting version $\boldsymbol{\mathcal X}_{[pab]}^{{\boldsymbol \sigma}\dagger}$, which enters $\boldsymbol{\mathcal X}^{I=5/2\,\dagger}$, is defined analogously,
\begin{align}
\left[f \circ \boldsymbol{\mathcal X}_{[pab]}^{\boldsymbol \sigma \dagger} \right] (\{ \bm k\})
= \sqrt{4\pi}\left[\sum_{\ell m} f_{p\ell m} Y^*_{\ell m}(\hat {\bm a}^\star)
\right]_{\bm p\to \bm k_{\sigma_1} , \, \bm a\to \bm k_{\sigma_2}, \, \bm b \to \bm k_{\sigma_3}}
\,.
\label{eq:XLdef}
\end{align}

Putting this all together, we see that the first entry in $\bcX^{I=5/2}$ corresponds to the nucleon being the spectator, since it sets $\bm p_3=\bm p$, while the second entry sums over the two choices in which the pion spectates. In this way the complete amplitude is built up.

One subtlety in this construction is that, in finite volume, the operators $\bcX^{\bm \sigma}_{[pab]}$ and $\bcX^{\bm \sigma \dagger}_{[pab]}$ lead to pair momenta that do not lie, in general, in the finite-volume set~\cite{\HSQCb,\BSnondegen,\tetraquark}. Thus the terms that are being combined in the matrix multiplication in \Cref{eq:M3I5} have different momentum arguments. These differences vanish, however, in the $L\to\infty$ limit, which is all that matters for obtaining the integral equations.


\section{Impact of \texorpdfstring{$N\pi \leftrightarrow N\pi\pi$}{Npi <-> Npipi} transitions for nonmaximal isospin}
\label{sec:dimer}

There is considerable interest in extending the formalism just presented to $N\pi\pi$ systems of nonmaximal isospin, i.e.~$I=1/2$ and $3/2$. This extension requires, however, taking into account the mixing with $N\pi$ states, which has to be addressed both in finite and infinite volume. A natural approach is to generalize the three-particle formalism to include two-to-three processes by explicitly incorporating two- and three-hadron states in the quantization condition. This has been carried out in the simplest setting with identical, spinless particles in ref.~\cite{Briceno:2017tce}. This extended formalism is, however, quite cumbersome, and furthermore requires generalization to include nondegenerate particles with spin.

In light of this, we have investigated an alternative way of dealing with $2\leftrightarrow3$ transitions, proposed and implemented in refs.~\cite{Romero-Lopez:2019qrt,Dawid:2021fxd,Dawid:2023jrj,Dawid:2023kxu,Briceno:2024txg,Hansen:2024ffk,Dawid:2024dgy}. In this approach, one of the particles forming the two-hadron state is a two-body bound state that appears as a subthreshold pole in the scattering amplitude of a two-particle subchannel. The three-body formalism with such a two-particle amplitude automatically incorporates $2 \leftrightarrow 3$ interactions. The particle--bound-state amplitude itself can be obtained by performing Lehmann-Symanzik-Zimmermann (LSZ) reduction on the three-particle amplitude after subthreshold analytic continuation in the bound-state channel. The method, which we refer to as the \emph{LSZ method}, holds both at the level of the finite-volume quantization condition, and for the integral equations describing infinite-volume scattering.

The LSZ method was first numerically investigated in toy systems of identical scalars~\cite{Romero-Lopez:2019qrt,Dawid:2021fxd,Dawid:2023jrj,Dawid:2023kxu,Briceno:2024txg}. Moreover, it has already been used for the $DD^*$ system at heavier-than-physical pion masses, for which the $D^*$ is stable~\cite{Dawid:2024dgy,Hansen:2024ffk,Dawid:2025wsn}. In such system, there are transitions of the form $DD\pi \leftrightarrow DD^*$, which can be accounted for by considering the $D^*$ meson as a bound-state pole in the $D\pi$ scattering amplitude.

The question we consider in this section is whether this alternative approach can be extended to the $N\pi\pi + N\pi$ system. The idea would make use of the fact that the nucleon is present as a subthreshold pole in the $I=1/2$ $p$-wave $N\pi$ amplitude. Since this amplitude is part of the three-particle $N\pi\pi$ formalism (as described in previous sections), the $N\pi\pi \to N\pi$ transitions would be present automatically if we ensured that this pole was included in the parametrization of the $N\pi$ amplitude. Furthermore, we could access the physical $N\pi\pi \to N\pi$, $N\pi \to N\pi\pi$, and $N\pi \to N\pi$ amplitudes from solutions to the integral equations by performing LSZ reduction at the subthreshold pole, as has been done successfully in refs.~\cite{Romero-Lopez:2019qrt,Hansen:2024ffk,Dawid:2024dgy,Dawid:2023jrj}.

\begin{figure}[h!]
\centering
\includegraphics[width=\textwidth]{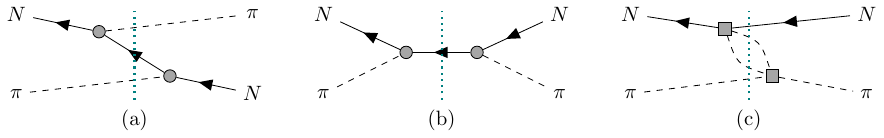}
\caption{Processes leading to left-hand singularities in the $N\pi$ amplitude: (a) $u$-channel nucleon exchange; (b) $s$-channel nucleon pole; (c) $t$-channel two-pion exchange. Nucleon lines are solid, while pion lines are dashed. Vertical dotted lines indicate the cuts leading to the singularities.}
\label{fig:LHC}
\end{figure}

Unfortunately, it turns out that this approach fails for the $N\pi\pi$ system. To explain the problem, it is useful to recall the leading left-hand singularities in the $N\pi$ amplitude, a clear discussion of which is given in the appendix of ref.~\cite{Doring:2009yv}. These singularities are due to the exchanges shown in \Cref{fig:LHC}. Starting at the threshold value of the $N\pi$ squared invariant mass, $\sigma_{N\pi}=(M_N+M_\pi)^2$, and moving to lower (subthreshold) values, one first encounters the left-hand cut due to $u$-channel nucleon exchange [\Cref{fig:LHC}(a)], which begins at $\sigma_{N\pi}= M_N^2+2M_\pi^2 $ and is present in all partial waves. Next there is the $s$-channel nucleon pole at $\sigma_{N\pi}=M_N^2$ [\Cref{fig:LHC}(b)], present only in the $I=1/2$ $p$ wave. Further below threshold one finds the left-hand cut due to $t$-channel two-pion exchange [\Cref{fig:LHC}(c)], present in all waves, which begins at $\sigma_{N\pi}= M_N^2-M_\pi^2$. Note that single-pion exchange is forbidden by G parity, given our assumption of exact isospin symmetry. Higher-order exchanges lead to additional singularities, but these lie even further below threshold.


\begin{figure}[h!]
\centering
\includegraphics[width=0.3\textwidth]{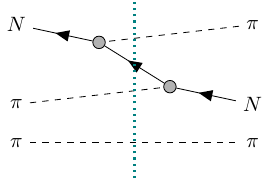}
\caption{$N\pi\pi$ diagram showing the need for a four-particle intermediate state (shown by the dotted vertical line) to account for the $u$-channel left-hand cut in an $N\pi$ subchannel. Notation as in \Cref{fig:LHC}.}
\label{fig:u-channel}
\end{figure}

The key issue is that the branch point of the $u$-channel left-hand cut lies closer to threshold than the nucleon pole. Thus, in order to reach the nucleon pole, as is required in the LSZ method, we must incorporate the cut into the formalism. We cannot simply circumvent the branch point, e.g.~by defining a contour from threshold to the nucleon pole in the complex plane, as the nonanalyticity leads to power-law finite-volume effects (for $\sigma_{N\pi} < M_N^2+2M_\pi^2$) that we must incorporate. Instead, we must explicitly include the corresponding diagram in the derivation. To do so in the $N\pi\pi$ system, however, one must consider intermediate four-particle $N\pi\pi\pi$ states, as shown in \Cref{fig:u-channel}. Thus, we conclude that the LSZ method fails for describing $N\pi \leftrightarrow N\pi\pi$ transitions.


To avoid the problem we must use a cutoff function that vanishes at or above the position of the $u$-channel left-hand branch point, as we have done in \Cref{sec:K2} above, and in the numerical studies of the following section. Thus the $s$-channel nucleon pole is hidden, and the alternative $2\leftrightarrow 3$ method is not available. Instead, the explicit inclusion of the $N\pi$ intermediate state is required, using a generalization of the methodology of ref.~\cite{Briceno:2017tce}.

In fact, there is a further, and arguably more fundamental, problem with applying the LSZ method to the $N\pi\pi$ system. In the course of the derivation sketched in \Cref{app:deriv}, one must assume that the two-particle TOPT Bethe-Salpeter (BS) kernels are nonsingular in the relevant kinematical region, since any singularities will lead to additional power-law finite-volume dependence. While in the case of a bound state, the BS kernel is nonsingular---the pole in the two-particle amplitude appearing only due to the inclusion of the contribution of two-particle intermediate states---the same does not hold when the singularity is due to one of the fundamental particles appearing in the $s$ channel. In the latter case, the singularity must be explicitly dealt with, which leads one back to a $2\leftrightarrow 3$ approach.

Although the restriction to $I=5/2$ avoids mixing with $N\pi$ states, diagrams involving $N\to N\pi$ vertices are present and introduce $N\pi\pi\pi$ intermediate states in the fully connected $3\to3$ amplitude even for $I=5/2$. Some of these diagrams contain subthreshold singularities, and it is necessary that the cutoff functions that we use avoid these singularities. This issue is discussed in \Cref{app:singularities}.

We close by noting that, in the $D D^* \leftrightarrow D D \pi$ case, the two-pion exchange is the source of the branch point nearest to threshold. This is further from threshold than the $D^*$ pole, and for this reason the LSZ method can be applied without issue~\cite{Dawid:2024dgy,Hansen:2024ffk,Dawid:2025wsn}.


\section{Numerical application}
\label{sec:num}

In this section, we present the results of a numerical implementation of the $I=5/2$ $N\pi\pi$ formalism. While this channel does not have a three-body resonance, it contains interesting dynamics as the $N\pi$ subchannel contains the $\Delta$ baryon. While the $\Delta$ is a resonance for physical quark masses, we choose a heavier-than-physical scenario in our illustrative implementation, which might be more accessible to near-term lattice QCD calculations. Specifically, we take $M_\pi = 300\;$MeV, $M_N = 1100\;$MeV, and $M_\Delta =1335\;$MeV. The latter two values are roughly based on recent lattice results for the given pion mass~\cite{Andersen:2017una,Brett:2018jqw,Bulava:2022vpq}. Note that, for these parameters, the $\Delta$ is stable with a binding energy of $65\;$MeV.

We have implemented the quantization condition using $\ell_{\rm max}=1$, which requires three different two-body subchannels: $s$-wave $I=2 \ \pi\pi$ scattering, and $s$- and $p$-wave $I=3/2 \ N\pi$ scattering. We choose the corresponding phase-shift parametrizations to be:
\begin{align}
\frac{k}{M_\pi} \cot \delta_0^{\pi\pi} & = \frac{M_\pi \sqrt{s}}{ s - 2M_\pi^2 } B_0, \\
\frac{k}{M_\pi} \cot \delta_0^{N\pi} & = - \frac{1}{M_\pi a_0^{N\pi}}, \\
\frac{k^3}{M^3_\pi} \cot \delta_1^{N\pi} & =\frac{6\pi \sqrt{s}}{M_\pi^3 g_{BW}^2} (s_\Delta- s),
\end{align}
with parameters:
\begin{equation}
B_0 = -4.73 , \quad M_\pi a^{N\pi}_0= 0.332, \quad g_{BW}=14.5 , \quad\frac{\sqrt{s_\Delta}}{M_\pi} = 4.5 .
\end{equation}
The first two parameters are taken from leading-order chiral perturbation theory, while the last two are inspired by lattice QCD works on the $\Delta$ resonance~\cite{Andersen:2017una,Brett:2018jqw,Bulava:2022vpq}. Given either $g_{BW}$ or $s_{\Delta}$, the other parameter is fixed by requiring that the $p$-wave $\pi N$ scattering amplitude has a pole at $M_\Delta$. The choice made here is consistent with the masses given above. We stress that the $\Delta$ is not included explicitly in the formalism, but rather appears as a bound-state pole in the $N\pi$ amplitude. This is analogous to previous work in the RFT approach on dimer-particle scattering~\cite{Romero-Lopez:2019qrt,Jackura:2020bsk,Dawid:2021fxd}, and the inclusion of the $D^*$ as a $D\pi$ bound state in the $DD\pi$ system~\cite{\tetraquark,\DRS}. As discussed in the previous section, this approach is valid as long as the $\Delta$ pole lies above the branch point due to $u$-channel nucleon exchange, a condition that is well satisfied here since the branch point lies at $1179 \;$MeV.

In the three-particle sector, we have no {\em a priori} information on the values of the parameters entering $\kdf$. We thus aim to explore the dependence of the spectrum on the three particle K matrix. We use the simple form given by the leading term in the threshold expansion, \Cref{eq:K0SS}. We vary the value of $\cK_0$ while setting $\cK_1=0$.

Details of the numerical implementation are similar to those for the $K\pi\pi$~\cite{Blanton:2021mih} and $DD\pi$~\cite{\DRS} systems, except for the complications arising from the additional spin index from the nucleon. This leads to changes that are nontrivial in the part of the quantization condition that involves the two-body K matrix, as described in \Cref{sec:K2,app:K2L}, and also for $\Kdf$, where the decomposition into the finite-volume variables is nontrivial. We describe this for the $\cK_0$ term in \Cref{app:K0decomp}.

We use here the smooth cutoff functions defined in eqs.~\eqref{eq:cutoff}, \eqref{eq:sigmaNmin}, and \eqref{eq:sigmapimin}. While these do not account for the additional singularities described in \Cref{app:singularities}, we do not expect that this will impact the qualitative features of the finite-volume spectrum described below. We leave a detailed study of the modified cutoff functions proposed in \Cref{app:newcutoff} for future work.

An additional complication is the projection to finite-volume irreducible representations (irreps). While the procedure is similar to that explained in section 3.1 of ref.~\cite{Blanton:2019igq}, the projectors have to be generalized to act in spin space. Specifically, we must use the doubled finite-volume groups~\cite{Morningstar:2013bda} and the corresponding fermionic irreps. The irreps containing energies that are affected by the $\cK_0$ term are described in \Cref{app:irreps}.


\begin{figure}[h!]
\centering
\includegraphics[width=\textwidth]{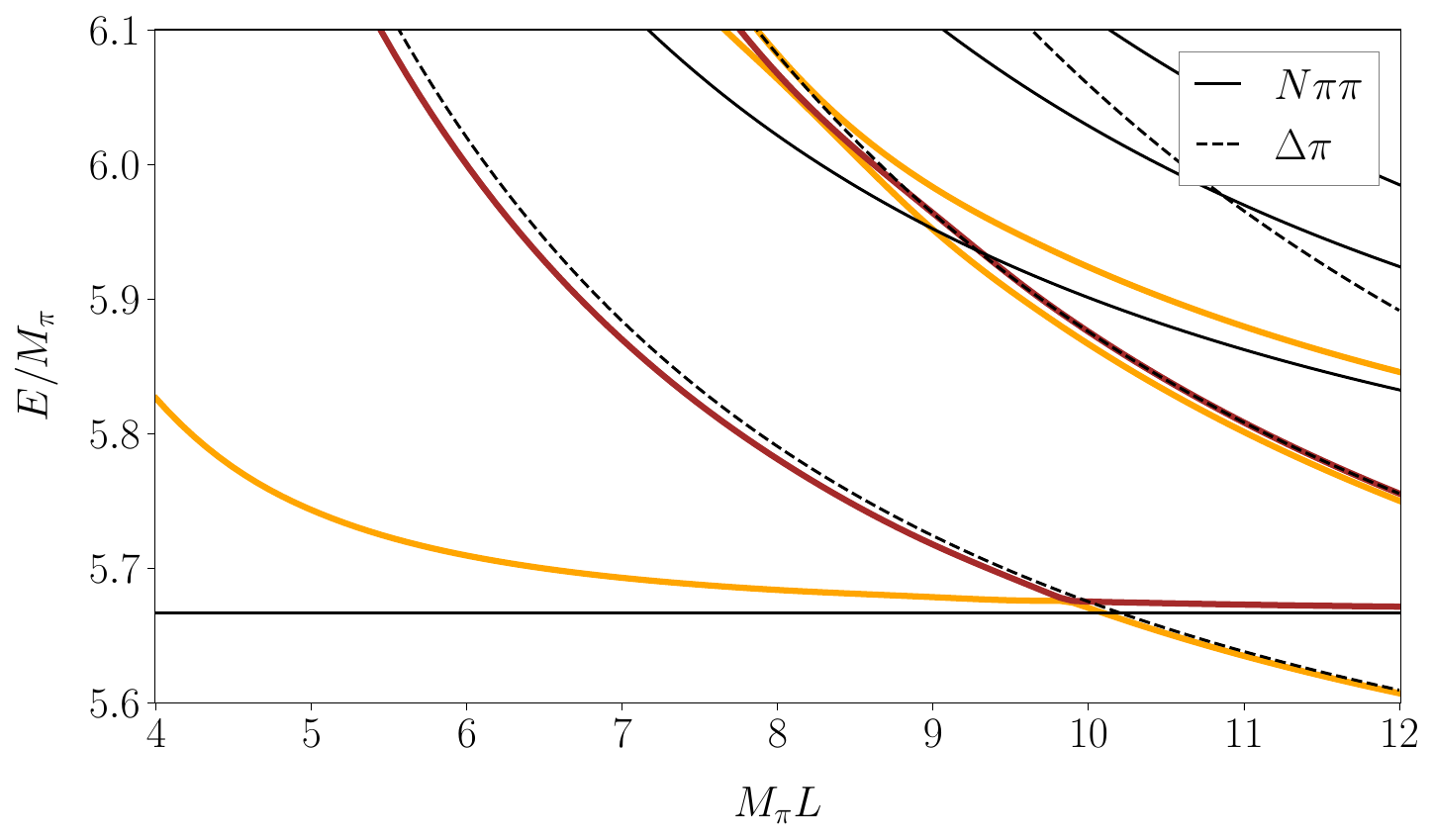}
\caption{Low-lying finite-volume spectrum of $I=5/2$ states with unit baryon number in the $G_{1g}$ irrep for $\boldsymbol P =0$.
Results are for the two-particle parameters described in the text and $\cK_0 = 0$. Results from the quantization condition are shown as solid, colored lines, with the alternating colors used to distinguish the levels but having no further significance. Noninteracting levels are shown by thin, black solid ($N\pi\pi$) and dashed ($\Delta\pi$) lines. The second $\Delta\pi$ noninteracting level is doubly degenerate.}
\label{fig:G1girrep}
\end{figure}

As a concrete example, we investigate the finite-volume spectrum in the $G_{1g}$ irrep for the rest frame ($\boldsymbol{P}=0$). In the absence of interactions, this irrep contains both $N\pi\pi$ states, including that in which all three hadrons are at rest, and $\Delta \pi$ states in a relative $p$ wave. This is also one of the irreps into which the $\cK_0$ term has a nonzero projection. The resulting five lowest interacting levels are shown in \Cref{fig:G1girrep} for $\cK_0=0$, along with the corresponding noninteracting energies. We display these for $M_\pi L \in [4,12]$, a range of box sizes that corresponds to $L \in [2.7,8]\;$fm.

Several interesting features arise. First, energy levels that qualitatively correspond to $N\pi\pi$ states experience a positive energy shift with respect to their non-interacting counterparts. This can be understood by the repulsive nature of $s$-wave $\pi\pi$ and $N\pi$ interactions. In addition, $\Delta\pi$-like levels have a negative shift that is smaller in magnitude. This, however, does not have a clear qualitative explanation as the $\Delta$ meson emerges dynamically as a bound state in the spectrum. The highest three levels correspond qualitatively to the second $N\pi\pi$ and $\Delta\pi$ noninteracting levels, with the latter having a multiplicity of two. This degeneracy is split by the interactions. Finally, we note the presence of avoided level crossings: one around $M_\pi L \sim 10$ for the two lowest states, and two additional crossings around $M_\pi L \sim 9$ involving three higher levels. These avoided crossings correspond to a change in the qualitative interpretation of these finite-volume levels from $N\pi\pi$-like to $\Delta \pi$-like and vice versa.

\begin{figure}[h!]
\centering
\includegraphics[width=\textwidth]{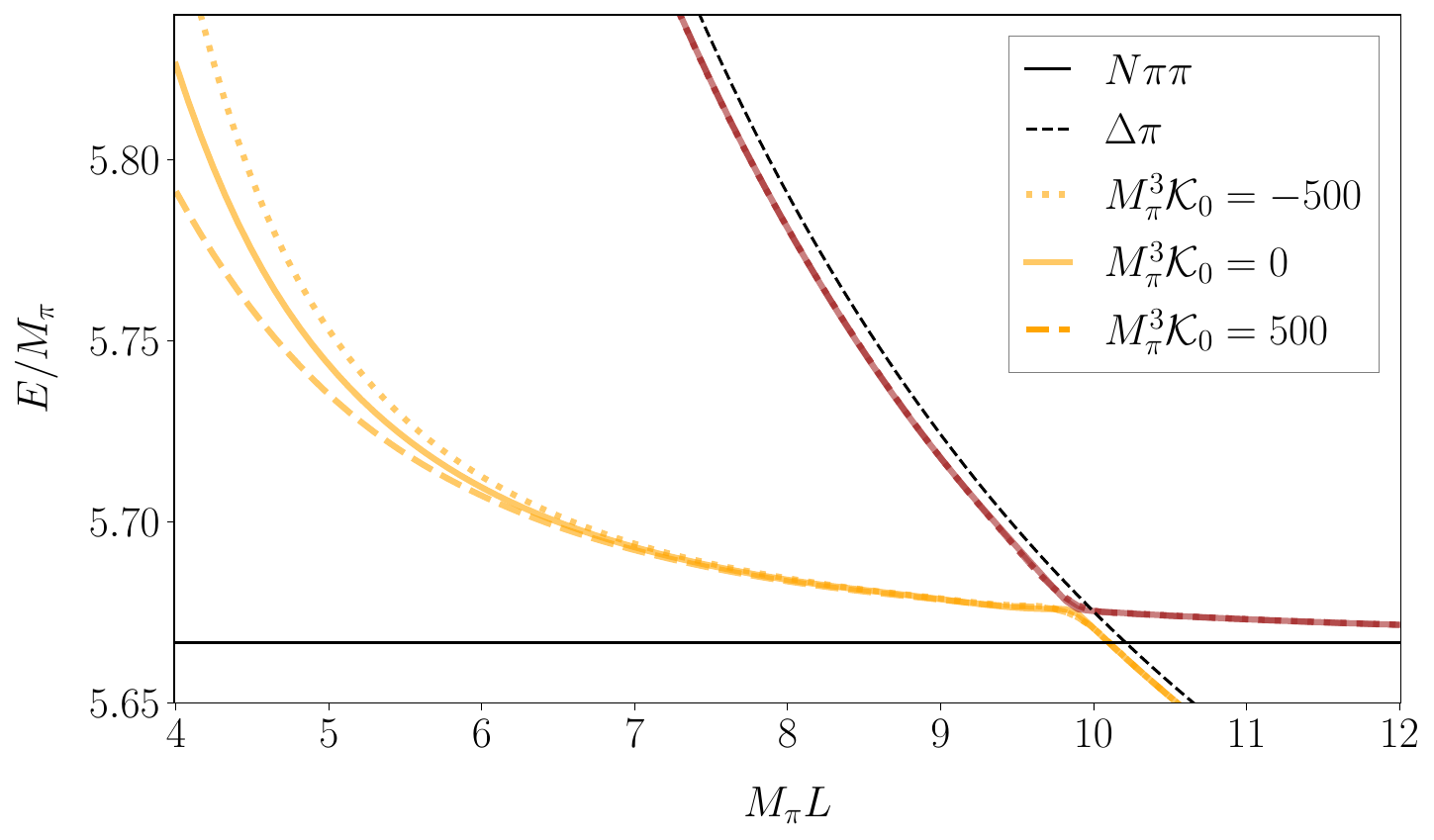}
\caption{Effect of $\cK_0$ in the lowest two states in the spectrum. Notation as in \Cref{fig:G1girrep}.}
\label{fig:G1girrepK0}
\end{figure}

In \Cref{fig:G1girrepK0} we examine the effect of $\kdf$ in the spectrum. To do so we zoom in on the two lowest energy levels. We observe that the effect of nonzero $\kdf$ is only visible at small volumes. This can be understood as a consequence of the relative $1/L^3$ suppression of three-body effects compared to two-body effects in finite volume. Moreover, only the lowest level is appreciably shifted by nonzero $\mathcal K_0$. This is expected as this level corresponds to $N \pi \pi$, all at rest, such that all particle pairs are in an $s$-wave. In contrast, $\Delta\pi$-like levels are dominated by a relative $p$-wave and thus the contribution of $\mathcal K_0$ is subleading. Higher order terms in $\kdf$ would be needed to shift this level more significantly.


\section{Summary and outlook}
\label{sec:concl}

In this work, we have taken an important step towards a finite-volume formalism for general $N\pi\pi$ systems. Specifically, we have derived and implemented the formalism for the maximal isospin sector, corresponding to $p \pi^+ \pi^+ \leftrightarrow \Delta^{++} \pi^+$ systems. Given the recent success in calculations involving the $\Delta$ resonance~\cite{Bulava:2022vpq,Alexandrou:2023elk}, this system appears to be a promising testbed for initial applications. We also note that, up to the differing particle masses, the formalism derived here applies to related baryon-meson-meson systems, such as $\Sigma^+ \pi^+ \pi^+$ or $p K^+ K^+$.

The implementation of our framework proceeds via the standard two-step procedure of the relativistic field-theoretic three-body approach. The first step, encoded in the quantization condition, relates the finite-volume energy spectrum to the K matrices. The second step uses integral equations to extract physical scattering amplitudes. To make these steps practical, parametrizations of the K matrices are required. Accordingly, in this work we have also determined the leading two terms in the threshold expansion of the three-body K matrix describing this system.

The inclusion of $N\pi \leftrightarrow N\pi\pi$ transitions remains the primary outstanding challenge in this framework. We have explored the possibility of incorporating the nucleon as a bound state in the $N\pi$ system; however, this approach fails due to the presence of a left-hand cut in the $N \pi$ amplitude with a branch point closer to threshold than the nucleon pole. This issue also rules out similar treatments for systems such as $NN\pi$. We therefore anticipate that the inclusion of such transitions will follow the strategy of ref.~\cite{Briceno:2017tce}, namely, the explicit inclusion an $N \pi$ flavor channel.

A first step towards such a complete analysis is included in \Cref{app:deriv}, where we derive the $N\pi\pi$ formalism for generic isospin under the assumption of no two-to-three mixing. While this is clearly an unphysical scenario, we expect that these equations will correspond to the $N\pi\pi \to N\pi\pi$ sub-block of the full $N\pi + N\pi\pi$ formalism.

As a demonstration, we have presented an implementation of the finite-volume $N\pi\pi$ spectrum in the maximal-isospin sector. To this end, we chose simple models for the two-body K matrix, guided by existing data and effective theories, to mimic a system with $M_\pi = 300$ MeV. Since the three-body K matrix is not known, we explored its impact on the finite-volume spectrum using the leading term in the threshold expansion of $\kdf$.

In the course of this study, we have realized that both the three-particle irreducible Bethe-Salpeter kernel and $\Kdf$ can have singularities, related to $N\pi\pi\pi$ intermediate states, that lie inside the allowed parameter space. This is possible because we allow pair invariant masses to go below their thresholds, and it is only in such subthreshold regions that the new singularities appear. We have made a study of the loci of these singularities in \Cref{app:singularities}, and suggested a modification to the cutoff functions that avoids them.

This previously unrecognized issue deserves further study, which we leave for future work. We do note that this is mainly an issue for nondegenerate systems, and, in particular, is absent for the case of three pions. It is present, however, for the $\pi\pi K$ and $KK\pi$ (and, for some values of $M_\pi/M_K$, also for $3K$) systems studied, for example, in refs.~\cite{Dawid:2025doq,Dawid:2025zxc}. In these cases involving pseudo-Goldstone bosons, however, the strength of the singularity is highly suppressed in chiral perturbation theory, since it involves at least two vertices derived from the Wess-Zumino-Witten term~\cite{Wess:1971yu,Witten:1983tw}. Thus it may not make a significant contribution in practice.

We have also realized that there are additional subtleties related to the nonanalyticity of the function $J(z_X)$ at $z_X=1$ [see \Cref{eq:cutoff}]. This is a general issue for the formalism, applicable to all processes, and we will address in a future publication.

Returning to $I=5/2$ $N\pi \pi$, we have not yet implemented the integral equations that relate the three-body K matrix to the infinite-volume scattering amplitude. This is left for future work and will require extending the partial-wave projection of the formalism's building blocks to systems involving spin.

This work provides a solid starting point for exploring baryon-meson-meson systems in finite volume. Given the recent success in lattice QCD three-body calculations~\cite{\ThreeBodyNumerics}, pioneering lattice QCD applications of the maximal-isospin $N\pi\pi$ system can now begin.
Looking ahead on the formal side, the general-isospin $N\pi\pi$ formalism, for which this work lays a clear path, will be instrumental in advancing our understanding of baryon resonances such as the $N(1440)$ (Roper). The nature of this state and the detailed mechanism by which its properties arise have been of great interest to the particle and nuclear theory communities for decades, and a fully systematic treatment in lattice QCD would represent a significant achievement.


\section*{Acknowledgments}

We thank Sebastian Dawid for extensive discussions about, and early collaboration on, this project, Sebastian Mizera for very helpful guidance on Landau equations, and Ra\'ul Brice\~no, Miguel Correia, Andrew Jackura, and Wyatt Smith for helpful discussions. SRS acknowledges financial support through the U.S. Department of Energy Contract No. DE-SC0011637. This work contributes to the goals of the USDOE ExoHad Topical Collaboration, contract DE-SC0023598. M.T.H.~is supported in part by UK STFC grants ST/X000494/1 and ST/T000600/1. M.T.H. is further supported by UKRI Future Leaders Fellowship MR/T019956/1.

\appendix


\section{Derivation of the formalism}
\label{app:deriv}

In this section we provide a sketch of the derivation of the formalism for the $I=5/2$, $N\pi\pi$ system, the results of which are presented in the main text in \Cref{sec:QC,sec:inteqs}. We additionally provide important partial results for the non-maximal isospin cases, $I=3/2$ and $I=1/2$. These are incomplete, because they ignore the coupling to $N\pi$ states. The full derivation of the formalism for these cases is left for future work.

We use the TOPT method introduced in ref.~\cite{\BSQC}, which was generalized to nondegenerate, spinless particles in ref.~\cite{\BSnondegen}. The addition of isospin degrees of freedom is similar to the case of $DD\pi$ studied in ref.~\cite{\tetraquark}, and we follow the logical development presented in appendix A of that work. We focus here on the new features that arise with $N\pi\pi$ in particular the fact that the nucleon has a spin degree of freedom, and the isospin wavefunctions that are specific to this system. We stress that, although the TOPT method involves intermediate Bethe-Salpeter kernels and K matrices that are not Lorentz invariant, the final quantization condition and integral equations (obtained using the symmetrization procedure described in \Cref{app:chtoiso}) do involve Lorentz-invariant K matrices. The argument for this is given in sec. XIII.F of ref.~\cite{\BSnondegen}.

This appendix is organized into six subsections. In the following subsection we summarize the relevant isospin states, then in section~\ref{app:FG} we introduce the finite-volume cuts and in sections~\ref{app:K2L}~and~\ref{app:Kdfflavor} the K matrices, $\cK_2$ and $\cK_{\df,3}$, respectively, that enter the quantization condition. In section~\ref{app:chtoiso} these building blocks are combined to give the final form of the quantization condition, including the rotation to the isospin basis. The key results are summarized in \cref{tab:FGK}. Finally, in section~\ref{app:inteqs} we review the derivation of the integral equations required to relate the K matrices to the scattering amplitudes.


\subsection{Overview and basis choices}
\label{app:bases}

The starting point in all RFT derivations is a finite-volume correlation function
\begin{equation}
C_{L,ij}(E, \bm P) = \int d^4 x \, e^{i E t - \bm P \cdot \bm x} \langle 0 | {\rm T} \cO_i(x) \cO^\dagger_j(0) | 0 \rangle_L\,,
\label{eq:CL}
\end{equation}
that involves the relevant three-particle creation and annihilation operators, $\cO^\dagger_j$ and $\cO_i$, respectively. We need a set of operators that couple to $N\pi\pi$ states with all choices of allowed isospin: $I=5/2$, $3/2$ and $1/2$ [see \Cref{eq:isospin_states}]. This is achieved by considering the three operators that destroy states with $m=1/2$. In momentum space, these are
\begin{multline}
\widetilde{\bm\cO} = \left\{\widetilde\cO_1,\widetilde\cO_2,\widetilde\cO_3\right\} = \left\{
\pi^+ (\bm k_1) \pi^- (\bm k_2) p(\bm k_3) ,\
\pi^0 (\bm k_1) \pi^0 (\bm k_2) p(\bm k_3) ,\
\pi^+ (\bm k_1) \pi^0 (\bm k_2) n(\bm k_3)
\right\}\,.\\[0pt]
\label{eq:underlying}
\end{multline}
These are composed of single-particle operators with definite finite-volume momenta $\bm k_i$, i.e. $\bm k_i \in (2\pi/L) \mathbb{Z}^3$ for a cubic periodic spatial volume with periodicity $L$ in each of the three spatial directions. The detailed form of these operators is not important in the following. In addition, for now we keep the spin degree of freedom in the nucleon fields implicit. We also note that the three operators correspond to the three choices of external states that can scatter into one another in the general $3\to 3$ scattering amplitude $\cM_3$. These are also readily identified as the three $N \pi \pi$ states with electromagnetic charge $Q=+1$. In the following, we sometimes refer to these three states as the ``underlying states''.

In the RFT formalism, three-particle states are characterized by choosing one of the particles as the ``spectator'' and the remaining two as the ``pair'' (sometimes denoted the ``dimer''). All the matrices that appear in the three-particle quantization condition have indices given by the (finite-volume) spectator momentum $\bm k$, and the spin $\ell$ and $z$-component $m$ of the dimer. In addition to the $\bm k \ell m$ indices, there is an index denoting the spin state of the nucleon, $m_s\in \{1/2, -1/2\}$. We define this is the lab frame, relative to a fixed $z$ axis, in a similar manner to the spin index of the spectator neutron in the three-neutron system studied in ref.~\cite{\threeneutron}.

The use of the spectator-pair decomposition implies that, for each of the states in \Cref{eq:underlying}, there are two or three choices of the flavor of the spectator, which introduces an enlarged flavor space. This extends the basis to $3+2+3=8$ dimensions, but here we choose to extend this further to the following 10-dimensional basis,
\begin{multline}
\bm v_{\rm ch} =
\Big\{ (\pi^+ \pi^-)_e \, p,\ \ (\pi^+ \pi^-)_o \, p,\ \ (p \pi^+)\,\pi^-,\ \ (p \pi^-)\, \pi^+,\ \
\\
(\pi^0 \pi^0) \, p,\ \ (p \pi^0) \,\pi^0,\ \
\\
(\pi^+ \pi^0)_e \, n,\ \ (\pi^+ \pi^0)_o \, n,\ \ (n \pi^+) \, \pi^0,\ \ (n \pi^0) \, \pi^+ \Big\} \,,
\label{eq:chargebasis}
\end{multline}
in which each line corresponds, respectively, to one of the three operators in \Cref{eq:underlying}. For each triplet, the two entries in parentheses form the dimer, with the third entry being the spectator. Within dimers of distinguishable particles the first entry is the ``primary'' member, and the direction of the primary's momentum in the CMF is used to define the decomposition into spherical harmonics. In addition, we decompose pairs containing two different pions into those having even and odd angular momenta (as indicated by the subscripts), which are, respectively, symmetric and antisymmetric under pion exchange. This is useful when connecting this basis, which we refer to as the charged basis, to that of isospin states, which we now discuss.

In the isospin basis we choose the pairs, as well as the triplets, to have definite isospin. The allowed choices are
\begin{multline}
\bm v_{\rm iso10} =
\Big\{[(\pi\pi)_2 N]_{5/2},\ \ [(N\pi)_{3/2} \pi]_{5/2}\,,
\\
[(\pi\pi)_2 N]_{3/2},\ \ [(\pi\pi)_1 N]_{3/2},\ \ [(N\pi)_{3/2} \pi]_{3/2},\ \ [(N\pi)_{1/2} \pi]_{3/2}\,,
\\
[(\pi\pi)_1 N ]_{1/2},\ \ [(\pi\pi)_0 N]_{1/2},\ \ [(N\pi)_{3/2} \pi]_{1/2},\ \ [(N\pi)_{1/2} \pi]_{1/2}
\Big\}\,,
\label{eq:isospinbasis}
\end{multline}
where subscripts indicate isospin. These triplets are defined for all allowed values of $m$, but we assume that $m=1/2$ in the following, as in \Cref{eq:underlying,eq:chargebasis}. Using standard isospin algebra, we find that the two bases are related as
\begin{equation}
\bm v_{\rm iso10} = C_{{\rm iso10} \leftarrow {\rm ch}} \bm v_{\rm ch}\,,
\end{equation}
with $C_{{\rm iso10} \leftarrow {\rm ch}}$ the orthogonal (therefore also unitary) matrix
\begin{equation}
C_{{\rm iso10} \leftarrow {\rm ch}} = \begin{pmatrix}
\frac{1}{\sqrt{5}} & 0 & 0 & 0 & \sqrt{\frac{2}{5}} & 0 & \sqrt{\frac{2}{5}} & 0 & 0 & 0 \\
0 & 0 & \frac{1}{\sqrt{10}} & \frac{1}{\sqrt{10}} & 0 & \sqrt{\frac{2}{5}} & 0 & 0 & \frac{1}{\sqrt{5}} & \frac{1}{\sqrt{5}} \\
-\sqrt{\frac{2}{15}} & 0 & 0 & 0 & -\frac{2}{\sqrt{15}} & 0 & \sqrt{\frac{3}{5}} & 0 & 0 & 0 \\
0 & \sqrt{\frac{2}{3}} & 0 & 0 & 0 & 0 & 0 & \frac{1}{\sqrt{3}} & 0 & 0 \\
0 & 0 & \sqrt{\frac{2}{5}} & -\frac{2}{3} \sqrt{\frac{2}{5}} & 0 & \frac13 \sqrt{\frac{2}{5}} & 0 & 0 & \frac{1}{3 \sqrt{5}} & -\frac{4}{3 \sqrt{5}} \\
0 & 0 & 0 & \frac{\sqrt{2}}{3} & 0 & \frac{\sqrt{2}}{3} & 0 & 0 & -\frac{2}{3} & -\frac{1}{3} \\
0 & -\frac{1}{\sqrt{3}} & 0 & 0 & 0 & 0 & 0 & \sqrt{\frac{2}{3}} & 0 & 0 \\
\sqrt{\frac{2}{3}} & 0 & 0 & 0 & -\frac{1}{\sqrt{3}} & 0 & 0 & 0 & 0 & 0 \\
0 & 0 & \frac{1}{\sqrt{2}} & \frac{1}{3 \sqrt{2}} & 0 & -\frac{\sqrt{2}}{3} & 0 & 0 & -\frac{1}{3} & \frac{1}{3} \\
0 & 0 & 0 & -\frac{2}{3} & 0 & \frac{1}{3} & 0 & 0 & -\frac{\sqrt{2}}{3} & \frac{\sqrt{2}}{3}
\end{pmatrix}\,.
\label{eq:Cisotoch}
\end{equation}

The derivation of the quantization condition now proceeds by writing down an all-orders expression in TOPT for the correlation function in~\Cref{eq:CL}, and then applying several resummations. An essential intermediate step is that the sum of all diagrams can be expressed as a geometric series involving two-particle irreducible (2PI) $2 \to 2$ Bethe--Salpeter kernels, as well as three-particle irreducible (3PI) $3 \to 3$ kernels. In both cases, irreducibility refers to cuts through either two- or three-particle states in the total energy channel (the $s$-channel) of the system.\footnote{%
In more detail, we define an $s$-channel set as any collection of $n$ internal propagators whose total energy and momentum sum to $(E,\bm{P})$ for every loop-momentum configuration. A diagram is said to be $n$-particle reducible in the $s$-channel if it falls into two or more disconnected components once an $s$-channel set is removed. If a diagram is neither one- nor two-particle reducible in the $s$-channel, we say it is two-particle irreducible (2PI) in that channel. Similarly, if a diagram is neither one-, two-, nor three-particle reducible in the $s$-channel, we say it is three-particle irreducible (3PI). In the original RFT formalism for three identical particles, it was necessary to include diagrams with a single $s$-channel propagator in the definition of the $3 \to 3$ kernel. That case does not arise here, however, since neither a single nucleon nor a single pion carries the quantum numbers of the $I=5/2$ $N\pi\pi$ system of interest.%
}
One can then use the fact that these kernels do not contain singularities over a range of kinematics to argue that they differ from their infinite-volume counterparts only by exponentially suppressed $L$-dependence. From this point, one analyzes the remaining finite-volume effects and expresses the correlation function as a geometric series that can be resummed into a closed form. The required algebraic steps are as in previous works---see, in particular, refs.~\cite{\BStwoplusone,\tetraquark,\multichannel}---and we do not repeat them. It is most straightforward to work first in the charged basis, and then convert at the end to the isospin basis. This approach also provides a built-in cross-check, as it must be the case that the final result is block-diagonal in isospin.

In particular, the correlation function in the charged basis can be expressed as
\begin{align}
C_{L,ij}(E, \boldsymbol P) - C_{\infty, ij}(E, \boldsymbol P) & = \widehat A_i \widehat F_3 \frac{1}{1+ \widehat{\mathcal K}_{\rm df,3} \widehat{F}_3} \widehat {A}^\dagger_j \,,
\label{eq:CLcharged}
\end{align}
where $\widehat F_3$ and $\widehat{\mathcal K}_{\rm df,3}$ are matrices in the $\boldsymbol k \ell m$ space as well as the charged-basis flavor space, and have dimension $10\times 10$ in the latter.\footnote{%
In this appendix, quantities with carets (``hats''), such as $\widehat F_3$, are matrices in the $10\times 10$ flavor space. This is in contrast to the main text, where the same symbols are used to represent $2\times 2$ matrices in the $I=5/2$ flavor space. No confusion should arise, as we do not use these conflicting notations in the same sections of the paper.%
}
The quantity $\widehat F_3$ is defined in terms of $\widehat F$, $\widehat G$ and $\widehat{\mathcal K}_{2,L}$ as in \cref{eq:F3} and each of these matrices are defined on the same space. Crucially, this result assumes that the coupling to $N \pi$ states is zero so that a $N \pi \to N \pi \pi$ transition is not included. This is the case for the $I=5/2$ channel, but not for $I=3/2$ or $I=1/2$. The vector $\widehat A_i$ gives the overlap of $\langle 0 \vert \mathcal O_i^\dagger$ onto the three-particle states and ${\widehat A}^\dagger_j$ is its adjoint. In the following we discuss the individual building blocks of the quantization condition in detail.


\subsection{\texorpdfstring{$\widehat F$ and $\widehat G$}{F and G} in the charged basis}
\label{app:FG}

We present $\widehat F$ and $\widehat G$ via their sum, $\widehat {\textbf C}_{FG} =\widehat F+\widehat G$. In some forms of the quantization condition, only this combination appears while in others, such as the one presented in the main text, $\widehat F$ and $\widehat G$ appear separately. In this appendix, while we give all equations in terms of $\widehat{\textbf C}_{FG}$, the individual terms can be easily separated by keeping track of the letters ($F$ and $G$) appearing in matrix entries in the various equations.

In the charged basis, the combined cut factor has the form
\begin{equation}
\widehat {\textbf C}_{FG} = {\rm diag}\left(
\textbf{C}_{FG}^{4}, \textbf{C}_{FG}^{2+1}, \textbf{C}_{FG}^{4} \right)\,,
\label{eq:FG0}
\end{equation}
with its block-diagonal form resulting from the fact that cuts cannot change the particle content. The entries in \Cref{eq:FG0} are
\begin{align}
\textbf C_{FG}^{4} &=
\begin{pmatrix}
\mathbb{P}_e F_{N} \mathbb{P}_e & 0 & \mathbb{P}_e G_{N\pi} & \mathbb{P}_e G_{N\pi}
\\
0 &\mathbb{P}_o F_{N} \mathbb{P}_o & - \mathbb{P}_o G_{N\pi} & \mathbb{P}_o G_{N\pi}
\\
G_{\pi N} \mathbb{P}_e & - G_{\pi N} \mathbb{P}_o & F_\pi & \mathbb{P}_\ell G_{\pi\pi} \mathbb{P}_\ell
\\
G_{\pi N} \mathbb{P}_e & G_{\pi N} \mathbb{P}_o & \mathbb{P}_\ell G_{\pi\pi} \mathbb{P}_\ell & F_\pi
\end{pmatrix}\,,
\label{eq:FG4d}
\\
\textbf C_{FG}^{2+1} &= \begin{pmatrix}
\mathbb{P}_e F_{N} \mathbb{P}_e & \sqrt{2} \mathbb{P}_e G_{N \pi}
\\
\sqrt{2} G_{\pi N} \mathbb{P}_e & F_\pi + \mathbb{P}_\ell G_{\pi\pi} \mathbb{P}_\ell \end{pmatrix}\,.
\label{eq:FG2p1}
\end{align}
where $\mathbb{P}_e$ and $\mathbb{P}_o$ are projectors onto even and odd partial waves of the pair, respectively, while $\mathbb{P}_\ell = \mathbb{P}_e-\mathbb{P}_o$ [and is given explicitly in \Cref{eq:Pell}].

The explicit forms of these matrices are as follows. $F_N$ denotes the non-switch cut factor in which the nucleon is the spectator, implying that the pair consists of two pions. It is given by
\begin{multline}
\big[ F_{N}(E, \boldsymbol P, L)\big]_{\boldsymbol p' \ell' m' m'_s ;\boldsymbol p \ell m m_s} =
\delta_{\bm p' \bm p} \delta_{m'_s m_s} \frac{H_N(\bm p)}{2\omega_{N}(\boldsymbol p) L^3}
\left[ \frac1{L^3} \sum_{\bm a} - \PV \int \frac{{\rm d}^3 \bm a}{(2\pi)^3} \right]
\\
\times \left[
\frac{\cY^*_{\ell' m'}(\bm a_N^\star(\boldsymbol p))}{\big(q_N^\star(\boldsymbol p)\big)^{\ell'}}
\frac{h(\bm a_N^{\star}(\bm p))}{4\omega_{\pi}(\boldsymbol a) \omega_{\pi}(\boldsymbol b)
\big(E\!-\!\omega_{N}(\boldsymbol p)\!-\!\omega_{\pi}(\boldsymbol a)\!-\!\omega_{\pi}(\boldsymbol b)\big)}
\frac{\cY_{\ell m}(\bm a_N^\star(\boldsymbol p))}{\big(q_N^\star(\boldsymbol p)\big)^{\ell}}
\right]
\,.
\label{eq:FNt}
\end{multline}
This is the form for nondegenerate particles given in refs.~\cite{\BSnondegen,\BStwoplusone}, with the addition of a Kronecker delta for the spin degree of freedom, except that we adopt the convention for complex conjugation of spherical harmonics proposed in ref.~\cite{\tetraquark}. The remaining notation is as follows: The sum over $\bm{a}$ extends over the finite-volume set $\{\, 2\pi \bm{n}/L \;\vert\; \bm{n} \in \mathbb{Z}^3 \,\}$, while the label $\mathrm{PV}$ denotes that the integral is evaluated using the principal-value pole prescription. $(E, \bm P)$ is the four-momentum running through the correlation function, with $\bm P$ drawn from the finite-volume set. The sum over $\bm a$ runs over the finite-volume set, and the ``third'' momentum is given by $\bm b= \bm P - \bm p -\bm a$.
On shell energies are given by
\begin{equation}
\omega_{N}(\boldsymbol p) = \sqrt{ \boldsymbol p^2 + M_N^2}\,,\quad
\omega_{\pi}(\boldsymbol p) = \sqrt{ \boldsymbol p^2 + M_\pi^2}\,.
\label{eq:omegadef}
\end{equation}
The momentum $\bm a_N^\star(\boldsymbol p)$ is the spatial part of the on-shell four-momentum $(\omega_\pi(\boldsymbol a), \boldsymbol a)$ after boosting to the CMF of the pair, which requires a boost velocity
\begin{equation}
\boldsymbol{\beta}_{\pi \pi}(\boldsymbol{p}) = -\frac{\boldsymbol{P} - \boldsymbol{p}}{E-\omega_{N}(\boldsymbol{p})} \,.
\end{equation}
The star in $\bm a_N^\star(\boldsymbol p)$ denotes a quantity evaluated in the pair CMF, while the subscript $N$ indicates that the nucleon is the spectator. Two other such quantities are
\begin{equation}
q_N^\star(\boldsymbol p) = \left(\frac{\lambda(\sigma_N(\boldsymbol p),M_\pi^2,M_\pi^2)}{4 \sigma_N(\boldsymbol p)}\right)^{1/2}\,, \qquad
\sigma_N(\boldsymbol p) = (E-\omega_N(\boldsymbol p))^2 - (\boldsymbol P-\boldsymbol p)^2 \,,
\label{eq:qpstarN}
\end{equation}
with $\lambda(a,b,c)=a^2+b^2+c^2-2ab-2ac-2bc$ the standard triangle function. $q_N^\star(\boldsymbol p)$ is the momentum of each member of the pair in their CMF if both are on shell, while $\sigma_N(\bm p)$ is the total squared invariant mass of the pair. Harmonic polynomials are defined as
\begin{equation}
\cY_{\ell m}(\bm a) = \sqrt{4\pi}\, \vert \bm a \vert^\ell\, Y_{\ell m}(\hat {\bm a}) \,.
\label{eq:harmonicpolynomials}
\end{equation}
We have also introduced $h(\bm a_N^{\star}(\bm p))$ as a smooth cutoff function for the sum-integral difference. One has a great deal of freedom in the definition of this cutoff. An example is given in ref.~\cite{Kim:2005gf}
\begin{equation}
h(\bm a_N^{\star}(\bm p)) = \exp \left (- \alpha \left [ \bm a_N^{\star}(\bm p)^2 - q_N^{\star}(\bm p)^2 \right ] \right )\,,
\end{equation}
where $\alpha$ is a small positive constant. The key requirements are that $h(\bm a_N^{\star}(\bm p))=1$ at the pole, and that it is a smooth function of $\boldsymbol a$ that is sufficiently slowly varying so that it does not lead to enhanced finite-volume effects. The smallness of $\alpha$ is relevant for this final point. Finally, $H_X(\bm p)$ is a smooth cutoff function given in \Cref{eq:cutoff} in the main text.

The result for the cut factor with a pion spectator is similar to that for $F_{N}$,
\begin{multline}
\big[ F_{\pi}(E, \boldsymbol P, L)\big]_{\boldsymbol p' \ell' m' m'_s ;\boldsymbol p \ell m m_s} =
\delta_{\boldsymbol p' \boldsymbol p} \delta_{m'_s m_s} \frac{H_{\pi}(\boldsymbol p)}{2\omega_{\pi}(\boldsymbol p) L^3}
\left[ \frac1{L^3} \sum_{\boldsymbol a} - \PV \int \frac{{\rm d}^3 \boldsymbol a}{(2\pi)^3} \right]
\\
\times \left[
\frac{\cY^*_{\ell' m'}(\boldsymbol a_\pi^\star(\boldsymbol p))}{\big(q_\pi^\star(\boldsymbol p)\big)^{\ell'}}
\frac{h(\bm a_\pi^{\star}(\bm p))}{4\omega_{N}(\boldsymbol a) \omega_{\pi}(\boldsymbol b)
\big(E\!-\!\omega_{\pi}(\boldsymbol p)\!-\!\omega_{N}(\boldsymbol a)\!-\!\omega_{\pi}(\boldsymbol b)\big)}
\frac{\cY_{\ell m}(\boldsymbol a_\pi^\star(\boldsymbol p))}{\big(q_\pi^\star(\boldsymbol p)\big)^{\ell}}
\right]
\,,
\label{eq:Fpit}
\end{multline}
except that flavor labels for the spectator and primary member of the pair are interchanged. The new quantities that appear are
\begin{equation}
q_\pi^{\star}(\boldsymbol p) = \left(\frac{\lambda(\sigma_\pi(\bm p),M_N^2,M_\pi^2)}{4 \sigma_\pi(\bm p)}\right)^{1/2}\,,\qquad
\sigma_\pi(\boldsymbol p) = (E-\omega_\pi(\boldsymbol p))^2 - (\boldsymbol P-\boldsymbol p)^2
\end{equation}
and $\bm a_\pi^{\star}(\boldsymbol p)$, the latter being the spatial part of the four momentum $(\omega_N(\boldsymbol a),\bm a)$, after boosting to the CMF of the pair.

We note that we are choosing the nucleon to be the primary member of the pair, i.e. it is the nucleon's (boosted) momentum that enters into the argument of the harmonic polynomials, rather than that of the pion. This choice matters for distinguishable particles if either $\ell'$ or $\ell $ is odd. We also stress that the Kronecker delta for the spin index is unchanged from $F_{N}$. This is because the spin index is defined in the lab frame, and is conserved by the free propagators that cross the cut.

The remaining quantities in \Cref{eq:FG4d,eq:FG2p1} are the cut factors associated with switches, the $G_{XY}$. Following Refs.~\cite{\BSnondegen,\BStwoplusone}, these are given by
\begin{align}
\hspace{-10pt}\big[ G_{\pi\pi}\big]_{\boldsymbol p \ell' m' m_s';\boldsymbol r \ell m m_s} &=
\frac{\delta_{m_s' m_s}}{2\omega_{\pi}(\boldsymbol p) L^3}
\frac{\cY^*_{\ell' m'}(\boldsymbol r_\pi^\star(\boldsymbol p))}{\big(q_{\pi}^\star(\boldsymbol p)\big)^{\ell'}}
\frac{H_{\pi}(\boldsymbol p) H_{\pi}(\boldsymbol r)}{b_{\pi\pi}^2-M_N^2}
\frac{\cY_{\ell m}(\boldsymbol p_\pi^\star(\boldsymbol r))}{\big(q_{\pi}^\star(\boldsymbol r)\big)^\ell}
\frac1{2\omega_{\pi}(\boldsymbol r) L^3}\,,
\label{eq:Gpipit}
\\
\hspace{-10pt}\big[ G_{\pi N}\big]_{\boldsymbol p \ell' m' m_s';\boldsymbol r \ell m m_s} &=
\frac{\delta_{m_s' m_s}}{2\omega_{\pi}(\boldsymbol p) L^3}
\frac{\cY^*_{\ell' m'}(\boldsymbol r_\pi^\star(\boldsymbol p))}{\big(q_{\pi}^\star(\boldsymbol p)\big)^{\ell'}}
\frac{H_{\pi}(\boldsymbol p) H_{N}(\boldsymbol r)}{b_{\pi N}^2-M_\pi^2}
\frac{\cY_{\ell m}(\boldsymbol p_N^\star(\boldsymbol r))}{\big(q_{N}^\star(\boldsymbol r)\big)^\ell}
\frac1{2\omega_{N}(\boldsymbol r) L^3}\,,
\label{eq:GpiNt}
\\
\hspace{-10pt}\big[ G_{N\pi}\big]_{\boldsymbol p \ell' m' m_s';\boldsymbol r \ell m m_s} &=
\frac{\delta_{m_s' m_s}}{2\omega_{N}(\boldsymbol p) L^3}
\frac{\cY^*_{\ell' m'}(\boldsymbol r_N^\star(\boldsymbol p))}{\big(q_{N}^\star(\boldsymbol p)\big)^{\ell'}}
\frac{H_{N}(\boldsymbol p) H_{\pi}(\boldsymbol r)}{b_{N\pi}^2-M_\pi^2}
\frac{\cY_{\ell m}(\boldsymbol p_\pi^\star(\boldsymbol r))}{\big(q_{\pi}^\star(\boldsymbol r)\big)^\ell}
\frac1{2\omega_{\pi}(\boldsymbol r) L^3}\,,
\label{eq:GNpit}
\end{align}
where the four-momentum of the exchanged particle is given, in general, by
\begin{equation}
b_{XY}=(E-\omega_{X}(\bm p)-\omega_{Y}(\bm r), \bm P - \bm p - \bm r)\,.
\end{equation}


\subsection{\texorpdfstring{$\widehat{\cK}_{2,L}$}{The two-particle K matrix} in the charged basis}
\label{app:K2L}

In the charged basis, we find the diagonal entries of $\widehat \cK_{2,L}$ to be
\begin{multline}
\bigg \{ \cK_{2,L} \big ([\pi^+\pi^-]_e\big ),\ \ \cK_{2,L} \big ([\pi^+\pi^-]_o\big ),\ \
\cK_{2,L} \big (p\pi^+\big ),\ \ \cK_{2,L} \big (p \pi^-\big ),\ \
\\
\tfrac12 \cK_{2,L} \big (\pi^0 \pi^0\big ),\ \ \cK_{2,L} \big (p\pi^0\big ),\ \
\\
\cK_{2,L} \big ([\pi^+\pi^0]_e\big ),\ \ \cK_{2,L} \big ([\pi^+\pi^0]_o\big ),\ \
\cK_{2,L} \big (n\pi^+\big ),\ \ \cK_{2,L} \big (n \pi^0\big )
\bigg \} \,,
\end{multline}
where the $1/2$ is a symmetry factor. The definition of $\cK_{2,L}$ is exemplified by~\cite{\tetraquark,\multichannel}
\begin{equation}
\left [ \cK_{2,L} \big (p\pi^- \big ) \right ]_{\bm p'\ell' m' m'_s; \bm p\ell m m_s} =
2 \omega_\pi(\bm p) L^3 \delta_{\bm p' \bm p}\, \cK_2 \big (p \pi^- \leftarrow p \pi^- \big )_{\ell' m' m'_s;\ell m m_s}\,,
\label{eq:K2Ldef}
\end{equation}
with $\cK_2$ the infinite-volume two-particle K matrix, the explicit form of which is discussed in \Cref{sec:K2}.

There are three pairs of offdiagonal elements
\begin{align}
\left[\widehat \cK_{2,L}\right]_{1,5} &= \left[\widehat \cK_{2,L}\right]_{5,1} = \tfrac{1}{\sqrt{2}} \cK_{2,L} \big ( [\pi^+\pi^-]_e \leftarrow \pi^0 \pi^0 \big ) \,,
\\
\left[\widehat \cK_{2,L}\right]_{6,9} &= \left[\widehat \cK_{2,L}\right]_{9,6} = \cK_{2,L } \big (p\pi^0 \leftarrow n \pi^+ \big )\,,
\\
\left[\widehat \cK_{2,L}\right]_{4,10} &= \left[\widehat \cK_{2,L}\right]_{10,4} = \cK_{2,L} \big (p\pi^- \leftarrow n \pi^0 \big )\,,
\end{align}
where the equality of transposed elements follows from PT symmetry.

It is useful to rewrite these quantities in terms of those in an isospin basis. From appendix D of ref.~\cite{\multichannel} we have the following results for the underlying K matrices,
\begin{align}
\cK_2([\pi^+\pi^-]_e) &= \tfrac{1}{6} \Big [ \cK_2(\pi\pi, I=2) + 2 \cK_2(\pi\pi, I=0) \Big ]\,,
\\[5pt]
\cK_2(\pi^0\pi^0) &= \tfrac{1}{3} \Big [2 \cK_2(\pi\pi, I=2) + \cK_2(\pi\pi, I=0) \Big ]\,,
\\[5pt]
\cK_2([\pi^+\pi^-]_o) &= \tfrac{1}{2} \cK_2(\pi\pi, I=1)\,,
\\[5pt]
\cK_2(\pi^0\pi^0 \leftrightarrow [\pi^+\pi^-]_e ) &= \tfrac{1}{3} \Big [ \cK_2(\pi\pi, I=2) - \cK_2(\pi\pi, I=0)\Big ]\,,
\\[5pt]
\cK_2(p \pi^+) &= \cK_2(N\pi, I=3/2) \,,
\\[5pt]
\cK_2(p \pi^0) &= \cK_2(n \pi^0) = \tfrac{1}{3} \Big [2 \cK_2(N\pi, I=3/2) + \cK_2(N\pi, I=1/2)\Big ]\,,
\\[5pt]
\cK_2(p \pi^-) &= \cK_2(n \pi^+) = \tfrac{1}{3} \Big [ \cK_2(N\pi, I=3/2) + 2 \cK_2(N\pi, I=1/2)\Big ]\,,
\\[5pt]
\begin{split}
\cK_2(p \pi^0 \leftrightarrow n\pi^+) &= \cK_2(p \pi^- \leftrightarrow n\pi^0)
\\
& =
\tfrac{\sqrt{2}}{3} \Big [ \cK_2(N\pi, I=3/2) - \cK_2(N\pi, I=1/2)\Big ]\,.
\end{split}
\end{align}
In addition we need the following related results,
\begin{align}
\cK_2([\pi^+\pi^0]_e) &= \tfrac{1}{2} \cK_2(\pi\pi, I=2)\,,
\\
\cK_2([\pi^+\pi^0]_o) &= \tfrac{1}{2} \cK_2(\pi\pi, I=1)\,,
\end{align}
which follows from the normalization of the states.


\subsection{\texorpdfstring{$\widehat{\cK}_{\rm df,3}$}{The three-particle K matrix} in the charged basis}
\label{app:Kdfflavor}

We can write down the structure of $\Kdf$ in the charged basis using previous work on distinguishable systems~\cite{\BSnondegen,\tetraquark} and 2+1 systems~\cite{\BStwoplusone}. The result is
\begin{equation}
\wh{\cK}_{\rm df,3} =
\begin{pmatrix}
\bcY^{(4)} \circ \cK_{11} \circ \bcY^{(4)\dagger} & \bcY^{(4)} \circ \cK_{12} \circ \bcY^{(2)\dagger} & \bcY^{(4)}\circ \cK_{13} \circ \bcY^{(4)\dagger}
\\
\bcY^{(2)}\circ \cK_{21}\circ \bcY^{(4)\dagger} & \bcY^{(2)}\circ \cK_{22} \circ\bcY^{(2)\dagger} & \bcY^{(2)}\circ \cK_{23}\circ \bcY^{(4)\dagger}
\\
\bcY^{(4)}\circ \cK_{31}\circ \bcY^{(4)\dagger} & \bcY^{(4)}\circ \cK_{32}\circ \bcY^{(2)\dagger} & \bcY^{(4)}\circ \cK_{33} \circ\bcY^{(4)\dagger}
\end{pmatrix}\,,
\label{eq:Kdfch}
\end{equation}
where all indices are implicit. The row vectors are given by
\begin{align}
\bcY^{(4)\dagger} &=
\begin{pmatrix}
\tfrac12(\YL312+\YL321),\ & \tfrac12( \YL312-\YL321),\ & \YL231,\ & \YL132
\end{pmatrix}\,,
\\[5pt]
\bcY^{(2)\dagger} &=
\begin{pmatrix}
\sqrt{\tfrac18}(\YL312+\YL321),\ & \tfrac12(\YL132+\YL231)
\end{pmatrix}\,,
\end{align}
in terms of the operators $\bcY^{[kab]}_{\boldsymbol\sigma}$ that are defined around \Cref{eq:YRdef}. The column vectors are obtained by hermitian conjugation. The $\cK_{ij}$ are underlying three-particle K matrices
\begin{equation}
\cK_{ij} \equiv \cK_{ij;\; m'_s, m_s}(\{\bm p\}, \{\bm k\})
\end{equation}
expressed as functions of the momenta and spin indices, using the notation introduced in \Cref{sec:Kdfflavor}. The labels $i,j=1,3$ refer to the three states in the underlying basis, \Cref{eq:underlying}. For $i\ne j$, $\cK_{ij}$ is related to $\cK_{ji}$ by PT symmetry. We stress that these underlying K matrices are ``symmetrized'', in the sense that they include all contributions and do not single out particular spectator particles.

Isospin symmetry implies relations between the $\cK_{ij}$, although we do not display these explicitly.


\subsection{Final quantization condition and converting to the isospin basis}
\label{app:chtoiso}

The TOPT approach leads to an explicit expression for the finite-volume correlation function---again, see refs.~\cite{\BStwoplusone,\tetraquark,\multichannel} for examples---from which one can read off the so-called asymmetric form of the three-particle quantization condition. This involves three-particle K matrices that are unsymmetrized, and are analogous to the unsymmetrized amplitudes $\cM_{\df,3}^{(u,u)}$ appearing in \Cref{sec:inteqs}. We do not quote this form of the quantization condition. Instead, we use the symmetrization procedure introduced in ref.~\cite{\BSQC}, and simplified in appendix G of ref.~\cite{\multichannel}, which leads to the so-called symmetric form of the quantization condition,
\begin{align}
\det_{i p \ell m m_s} \left( 1+ \widehat{\mathcal K}_{\rm df,3} \widehat{F}_3 \right) &= 0\,,
\label{eq:QC3ch}
\end{align}
with $\widehat F_3$ defined as in \Cref{eq:F3}. We stress again that, although this equation looks identical to the quantization condition given in the main text, \Cref{eq:QC3}, the matrices here act in the ten-dimensional charged basis of \Cref{eq:chargebasis} , whereas those in the main text act in the two-dimensional $I=5/2$ basis of \Cref{eq:I5basis}. We also reiterate that this quantization condition holds in the limit in which the $N\pi\to N\pi\pi$ coupling vanishes.

If we represent a generic matrix in the charged basis as $\widehat M_{{\rm ch}}$, then in the isospin basis it is given by
\begin{equation}
\widehat M_{ {\rm iso10}} = C_{{\rm iso10} \leftarrow {\rm ch}} \cdot \widehat M_{{\rm ch}} \cdot C_{{\rm iso10} \leftarrow {\rm ch}}^\dagger\,.
\end{equation}
Thus, since $C_{{\rm iso10} \leftarrow {\rm ch}}$ is unitary, all entries on the right-hand side of the expression for the correlation function, \cref{eq:CLcharged}, can be expressed in the isospin basis via insertions of the identity. All matrices in the isospin basis have a block-diagonal form, with one block for each value of total isospin.\footnote{%
In fact, for $\Kdf$, we determine the relations between the $\cK_{ij}$ by enforcing this block-diagonal structure.%
}
It follows that the quantization condition, \Cref{eq:QC3ch}, when converted into the isospin basis, separates into a different condition for each choice of isospin $I$,
\begin{align}
\det_{i p \ell m m_s} \left( 1+ \widehat{\mathcal K}_{\rm df,3}^I \widehat{F}_3^I \right) &= 0\,.
\label{eq:QC3I}
\end{align}
Here the index $i$ runs over those entries in the isospin basis, \Cref{eq:isospinbasis}, corresponding to the choice of $I$. Thus $i$ runs over two values for $I=5/2$, and four values for $I=3/2$ and $1/2$. The quantization condition \Cref{eq:QC3I} holds without approximation for $I=5/2$, where mixing with $N\pi$ is forbidden, but holds only for $I=3/2$ and $1/2$ in the limit that there are no $N\pi\to N\pi\pi$ transitions. The $I=5/2$ form is reproduced in \Cref{eq:QC3} in the main text.

To complete the description of the quantization condition, we need to provide the isospin blocks for the matrices that enter $\widehat{F}_3^I$. The former are collected in \Cref{tab:FGK}, with the $I=5/2$ blocks also reproduced in the main text in \Cref{eq:FI5,eq:GI5,eq:K2L5}.

A final discussion of the form of $\widehat{\mathcal K}_{\rm df,3}$ for the different isospin channels is also required. For $I=5/2$, we find the result given in \Cref{eq:Kdf3I5form} in the main text. The results for other isospins take a similar outer-product form. For $I=3/2$, we find
\begin{equation}
\left[ \widehat{\cK}_{\rm df,3}^{I=3/2}\right]_{p \ell' m' m'_s; k \ell m m_s} =
\sum_{x,y\in S,A}
\boldsymbol {\mathcal Y}^{I=3/2; x}_{p\ell' m'} \circ \cK_{\df,3;\;m'_s, m_s}^{I=3/2;\; x, y}(\{\bm p\},\{\bm k\})
\circ \left[\boldsymbol {\mathcal Y}^{I=3/2; y}_{k \ell m}\right]^\dagger
\,,
\label{eq:Kdf3I3form}
\end{equation}
where
\begin{align}
\begin{split}
\left[\bcY^{I=3/2;S}\right]^\dagger &= \left( \sqrt{\tfrac12}\YL312,\ 0,\ -\sqrt{\tfrac16}\YL132,\ -\sqrt{\tfrac56} \YL132 \right)\,,
\\
\left[\bcY^{I=3/2;A}\right]^\dagger &= \left( 0,\ \sqrt{\tfrac12} \YL312,\ -\sqrt{\tfrac56}\YL132,\ \sqrt{\tfrac16} \YL132 \right)\,,
\end{split}
\label{eq:Y3SA}
\end{align}
and similarly for the column vectors $\bcY^{I=3/2;x}$. The action of the $\cY$ operators is defined in \Cref{eq:YRdef}, and explained in the surrounding text. The superscripts $S$ and $A$ refer, respectively, to the symmetry or antisymmetry of the underlying K matrix under pion exchange, a transformation that can be done independently in initial and final states. Thus there are four underlying three-particle K matrices, although the off-diagonal matrices are related by $PT$ symmetry.

The result for $I=1/2$ takes the same form as \Cref{eq:Kdf3I3form}, with $3/2$ replaced by $1/2$, although the vectors of operators that make up the outer product differ, and are given by
\begin{align}
\begin{split}
\bcY^{I=1/2;S\dagger} &= \left( 0,\ \sqrt{\tfrac12} \YL312,\ \sqrt{\tfrac23}\YL132,\ -\sqrt{\tfrac13} \YL132 \right)\,,
\\
\bcY^{I=1/2;A\dagger} &= \left( \sqrt{\tfrac12}\YL312,\ 0,\ \sqrt{\tfrac13}\YL132,\ \sqrt{\tfrac23} \YL132 \right)\,.
\end{split}
\label{eq:Y1SA}
\end{align}
Note that $I=1/2$ also has four underlying K matrices (or three if $PT$ is used).


\begin{table}[h!]
\begin{center}
\begin{tabular}{|l|}
\hline
\shortstack{
\vspace{2pt}\\
$\displaystyle \widehat F^{I=5/2} =
\begin{pmatrix}
F_N & 0 \\
0 & F_\pi
\end{pmatrix}$\\
\vspace{-2pt}
}
\\
\hline
\shortstack{
\vspace{2pt}\\
$\displaystyle \widehat F^{I=3/2} =
\begin{pmatrix}
F_N & 0 & 0 & 0 \\
0 & F_N & 0 & 0 \\
0 & 0 & F_\pi & 0 \\
0 & 0 & 0 & F_\pi
\end{pmatrix}$
\\
\vspace{-2pt}
}
\\
\hline
\shortstack{
\vspace{2pt}\\
$\displaystyle \widehat F^{I=1/2} =
\begin{pmatrix}
F_N & 0 & 0 & 0 \\
0 & F_N & 0 & 0 \\
0 & 0 & F_\pi & 0 \\
0 & 0 & 0 & F_\pi
\end{pmatrix}$
\\
\vspace{-2pt}
}
\\
\hline \hline
\shortstack{
\vspace{2pt}\\
$\displaystyle \widehat G^{I=5/2} = \begin{pmatrix}
0 & \sqrt{2} \mathbb{P}_{\ell} G_{N \pi} \\
\sqrt{2} G_{\pi N }\mathbb{P}_{\ell} & \mathbb{P}_\ell G_{\pi \pi} \mathbb{P}_\ell \\
\end{pmatrix}
$\\
\vspace{-2pt}
}
\\
\hline
\shortstack{
\vspace{2pt}\\
$\displaystyle \widehat G^{I=3/2} =
\begin{pmatrix}
0 & 0 & -\sqrt{\tfrac13} G_{N\pi} & -\sqrt{\tfrac53} G_{N\pi}
\\
0 & 0 & -\sqrt{\tfrac53} G_{N\pi} & \sqrt{\tfrac13} G_{N\pi}
\\
-\sqrt{\tfrac13} G_{\pi N} & -\sqrt{\tfrac53} G_{\pi N} &
- \tfrac23 \mathbb{P}_\ell G_{\pi\pi} \mathbb{P}_\ell & \tfrac{\sqrt5}3 \mathbb{P}_\ell G_{\pi\pi} \mathbb{P}_\ell
\\
- \sqrt{\tfrac53} G_{\pi N} & \sqrt{\tfrac13} G_{\pi N} &
\tfrac{\sqrt5}3 \mathbb{P}_\ell G_{\pi\pi} \mathbb{P}_\ell & \tfrac23 \mathbb{P}_\ell G_{\pi\pi} \mathbb{P}_\ell
\end{pmatrix}
$
\\
\vspace{-2pt}
}
\\
\hline
\shortstack{
\vspace{2pt}\\
$\displaystyle \widehat G^{I=1/2} =
\begin{pmatrix}
0 & 0 & \sqrt{\tfrac23} G_{N\pi} & \sqrt{\tfrac43} G_{N\pi}
\\
0 & 0 & \sqrt{\tfrac43} G_{N\pi} & -\sqrt{\tfrac23} G_{N\pi}
\\
\sqrt{\tfrac23} G_{\pi N} & \sqrt{\tfrac43} G_{\pi N} &
\tfrac13 \mathbb{P}_\ell G_{\pi\pi} \mathbb{P}_\ell & - \tfrac{2\sqrt2}3 \mathbb{P}_\ell G_{\pi\pi} \mathbb{P}_\ell
\\
\sqrt{\tfrac43} G_{\pi N} & -\sqrt{\tfrac23} G_{\pi N} &
- \tfrac{2\sqrt2}3 \mathbb{P}_\ell G_{\pi\pi} \mathbb{P}_\ell & - \tfrac13 \mathbb{P}_\ell G_{\pi\pi} \mathbb{P}_\ell
\end{pmatrix}
$
\\
\vspace{-2pt}
}
\\
\hline \hline
\shortstack{
\vspace{2pt}\\
$\displaystyle \widehat \cK^{I=5/2}_{2,L} =
\begin{pmatrix}
\text{\footnotesize{$\tfrac{1}{2} \cK_{2,L}(\pi\pi, I=2)$}} & 0 \\
0 & \text{\footnotesize{$\cK_{2,L}(N\pi, I=3/2)$}}
\end{pmatrix}$
\\
\vspace{-2pt}
}
\\
\hline
\shortstack{
\vspace{2pt}\\
$\displaystyle \widehat \cK^{I=3/2}_{2,L} =
\begin{pmatrix}
\text{\footnotesize{$\tfrac{1}{2} \cK_{2,L}(\pi\pi, I=2)$}} & 0 & 0 & 0 \\
0 & \text{\footnotesize{$\tfrac{1}{2} \cK_{2,L}(\pi\pi, I=1)$}} & 0 & 0 \\
0 & 0 & \text{\footnotesize{$\cK_{2,L}(N\pi, I=3/2)$}} & 0 \\
0 & 0 & 0 & \text{\footnotesize{$\cK_{2,L}(N\pi, I=1/2)$}}
\end{pmatrix}$
\\
\vspace{-2pt}
}
\\
\hline
\shortstack{
\vspace{2pt}\\
$\displaystyle \widehat \cK^{I=1/2}_{2,L} =
\begin{pmatrix}
\text{\footnotesize{$\tfrac{1}{2} \cK_{2,L}(\pi\pi, I=1)$}} & 0 & 0 & 0 \\
0 & \text{\footnotesize{$\tfrac{1}{2} \cK_{2,L}(\pi\pi, I=0)$}} & 0 & 0 \\
0 & 0 & \text{\footnotesize{$\cK_{2,L}(N\pi, I=3/2)$}} & 0 \\
0 & 0 & 0 & \text{\footnotesize{$\cK_{2,L}(N\pi, I=1/2)$}}
\end{pmatrix}$
\\
\vspace{-2pt}
}
\\
\hline
\end{tabular}
\end{center}
\caption{
Isospin blocks for the matrices $\widehat F$, $\widehat G$ and $\widehat \cK_{2,L}$.
}
\label{tab:FGK}
\end{table}


\subsection{Integral equation for \texorpdfstring{$\cM_3$}{the three-particle scattering amplitude}}
\label{app:inteqs}

In this section we fill in further details of the argument that leads to the integral equations described in \Cref{sec:inteqs}. The steps are the same as those described in detail in appendix A.6 of ref.~\cite{\tetraquark}, which generalizes the approach of refs.~\cite{\HSQCb,\BSQC,\BSnondegen,\BStwoplusone}. We do not repeat the technical details provided in ref.~\cite{\tetraquark}, instead only pointing out the changes necessary here because of the different flavor structure.

One starts by considering the finite-volume scattering amplitude, $\cM_{3,L}$, which is a $3\times 3$ matrix in the basis of operators \Cref{eq:underlying}. This is then extended to a $10\times 10$ matrix in the spectator-pair charge basis \Cref{eq:chargebasis}, with momentum dependence converted to the $\bm p \ell m$ matrix basis. The connection between these bases is provided by the $3\times 10$ matrix of operators [replacing eq.~(A.71) of ref.~\cite{\tetraquark}]
\begin{equation}
\CR = \begin{pmatrix}
\bcX^{(4)} & 0 & 0
\\
0 & S_{12} \bcX^{(2)} & 0
\\
0 & 0 & \bcX^{(4)}
\end{pmatrix}\,,
\label{eq:CRdef}
\end{equation}
where
\begin{align}
\begin{split}
\bcX^{(4)} &\equiv \left( \XR312 \mathbb P_e,\ \XR312 \mathbb P_o , \XR231, \XR132 \right) \,,
\\
\bcX^{(2)} &\equiv \left(\sqrt{\tfrac12} \XR312, \XR132\right) \,.
\end{split}
\label{eq:bradefs}
\end{align}
Here the operators $\boldsymbol{\mathcal X}^{\boldsymbol \sigma}_{[pab]}$, defined in \Cref{eq:XRdef}, convert $\bm p\ell m$ indices into dependence on three on-shell momenta. $S_{12}$ is the symmetrization operator acting on the momenta $\bm p_1$ and $\bm p_2$ of the two pions, normalized as $S_{12}=1 + P_{12}$, with $P_{12}$ the permutation operator. $\mathbb P_e$ and $\mathbb P_o$ are defined following \Cref{eq:FG2p1}. We note that $\bcX^{(2)}$ is the same as the object $\boldsymbol{\mathcal V}_\alpha$ introduced in ref.~\cite{\tetraquark}.

Using the same algebraic steps as in ref.~\cite{\tetraquark}, one finds
\begin{equation}
\cM_{3;m'_s,m_s}(\{\bm p\},\{\bm k\}) = \lim_{\epsilon\to 0^+} \lim_{L\to \infty} \CR \circ \widehat{\cM}_{3,L; m'_s,m_s}^{(u,u)} \circ \CR^\dagger \,,
\label{eq:M3underlying}
\end{equation}
where $\widehat{\cM}_{3,L}^{(u,u)}$ is given by the same expressions as in the main text, \Cref{eq:MhatuuI,eq:DhatuuI,eq:Mhatdf3LI}, except that here the matrices are ten-dimensional. Here $\cM_3$ is the appropriate infinite-volume limit of the on-shell projection of $\cM_{3,L}$. The nature of this limit is discussed in the main text around \Cref{eq:M3hatI}.

The final step is to convert the result \Cref{eq:M3underlying} into the isospin basis for $N\pi\pi$ states. As shown by \Cref{eq:isospin_states}, there are five such states, which we order as
\begin{equation}
\left\{ [(\pi \pi)_{2} N]_{5/2}, [(\pi\pi)_{2} N]_{3/2}, [(\pi\pi)_1 N]_{3/2}, [(\pi\pi)_0 N]_{1/2}, [(\pi\pi)_1 N]_{1/2} \right\},
\label{eq:isospin_d5}
\end{equation}
corresponding to the 1${}^{\text{st}}$, 3${}^{\text{rd}}$, 4${}^{\text{th}}$, 8${}^{\text{th}}$ and 7${}^{\text{th}}$ entries, respectively, of $\boldsymbol v_{\text{iso}}$, defined in \Cref{eq:isospinbasis}. It is understood that we are considering the $m_I=1/2$ component, and that the momentum arguments are $\bm k_1, \bm k_2, \bm k_3$ for all triplets. The $5\times 3$ matrix connecting this isospin basis to that of \Cref{eq:underlying} is
\begin{equation}
C_{{\rm iso}5 \leftarrow {\rm ch}}
=
\begin{pmatrix}
\sqrt{\tfrac1{10}} S_{12} & \sqrt{\tfrac25} & \sqrt{\tfrac15} S_{12}
\\
-\sqrt{\tfrac1{15}} S_{12} & -\sqrt{\tfrac4{15}} & \sqrt{\tfrac3{10}} S_{12}
\\
\sqrt{\tfrac13} A_{12} & 0 & \sqrt{\tfrac16} A_{12}
\\
\sqrt{\tfrac1{3}} S_{12} & -\sqrt{\tfrac13} & 0
\\
-\sqrt{\tfrac16} A_{12} & 0 & \sqrt{\tfrac13} A_{12}
\end{pmatrix}\,,
\label{eq:Cisotoch5}
\end{equation}
where $A_{12}=1-P_{12}$.

To express the integral equations in the isospin basis, we need the $5\times10$ dimensional matrix of operators that convert from the 10-d isospin basis, \Cref{eq:isospinbasis}, which appears in the quantization condition described in the main text, to the 5-d basis, \Cref{eq:isospin_d5}. Combining the above results, we find this matrix to be
\begin{equation}
\CR_{{\rm iso5} \leftarrow {\rm iso10}} = C_{{\rm iso5} \leftarrow {\rm ch}} \cdot \CR \cdot
C^{\dagger}_{{\rm iso10} \leftarrow {\rm ch}} \,.
\end{equation}
Applying manipulations similar to those in appendix A.7 of ref.~\cite{\tetraquark}, we find that this matrix is block diagonal in isospin, as expected, and can be written as
\begin{equation}
\CR_{{\rm iso5} \leftarrow {\rm iso10}} =
\begin{pmatrix} \bcX^{I=5/2} & 0 & 0 \\
0 & \bcX^{I=3/2;S} & 0 \\
0 & \bcX^{I=3/2;A} & 0 \\
0 & 0 & \bcX^{I=1/2;S} \\
0 & 0 & \bcX^{I=1/2;A}
\end{pmatrix}\,,
\end{equation}
where
\begin{align}
\bcX^{I=5/2} & =
\left(\sqrt2 \XR312,\ \XR132+\XR231\right) \,,
\label{eq:XI5_appendix}
\\[5pt]
\bcX^{I=3/2;S} &= \left(\sqrt2 \XR312,\ 0,\ -\sqrt{\tfrac16}(\XR132+\XR231),
\ -\sqrt{\tfrac56}(\XR132+\XR231) \right)\,,
\\
\bcX^{I=3/2;A} &= \left(0,\ \sqrt2 \XR312,\ -\sqrt{\tfrac56}(\XR132-\XR231),
\ \sqrt{\tfrac16}(\XR132-\XR231) \right)\,,
\\[5pt]
\bcX^{I=1/2;S} &= \left(0,\ \sqrt2 \XR312,\ \sqrt{\tfrac23}(\XR132+\XR231),
\ -\sqrt{\tfrac13}(\XR132+\XR231) \right)\,,
\\
\bcX^{I=1/2;A} &= \left(\sqrt2 \XR312,\ 0,\ \sqrt{\tfrac13}(\XR132-\XR231),
\ \sqrt{\tfrac23}(\XR132-\XR231) \right)\,,
\end{align}
Here \Cref{eq:XI5_appendix} is the same as \Cref{eq:XI5} in the main text, and is repeated here for ease of comparison.

For the $I=5/2$ case the result for the scattering amplitude, \Cref{eq:M3I5}, immediately follows, while for $I=3/2$ and $I=1/2$, the result can instead be written as
\begin{equation}
\cM_{3;\; m'_s,m_s}^{I;X, Y}(\{\bm p\},\{\bm k\}) = \lim_{\epsilon\to 0^+}\lim_{L\to\infty}
\boldsymbol{\mathcal X}^{I; X} \circ
\widehat{\cM}_{3,L;\; m'_s; m_s }^{(u,u), I}
\circ \left[\boldsymbol{\mathcal X}^{I; Y }\right]^\dagger \,,
\label{eq:M3I3}
\end{equation}
where $X,Y\in \{S,A\}$. We stress for the final time that this holds only in the fictitious world in which there are no $N\pi\to N\pi\pi$ transitions. Nevertheless, we expect that these results will be useful as building blocks in the complete analysis including such transitions.


\section{Singularities in \texorpdfstring{$3\to3$}{3 to 3} kernels}
\label{app:singularities}

This appendix addresses the issue raised briefly in \Cref{sec:dimer}, namely whether the smooth cutoff functions that are intrinsic to the formalism avoid singularities in the $3\to3$ kernels. This must be the case for the derivation of \Cref{app:deriv} to hold, as singularities in the ${3\to3}$ Bethe-Salpeter kernel would lead to uncontrolled power-law finite-volume dependence. Furthermore, we have assumed in \Cref{sec:inteqs} that the $3\to 3$ K matrix $\Kdf$, which appears in the quantization condition and the integral equations, is free of singularities below the inelastic threshold. $\Kdf$ inherits any singularities present in the ${3\to3}$ Bethe-Salpeter kernel, but also contains additional contributions that can introduce new singularities. Thus additional analysis is required.

We note that although the Bethe-Salpeter kernels are defined in TOPT, and are thus not Lorentz invariant, their singularities, which are due to cuts in which all intermediate particles are on shell, occur at the same positions as those of the corresponding Feynman diagrams, and thus are at Lorentz-invariant positions. We can therefore study their positions using techniques developed for Feynman diagrams. We additionally note that, because the $3\to 3$ kernel is 3PI with respect to $s$-channel cuts of $N \pi \pi$ states, singularities due to these states are absent by construction. The same holds for $\Kdf$, for which the potential unitarity cut from $N\pi\pi$ is removed by using the PV prescription.

We recall that the cutoff functions $H_N(\bm p)$ and $H_\pi(\bm p)$, defined in \Cref{eq:cutoff,eq:sigmaNmin,eq:sigmapimin}, are chosen such that, if the spectator has a real, on-shell momentum, then the corresponding pair can only go a certain distance below its threshold, with the allowed range set by the requirement of avoiding $t$- and $u$-channel singularities. For example, the cutoff $H_\pi(\bm p)$, where $\bm p$ is the momentum of the spectator pion, avoids the left-hand singularity due to $u$-channel nucleon exchange in the $N\pi$ subchannel, shown in \Cref{fig:u-channel}.

The diagram in \Cref{fig:u-channel} involves only $2\to 2$ scattering. The issue we address here is whether singularities in fully-connected $3\to3$ kernels are also avoided. We do so in four stages. First, we map out the three-particle parameter space allowed by our cutoff functions. From this we learn that there are subregions in which additional singularities may arise. Second, we study whether singularities arise in tree diagrams that contribute to the $3\to3$ kernels, finding that they do not. Third, we consider loop diagrams, and by analyzing their Landau singularities discover that certain regions of the allowed parameter space must be avoided. Finally, we comment briefly on how we can change the cutoff functions to avoid the singularities.

Unless stated otherwise, we work in this appendix in the overall CMF, i.e.~with $\bm P= \bm 0$. This leads to no loss of generality since, as noted above, the position of the singularities is Lorentz invariant. For the sake of brevity, we denote the energy in this frame $E$ rather than $E^\star$.


\subsection{Allowed parameter space}
\label{app:paramspace}

\begin{figure}[th!]
\centering
\includegraphics[width=\textwidth]{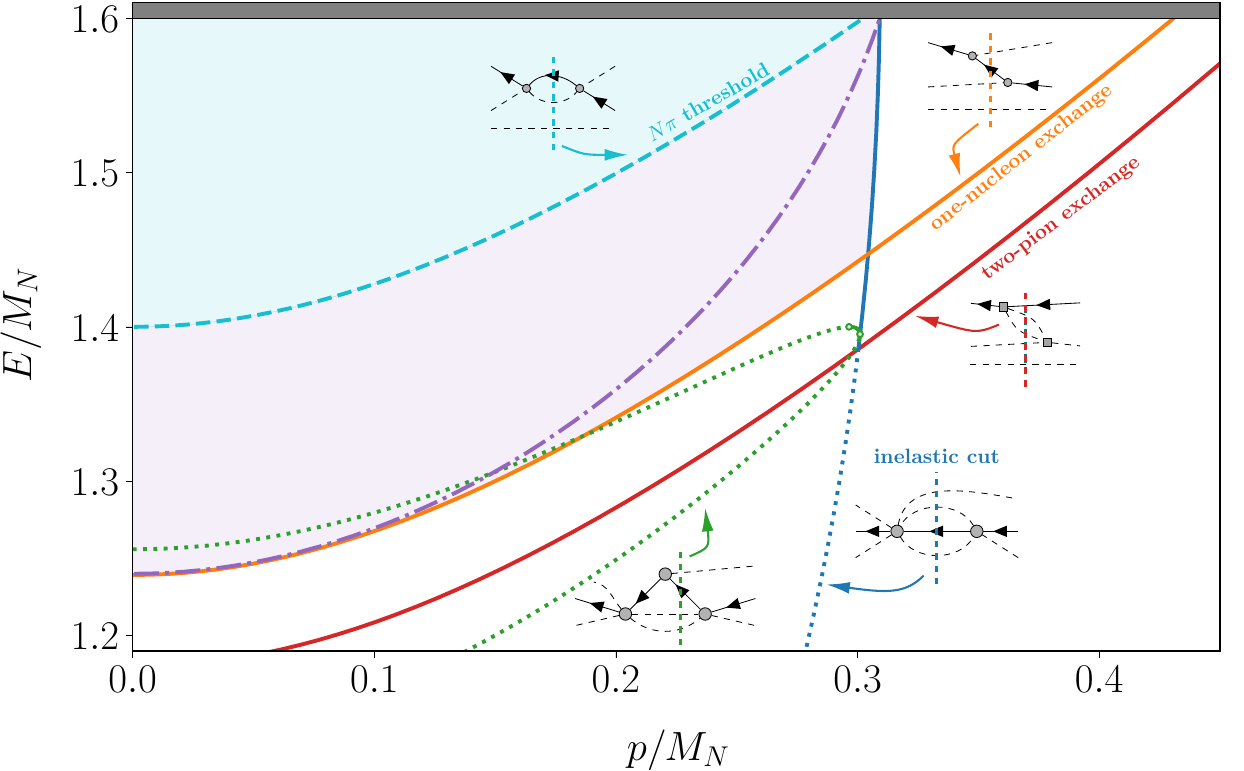}
\caption{Allowed parameter space for the $N\pi\pi$ system with total energy $E$, pion spectator momentum $p = |\bm p|$ (both defined in the $N\pi\pi$ CMF), and $M_\pi/M_N = 0.2$. Solid, dashed and dotted curves correspond to singularities associated with the accompanying Feynman diagram, and are discussed in the text. Solid curves indicate singularities on the physical sheet, while dotted curves correspond to singularities on unphysical sheets. The cyan dashed curve separates the physical region (also shaded in cyan), in which all momenta are real, from the region of subthreshold $N \pi$ kinematics. The latter is shaded (in purple) until the first singularity is encountered, which marks the edge of the physical sheet. The dashed curve is not solid because the associated threshold singularity is absent from the $3\to3$ Bethe-Salpeter kernel and from $\Kdf$. The standard RFT cutoff function $H_\pi(\bm p)$ (given by \Cref{eq:cutoff,eq:sigmapimin}) has a lower end of support on the orange curve, and thus includes the upper-right triangular region (to the right of the blue nearly vertical curve) which is not on the physical sheet. The purple dot-dashed curve illustrates a possible choice for the lower end of support for a modified cutoff function, one that ensures that the parameter space is free from subthreshold nonanalyticities (see \Cref{app:newcutoff}).}
\label{fig:paramspace}
\end{figure}

We find that new constraints only arise when the pion is the spectator, so we focus on that case. We call the spectator four-momentum $p_1$, while those for the nucleon and pion in the remaining pair are denoted $p_2$ and $p_3$, respectively. We use $\bm p$ for the spectator three-momentum (dropping subscripts to lighten the notation), so that $p_1^\mu=(\omega_\pi(\bm p), \bm p)$. In \Cref{fig:paramspace} we show the allowed region in the space of $p=|\bm p|$ and the total CMF energy $E$, choosing $M_\pi/M_N = 0.2$ for definiteness.\footnote{%
{In the final implementation of the formalism, we will need to make reference to the finite-volume frame and allow for $\bm P\neq \bm 0$ in this frame. In this situation one defines $(E, \bm P)$ as the total four-momentum in the finite-volume frame, while $E^\star$ is the CMF energy. The spectator momentum $\bm p$ is then defined in the finite-volume frame and we introduce $p^\star$ as the magnitude of the spatial part of the four-vector obtained by boosting $p_1$ to the overall CMF. All expressions in this appendix are then generalized by replacing $E$ with $E^\star$ and $p$ with $p^\star$. In practice this is implemented via the identity $(E - \omega_\pi(\bm p))^2 - (\bm P - \bm p)^2 = (E^\star - \omega_\pi( p^\star))^2 - p^{\star2}$.}%
}
$E$ is bounded from above by the inelastic threshold at $E_{\rm max} = M_N + 3 M_\pi$, shown as the (solid) black horizontal line in the figure. The lower bound on $E$ for a given $p$ is given by the solid orange curve (labelled ``one-nucleon exchange''). This is enforced by the vanishing of the standard RFT cutoff function $H_\pi(\bm p)$ [defined in \Cref{eq:cutoff,eq:sigmapimin}] at the position of the left-hand cut in the $N\pi$ channel, which occurs at
\begin{equation}
M_{23}^2 = \sigma_\pi^{\rm min} \equiv M_N^2 + 2 M_\pi^2\,.
\end{equation}
Here the pair squared invariant mass is given by
\begin{equation}
M_{23}^2 = (p_2+p_3)^2 = (P-p_1)^2 = E^2 - 2 E \omega_\pi(\bm p) + M_\pi^2
\,.
\label{eq:M23sq}
\end{equation}
The dashed cyan curve (labelled ``$N \pi$ threshold'') corresponds to $M_{23}=M_N+M_\pi$, i.e.~the $N\pi$ threshold. Above this curve all momenta are real (the cyan shaded region). By contrast, below this curve the three-momenta of the pair in their CMF are purely imaginary, and, upon boosting to the overall CMF, the four-momenta are in general complex. As noted above, neither the $3\to3$ Bethe-Salpeter kernel nor $\Kdf$ is singular along the threshold curve, which is why it is dashed in the figure.


We now turn to the remaining curves in the figure. The dotted green curve and the dot-dashed purple curve will be discussed in \Cref{app:loopsings,app:newcutoff}, respectively. What is important here is the solid blue curve that runs almost vertically for $p\approx 0.3$ and its dotted extension (labelled ``inelastic cut'') as well as the red curve (labelled ``two-pion exchange''). The solid blue curve depends on the value of the invariant mass of the $N\pi$ pair composed of the spectator pion and the nucleon from the pair, ${M_{12}=\sqrt{(p_1+p_2)^2}}$. Below the dashed cyan $N \pi$-threshold curve, $M_{12}$ is, in general, complex. But if one chooses the direction of the (imaginary) three-momentum of $p_2$ in the $\{23\}$ pair rest frame to be orthogonal to $\bm p$, then $M_{12}$ becomes real, and is given by
\begin{align}
\begin{split}
&M_{12}^2 = (P-p_3)^2 = E^2 - 2 E \omega_\pi(\bm p_3) + M_\pi^2\,, \\ &\omega_\pi(\bm p_3) = \frac{E - \omega_\pi(\bm p)}{M_{23}} \omega_\pi(\bm p_3^\star), \quad \bm p_3^{\star 2} = \frac{\lambda(M_{23}^2, M_N^2, M_\pi^2)}{4 M_{23}^2} \label{eq:M12sq}
\end{split}
\end{align}
This real value increases with $p$, and the solid blue curve corresponds to when it reaches the inelastic threshold for the $N\pi$ pair, $M_{12}=M_N+2 M_\pi$. To the right of this curve, $M_{12}$ exceeds this inelastic threshold, while the dotted extension represents its smooth analytic continuation, corresponding to a singularity on an unphysical sheet.\footnote{%
We have also investigated the invariant mass of the two-pion pair, $M_{13}$, which is real under the same conditions as $M_{12}$. We find, however, that it always lies below its inelastic threshold at $M_{13}=4M_\pi$ within the allowed parameter space (between the solid black and orange lines).%
}
As $M_\pi/M_N$ increases, the blue curve moves away from the elastic scattering region (bordered by the dashed cyan curve). Using the coordinate $M_{12}^2$ as a measure the minimum distance between the dashed cyan and solid blue curves occurs for $E = M_N + 3M_\pi$ and is given by $3 M_\pi^3/(M_N+M_\pi)$.

The conclusion is that, despite the constraint $E< M_N+3 M_\pi$, it is possible for one of the $N\pi$ pairs to go above the its {\em inelastic} ($N+2\pi$) threshold when the other pair lies below its {\em elastic} ($N+\pi$) threshold, as long as the spectator momentum is large enough. Given that a $2 \to 2$ subprocess can go above the $N + 2 \pi$ threshold, the concern arises whether there can be kinematic singularities in the $3\to 3$ kernels associated with $N+3\pi$ threshold. We investigate this possibility in the remaining sections of this appendix and find that the solid blue curve indeed represents such a singularity, although other possible singularities turn out not to lie on the physical sheet in the allowed subthreshold region.

We conclude this section by explaining the solid red curve (labelled ``two-pion exchange''). This is given by
\begin{equation}
M_{23}^2 = M_N^2 - M_\pi^2\,,
\end{equation}
and can be understood in two ways. First, it is the value of $M_{23}$ below which the expression for $\bm p_3^{\star 2}$ in \Cref{eq:M12sq} breaks down, i.e. no longer satisfies $\omega_N(\bm p_3^\star)+\omega_\pi(\bm p_3^\star) = M_{23}$, unless one shifts to a different branch of the square-root in $\omega_\pi$. Thus the expression for $M_{12}$ given in \Cref{eq:M12sq} is nonanalytic when it crosses the red curve, which is why the blue line is shown as dotted below the red curve. Second, the red curve is the position of the beginning of the left-hand cut in the $N\pi$ amplitude due to $t$-channel two-pion exchange.


\subsection{Contributions from tree-level diagrams}
\label{app:treesings}

\begin{figure}[h!]
\centering
\includegraphics[width=0.35\textwidth]{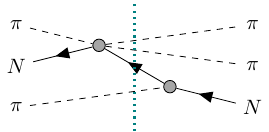}
\caption{Tree-level contribution to the $3\to 3$ 3PI TOPT Bethe-Salpeter kernel involving an $N\pi\pi\pi$ intermediate state. Notation as in \Cref{fig:u-channel}.}
\label{fig:singularities}
\end{figure}

We first consider the possibility of singularities in 3PI tree diagrams. To obtain an $N+3\pi$ intermediate state, we need an $N\to N\pi$ vertex, and the only tree diagram allowed for $I=5/2$ is that shown in \Cref{fig:singularities}(a), together with its time reflection. This diagram is analogous to the $u$-channel exchange in two-particle amplitudes, \Cref{fig:u-channel}. We stress that, while this diagram involves an $N\to N\pi$ vertex, it is allowed for $I=5/2$, unlike the $2\to 3$ transitions discussed in the main text. This is because $N\pi\pi\pi$ states can have $I=1/2$, $3/2$, $5/2$, and $7/2$.


The left-hand vertex in \Cref{fig:singularities} involves an $N\pi \leftarrow N\pi\pi$ vertex, and thus provides a possible instantiation of the singularity discussed in the previous subsection. Indeed, if $M_{12}$ corresponds to the invariant mass of the upper $\pi N$ pair in the final state, then the analysis in \Cref{app:paramspace} indicates that a physical cut (indicated by the vertical teal dotted line) is possible. To see whether it occurs, however, one must take into account the fact that the initial state consists of one real momentum and a pair that can lie below its threshold. Thus the problem is more constrained than considering either the initial or the final state separately, as was done in \Cref{app:paramspace}.

To investigate this further, we have shown analytically that, if the initial state nucleon has a real momentum, then the singularity can only be accessed if $E > 2 M_N$, which exceeds the upper limit of applicability of the formalism, $E < M_N + 3 M_\pi$, as long as $M_\pi < M_N/3$. The latter condition holds for physical masses as well as those used in most simulations. It is also straightforward to show that the singularity cannot be accessed if the final-state pion attached to the lower vertex has a real momentum. Finally, we have considered the possibility that one of the pions in the initial-state $N\pi$ pair, together with one of the particles in the final-state two-pion pair, have real momenta. Here we have resorted to a brute force scan over the eight-dimensional parameter space allowed by our cutoff functions and kinematics. We find that the singularity does not appear anywhere in this space for pion masses of interest.

Thus we conclude that tree diagrams do not provide an example of the potential singularity suggested by \Cref{app:paramspace}.


\subsection{Singularities in loop diagrams}
\label{app:loopsings}

Loop diagrams allow access to a wider range of kinematical configurations. The infrared singularities of interest can be studied using the Landau equations and their generalizations~\cite{Bjorken:1959fd,Landau:1959fi,Nakanishi:1959jzx}. This is a large and very active field, and we have found the {\tt Mathematica} package SOFIA to be particularly useful~\cite{Correia:2025yao}, as well as the work described in refs.~\cite{Eden:1966dnq,Mizera:2021fap,Fevola:2023kaw,Fevola:2023fzn}.

\begin{figure}[h!]
\centering
\includegraphics[width=0.9\textwidth]{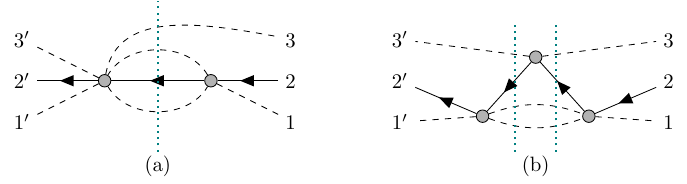}
\caption{Examples of loop contributions to 3PI TOPT Bethe-Salpeter kernel. Notation as in \Cref{fig:u-channel}. Numbers indicate momentum labels. Time runs from right to left.}
\label{fig:loop1}
\end{figure}

The simplest loop diagram that leads to the singularity postulated in \Cref{app:paramspace} is that of \Cref{fig:loop1}(a) (and its time-reversed partner). It is immediately apparent that this diagram has a threshold singularity at $M_{12}=M_N+2 M_\pi$, corresponding to the blue line in \Cref{fig:paramspace}. Note that the fact that $p_3$ is complex does not impact the presence of the singularity, since the third initial-state particle is a spectator to the loop.\footnote{%
In the TOPT derivation of the quantization condition, all possible choices of spectator contribute for a given $3\to 3$ Bethe-Salpeter kernel~\cite{\BSnondegen}, and one of the choices has $p_1$ as the spectator momentum for this diagram.%
}
The existence of this diagram demonstrates that, for a rigorous derivation of the finite-volume quantization condition, one must modify the cutoff function to avoid the region to the right of the blue line in \Cref{fig:paramspace}.


Another loop diagram with the same singularity is shown in \Cref{fig:loop1}(b). This has two $N\pi\pi\pi$ cuts, and if one solves the Landau equations using SOFIA for the symmetric configuration of external momenta in which $p'_i=p_i$, then there is a singularity at $M_{12}=M_N+2 M_\pi$.

\begin{figure}[h!]
\centering
\includegraphics[width=0.9\textwidth]{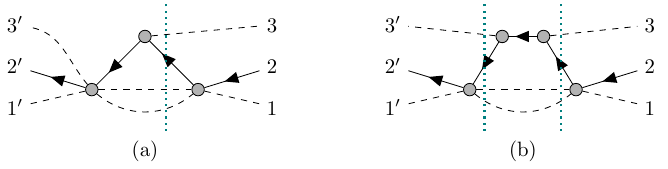}
\caption{Examples of loop contributions to $\Kdf$ that are not present for the 3PI Bethe-Salpeter kernel. Notation as in \Cref{fig:loop1}.}
\label{fig:loop2}
\end{figure}

We now consider loop diagrams contributing to $\Kdf$. These include all diagrams contributing to the 3PI Bethe-Salpeter kernel, so that the singularity in $M_{12}$ from the diagrams of \Cref{fig:loop1} applies to $\Kdf$ as well. However, there are also diagrams with three-particle cuts, in which these cuts are regulated by a PV prescription, so that integrals over them do not lead to nonanalyticities. Examples of these diagrams are shown in \Cref{fig:loop2}. The question we aim to address is whether these diagrams, both of which have $N\pi\pi\pi$ cuts, lead to additional restrictions on the parameter space for which $\Kdf$ is analytic.


We have analyzed these diagrams using SOFIA.\footnote{%
SOFIA assumes that all propagators are regulated with the $i\epsilon$ prescription, while in the three-particle cuts one of the propagators in $\Kdf$ is regulated using the PV prescription. However, in the subthreshold regime, the RFT formalism transitions smoothly between the PV prescription (multiplied by $H(\bm p)$) and the $i\epsilon$ prescription (multiplied by $1-H(\bm p)$). Thus singularities in the subthreshold region can be studied assuming all propagators use the $i\epsilon$ prescription.
}
This program provides a list of the solutions to the Landau equations and their generalizations both for the original diagram (corresponding to classical on-shell propagation of all particles in the loop~\cite{Coleman:1965xm}) and for contracted diagrams in which one or more propagators are contracted to a point (corresponding to having a vanishing Feynman parameter). We focus on the uncontracted diagrams (``leading Landau singularities''), since the contracted diagrams do not lead to new singularities. For example, if we contract the left-hand nucleon propagator in \Cref{fig:loop2}(a), we obtain \Cref{fig:loop1}(a), which we have already analyzed.

For \Cref{fig:loop2}(a) SOFIA identifies two possible loci of singularities that could lie in our parameter space,
\begin{multline}
E^4 M_N^2 - 2 E^2 M_{12}^2 M_N^2 + M_{12}^4 M_N^2 + E^2 M_{12}^2 M_\pi^2
\\
- E^2 M_N^2 M_\pi^2 - M_{12}^2 M_N^2 M_\pi^2 + M_N^4 M_\pi^2 = 0\,,
\end{multline}
and
\begin{multline}
E^4 M_N^2 - 2 E^2 M_{12}^2 M_N^2 + M_{12}^4 M_N^2 + E^2 M_{12}^2 M_\pi^2 -
E^2 M_N^2 M_\pi^2
\\
- M_{12}^2 M_N^2 M_\pi^2 + M_N^4 M_\pi^2 - 4 E^2 M_\pi^4 -
4 M_{12}^2 M_\pi^4 - 8 M_N^2 M_\pi^4 + 20 M_\pi^6 = 0 \,.
\end{multline}
Inserting the expression for $M_{12}(E,p)$ for the configuration in which $M_{12}$ is real, we find that the former locus requires complex $M_{12}$ and is thus not relevant, while the latter does impinge upon our parameter space, and is shown in \Cref{fig:paramspace} as the dotted green line. Its position is qualitatively similar for all values of $M_\pi/M_N$ ranging from the physical ratio to unity.

Given a locus of singularities, the next question is whether it lies on the physical sheet, i.e. can be reached from the region of physical kinematics (which lies above the dashed cyan line in \Cref{fig:paramspace}) by analytic continuation without the need to pass onto a different sheet at a branch point. This can be answered by determining the sign of the Schwinger parameters associated with each of the edges of the diagram: all must be nonnegative for the singularity to lie on the physical sheet~\cite{Coleman:1965xm,Eden:1966dnq}. We find, using standard methods,\footnote{%
We are indebted to Sebastian Mizera for a tutorial on these methods.%
}
that this is only the case for a short segment of the green line shown as solid (and lying between the two open circles) in the figure. Thus the segment that lies within our parameter space is not on the physical sheet, and does not reduce the allowed parameter space.

Turning to \Cref{fig:loop2}(b), we find that, if we consider the symmetric configuration, with $p'_i=p_i$, then no additional constraints beyond that from \Cref{fig:loop2}(a) appear. We suspect that the symmetric configuration leads to the strongest constraints, because then both $N\pi\pi\pi$ cuts can go on shell simultaneously, but we have not done a thorough study of this conjecture.

Higher-loop diagrams may lead to further constraints. However, our expectation is that the triangle diagram will provide the strongest constraints, since it becomes increasingly difficult to find classical on-shell configurations as more particles and interactions are involved. We leave a study of this issue to future work.


\subsection{Adjusting the cutoff functions}
\label{app:newcutoff}

We conclude this appendix with some brief comments on how the singularities discussed above can be avoided by adjusting the cutoff functions. Specifically, we need to redefine $H_\pi(\bm p)$, previously given by \Cref{eq:sigmaNmin,eq:sigmapimin}. We can do so by choosing $\sigma_\pi^{\rm min}$ to be a function of $E$.\footnote{
In a general frame, $\sigma_\pi^{\rm min}$ is a function of $E^*$.
}
A simple example which avoids the singularities is shown in \Cref{fig:paramspace} as the dotted blue line, and is given for $M_\pi= 0.2\, M_N$ by
\begin{equation}
\sigma_\pi^{\rm min}(E) = \sigma_A + (\sigma_B-\sigma_A) \frac{(E-E_A)^2}{(E_B-E_A)^2}\,,
\end{equation}
with
\begin{equation}
\sigma_A = (1.04\, M_N)^2,\ \ \sigma_B = (1.192\, M_N)^2,\ \
E_A = 1.24\, M_N,\ \ E_B=1.6\, M_N\,.
\end{equation}
In practice, these parameters will need to be adjusted depending on the specific values of the ratio $M_\pi/M_N$ under consideration.

It is clear from the proximity of the dashed cyan curve (where $H_\pi=1$) to the dot-dashed purple curve (where $H_\pi=0$), that the cutoff function necessarily varies rapidly for $E$ close to its maximum value. This appears unavoidable, and may lead to an enhancement of exponentially-suppressed volume effects, although the narrowness occurs in a relatively restricted region. In this regard, we note that the minimum gap between the dashed cyan and dot-dashed purple curves is given (for $M_\pi=0.2\,M_N$) by
\begin{equation}
\Delta \sigma_\pi \equiv \Delta M_{23}^2 = 0.46 M_\pi^2\,,
\end{equation}
while the maximum distance is $\sim 8 M_\pi^2$. We also observe that, as one approaches the inelastic threshold at $E=M_N+3 M_\pi$, exponentially-suppressed corrections associated with the distance to this threshold are enhanced, leading to the breakdown of the quantization condition. Thus, strictly speaking, we are automatically kept away from the dangerous corner of the restricted parameter space.


\section{Integrals over Wigner rotations}
\label{app:intD}

In this appendix we evaluate the integrals \Cref{eq:intD,eq:intDbar}, which we repeat for the sake of clarity,
\begin{align}
[D(\bm p)]_{\ell m m_s; \ell^\star m^\star m^\star_s} &= \int d\Omega_{a^\star}
Y_{\ell m}^*(\hat {\bm a}^\star) [\bm \cD^{(1/2)}( \bm a, \bm p)^{-1}]_{m_s m^\star_s} Y_{\ell^\star m^\star}(\hat {\bm a}^\star)\,,
\label{eq:intDa}
\\
[\overline D(\bm p)]_{\ell^\star m^\star m^\star_s; \ell m m_s} &= \int d\Omega_{a^\star}
Y_{\ell^\star m^\star}^*(\hat {\bm a}^\star) [{\bm \cD}^{(1/2)}(\bm a, \bm p)]_{m^\star_s m_s} Y_{\ell m}(\hat {\bm a}^\star)\,.
\label{eq:intDbara}
\end{align}

The first step is to rewrite the expression for the Wigner rotation, \Cref{eq:axisangleWD}, in terms of the integration variable $\bm a^\star$ rather than $\bm a$. Using the following relations,
\begin{gather}
\gamma^{\prime}(\bm a, \bm p) = \gamma_N(\bm a^\star) = \gamma_{N \pi}(\bm p) \gamma_N(\boldsymbol a)(1+\boldsymbol{\beta}_{N\pi}(\bm p)\cdot\boldsymbol{\beta}_N(\bm a))\,,
\\[5pt]
\gamma_N(\bm a) =
\gamma_{N \pi}(\bm p) \gamma_N(\bm a^\star) (1 -\boldsymbol \beta_{N \pi}(\bm p) \cdot\boldsymbol\beta_N(\bm a^\star))\,,
\\[5pt]
\frac{\boldsymbol \beta_{N \pi}(\bm p) \times \boldsymbol\beta_N(\bm a^\star)}{|\boldsymbol \beta_{N \pi}(\bm p) \times \boldsymbol\beta_N(\bm a^\star)|}
=
\frac{\boldsymbol \beta_{N \pi}(\bm p) \times \boldsymbol\beta_N(\bm a)}{|\boldsymbol \beta_{N \pi}(\bm p) \times \boldsymbol\beta_N(\bm a)|}\,,
\end{gather}
and recalling that the boost velocity is given by \Cref{eq:boostvelocity}, we find that the required Wigner matrix is
\begin{align}
\bm \cD^{(1/2)}( \bm a, \bm p)^{-1} &= \bar c + i \bar s \left [ \frac{ (\bm p-\bm P)\times \bm a^\star}{\vert (\bm p-\bm P)\times \bm a^\star \vert }\cdot \boldsymbol\sigma \right ]\,,
\end{align}
where
\begin{equation}
\bar c = \sqrt{\frac{1+c}2}\,,\quad \bar s = \sqrt{\frac{1-c}2}\,,
\end{equation}
and we have introduced the shorthand $ c = \cos \theta(\bm a, \bm p)$, with the latter defined in \Cref{eq:axisangleWD}. Note that we are using the fact that the nucleon is primary in the $N \pi$ pair to fix the sign of the second term in the Wigner rotation.

We would like to perform the integral choosing the $z$ axis to lie along the direction $ \bm P-\bm p$, but this requires rotating the basis states used in the definitions of the spherical harmonics and Wigner matrices. If $R_{N \pi}(\bm p)$ is a rotation that acts as
\begin{equation}
R_{N \pi}(\bm p) \frac{{\bm P - \bm p}}{\vert \bm P - \bm p \vert} = \widehat {\bm z}\,,
\end{equation}
then we find that
\begin{align}
D(\bm p) & = \wt{\cD}(R_{N\pi}(\bm p)^{-1}) \cdot D^{(0)}(\bm p) \cdot \wt{\cD}(R_{N\pi}(\bm p))
\label{eq:DtoD0}
\\
\overline D(\bm p) & = \wt{\cD}(R_{N\pi}(\bm p)^{-1}) \cdot \overline D^{(0)}(\bm p) \cdot \wt{\cD}(R_{N\pi}(\bm p))
\label{eq:DbartoD0bar}
\end{align}
where
\begin{align}
\wt{\cD}(R_{N\pi}(\bm p))_{\ell' m' m_s'; \ell m m_s}
&= \delta_{\ell' \ell} \cD(R_{N\pi}(\bm p))^{(\ell)}_{m' m} \cD(R_{N\pi}(\bm p))^{(1/2)}_{m_s' m_s} \,.
\label{eq:DR0}
\end{align}
Here $D^{(0)}(\bm p)$ and $\overline D^{(0)}(\bm p)$ are defined as in \Cref{eq:intDa,eq:intDbara}, respectively, except now the $z$ axis lies along $\bm P - \bm p$. There is an ambiguity in $R_{N \pi}(\bm p)$ corresponding to a final rotation about the $z$ axis, but the final results presented below do not depend on this choice.

Using $(\theta,\phi)$ for the polar and azimuthal angles of $\bm a^\star$ relative to $\bm P - \bm p$, the only quantities that depend on the orientation of $\hat {\bm a}^\star$ are
\begin{equation}
\frac{(\bm p-\bm P)\times \bm a^\star}{\vert (\bm p-\bm P)\times \bm a^\star \vert}= (s_\phi, -c_\phi,0)\,,\qquad
(\bm P - \bm p) \cdot \bm a^\star = \vert \bm P - \bm p \vert \, q_\pi^\star(\bm p) \, c_\theta\,,
\end{equation}
where $c_\phi=\cos\phi$, etc. The quantities entering the Wigner D matrix can be rewritten as
\begin{align}
\bar c &= \frac{(1+\gamma_{N\pi}(\bm p))(1+\gamma_N(\bm a^\star)) + d(\bm p) c_\theta}{\sqrt{2(1+\gamma_{N\pi}(\bm p))(1+\gamma_N(\bm a^\star))}}
\frac1{\sqrt{1+ \gamma_{N\pi}(\bm p)\gamma_N(\bm a^\star) + d(\bm p) c_\theta}}\,,
\\
\bar s &= \frac{d(\bm p) s_\theta}{\sqrt{2(1+\gamma_{N\pi}(\bm p))(1+\gamma_N(\bm a^\star))}}
\frac1{\sqrt{1+ \gamma_{N\pi}(\bm p)\gamma_N(\bm a^\star) + d(\bm p) c_\theta}}\,,
\end{align}
where we have also introduced $d(\bm p) = \vert \bm \beta_{N\pi}(\bm p) \vert \vert \bm \beta_N(\bm a^\star) \vert \gamma_{N\pi}(\bm p) \gamma_N(\bm a^\star)$, which can be written in various ways, e.g.
\begin{align}
d(\bm p)
& = \frac{\vert \bm P - \bm p \vert q_\pi^{\star}(\bm p)}{\sqrt{\sigma_\pi(\bm p)} M_N}
= \sqrt{(\gamma_{N\pi}(\bm p)^2-1)(\gamma_N(\bm a^\star)^2-1)}\,.
\end{align}

We now restrict $\ell, \ell^\star \le \ell_{\rm max} = 1$, so that all matrices become eight dimensional. We stress that there is no difficulty in principle in raising the maximum value, but we expect $\ell_{\rm max}=1$ to be used in initial implementations, and indeed use this value is used in the example presented in the main text. Using the expressions for spherical harmonics, the $\phi$ integral can be done trivially, leading to
\begin{equation}
D^{(0)}(\bm p) = \int_{c_\theta}
\begin{pmatrix}
\bar c & \sqrt{\tfrac38} \bar s s_\theta \sigma_+ & \sqrt3 \bar c c_\theta & \sqrt{\tfrac38} \bar s s_\theta \sigma_-
\\
-\sqrt{\tfrac38} \bar s s_\theta \sigma_- & \tfrac32 \bar c s_\theta^2 &
- \tfrac{3}{\sqrt8} \bar s s_\theta c_\theta \sigma_- & 0
\\
\sqrt3 \bar c c_\theta & \tfrac3{\sqrt8} \bar s s_\theta c_\theta \sigma_+ & 3 \bar c c_\theta^2 &
\tfrac3{\sqrt8} \bar s s_\theta c_\theta \sigma_-
\\
-\sqrt{\tfrac38} \bar s s_\theta \sigma_+ & 0 & -\tfrac3{\sqrt8} \bar s s_\theta c_\theta \sigma_+ &
\tfrac32 \bar c s_\theta^2
\end{pmatrix}\,,
\label{eq:D0matrix}
\end{equation}
where $\displaystyle \int_{c_\theta} \equiv \frac12 \int_{-1}^1 dc_\theta$, and
\begin{equation}
\sigma_\pm = \sigma_x \pm i \sigma_y\,.
\end{equation}
Here we are displaying the $8\times 8$ matrix as a Kronecker product of the $(\ell, m)$ space, which is shown explicitly using the ordering $(\ell, m)=\{(0,0),\ (1,1),\ (1,0),\ (1,-1)\}$, and the spin space.

The required integrals are
\begin{equation}
I_1 = \int_{c_\theta} \bar c \,,\qquad
I_2 = \int_{c_\theta} \bar c c_\theta \,,\qquad
I_3 = \int_{c_\theta} \bar c c_\theta^2 \,,\qquad
I_4 = \int_{c_\theta} \bar s s_\theta \,,\qquad
I_5 = \int_{c_\theta} \bar s s_\theta c_\theta \,,
\end{equation}
and are evaluated in \cref{tab:integrals}. In terms of these integrals, we have
\begin{equation}
D^{(0)}(\bm p) = \begin{pmatrix}
I_1 & \sqrt{\tfrac38} I_4 \sigma_+ & \sqrt3 I_2 & \sqrt{\tfrac38} I_4 \sigma_-
\\
-\sqrt{\tfrac38} I_4 \sigma_- & \tfrac32 (I_1-I_3) & - \tfrac{3}{\sqrt8} I_5 \sigma_- & 0
\\
\sqrt3 I_2 & \tfrac3{\sqrt8} I_5 \sigma_+ & 3 I_3 & \tfrac3{\sqrt8} I_5 \sigma_-
\\
-\sqrt{\tfrac38} I_4 \sigma_+ & 0 & -\tfrac3{\sqrt8} I_5 \sigma_+ & \tfrac32 (I_1-I_3)
\end{pmatrix}\,.
\end{equation}
The result for $\overline D^{(0)}(\bm p)$ has the same form up to the replacements $I_4 \to -I_4$ and $I_5 \to -I_5$.


\begin{table}
\centering
\renewcommand{\arraystretch}{2.7}
\begin{tabular}{|l|}
\hline
\shortstack{
\vspace{-5pt}\\
$I_k = \displaystyle \frac{1}{\sqrt{2(A+B)}} \frac1{3C} F_k$
\\
\vspace{-10pt}
}
\\
\hline \hline
$
F_1 = \displaystyle \sqrt{A + C}(A + 3B + C) - \sqrt{A - C}(A + 3B - C)
$ \\
\hline
$
F_2 = \displaystyle \frac{1}{5C} \left[
\sqrt{A - C} \left(2A(A + 5B) + (A + 5B)C - 3C^2\right) \right.
$
\\[-15pt]
\hspace{4cm}
$\left. - \sqrt{A + C} \left(2A(A + 5B) - (A + 5B)C - 3C^2\right) \right]
$ \\
$
F_3 = \displaystyle \frac{1}{35C^2} \left[
\sqrt{A + C} \left(8A^2(A + 7B) - 4CA(A + 7B) + 3C^2(A + 7B) + 15C^3\right) \right.
$
\\[-15pt]
\hspace{2cm}
$\left. - \sqrt{A - C} \left(8A^2(A + 7B) + 4CA(A + 7B) + 3C^2(A + 7B) - 15C^3\right) \right]
$ \\
\hline
$
F_4 = \displaystyle \frac{4}{5C} \left[ (2A + 3C)(A - C)^{3/2} - (2A - 3C)(A + C)^{3/2} \right]
$ \\
\hline
$
F_5 = \displaystyle \frac{4}{35C^2} \left[ (A + C)^{3/2}(12A^2 - 18AC + 5C^2) \right.
$
$
\left. - (A - C)^{3/2}(12A^2 + 18AC + 5C^2) \right]
$
\\
\hline \hline
\shortstack{
\vspace{0pt}\\
$ A = 1 + \gamma_{N\pi}(\bm p) \gamma_{N}(\bm a^\star)\,, \quad B = \gamma_{N\pi}(\bm p) + \gamma_{N}(\bm a^\star)\,, \quad C = d(\bm p)$
\\
\vspace{2pt}
}
\\
\hline
\end{tabular}
\renewcommand{\arraystretch}{1.0}
\caption{Analytic expressions for the integrals $I_1$--$I_5$ appearing in the evaluation of $D^{(0)}(\bm p)$ and $\overline D^{(0)}(\bm p)$.\label{tab:integrals}}
\end{table}

This completes the aim of this appendix, namely to evaluate $D(\bm p)$ and $\overline D(\bm p)$ for $\ell_{\rm max} = 1$, and illustrate the general procedure for higher values of $\ell_{\rm max}$. We close by considering the low energy limit of our results. As $q_\pi^{\star}(\bm p)\to 0$, the Wigner rotation becomes trivial, and $D^{(0)}(\bm p)$ should limit to the identity matrix. Noting that, in this limit $A = B + \cO(q_\pi^{\star}(\bm p)^2) = 1 + \gamma_{N\pi}(\bm p)$, we find the following behaviors
\begin{gather}
I_1 = 1 + \cO(q_\pi^{\star}(\bm p)^2)\,, \quad
I_2 = \cO(q_\pi^{\star}(\bm p)^3)\,, \quad
I_3 = 1/3 + \cO(q_\pi^{\star}(\bm p)^2)\,, \\
I_4 = \cO(q_\pi^{\star}(\bm p))\,, \quad
I_5 = \cO(q_\pi^{\star}(\bm p)^2)\,,
\end{gather}
which indeed lead to the correct limit for $D^{(0)}(\bm p)$. Furthermore, were we to restrict $\cK_2$ to be purely $s$-wave, then the induced $p$-wave contribution would be proportional to $q_\pi^{\star}(\bm p)$, consistent with the claim in the main text that the threshold behavior is preserved under multiplication by $D(\bm p)$ and $\overline D(\bm p)$.


\section{Implementing the \texorpdfstring{$\cK_0$}{leading} term in \texorpdfstring{$\Kdf$}{the three-particle K matrix}}
\label{app:K0decomp}

In this final appendix we describe the implementation of the leading term in $\Kdf$, given in \Cref{eq:K0SS}. We do so explicitly for $I=5/2$, using \Cref{eq:Kdf3I5form,eq:Y5}, although the corresponding results for $I=3/2$ and $1/2$ can be obtained straightforwardly from the results below (up to the usual caveat that the $I=3/2$ and $1/2$ cases are only applicable in the unrealistic limit that the $N \pi \to N \pi \pi$ interaction has been tuned to vanish).

Compared to corresponding implementations in mesonic systems, e.g.~that for three pions given in \cite{\dwave,\isospin,Baeza-Ballesteros:2024mii}, the presence of spin introduces additional complications, both through the presence of an enlarged matrix space and the kinematic contributions from the Dirac spinors. In the following we assume the Dirac representation of the gamma matrices. This also dictates the form of the Dirac spinors,
\begin{equation}
u(\bm p,m_s) = \sqrt{\omega(\bm p)+M_N} \begin{pmatrix}
\chi(m_s) \\ \frac{\boldsymbol \sigma \cdot \boldsymbol p}{\omega(\bm p) + M_N}\chi(m_s)
\end{pmatrix}\,,
\end{equation}
with $\chi(m_s)$ a nonrelativistic two-component spinor with $z$ component $m_s$, and, in this appendix, $\omega(\bm p) = \omega_N(\bm p) = \sqrt{\bm p^2 + M_N^2}$. The relevant term is $\Kdf \supset \cK_0\; \cK_{\rm df,3}^{(0)}$ where
\begin{multline}
\cK_{\rm df,3}^{(0)}(\{\bm
p\},\{\bm k\})_{m'_s,m_s} \equiv \bar u(\bm p_3,m'_s) u(\bm k_3,m_s)
\\
= \sqrt{(\omega(\bm p_3) \!+\! m_N) (\omega(\bm k_3) \!+\! m_N)}
\left\{
\delta_{m'_s m_s} - \frac{[\bm p_3 \bm k_3]_{m'_s m_s}}
{(\omega(\bm p_3)\!+\!M_N)(\omega(\bm k_3)\!+\!M_N)}
\right\}\,.
\end{multline}
We recall that $\bm p_3$ and $\bm k_3$ are, respectively, the momenta of the final and initial nucleons, and have introduced the compact notation
\begin{align}
[\bm a \bm b] &\equiv (\bm a\cdot \boldsymbol \sigma) (\bm b \cdot \boldsymbol \sigma)
= a_j r_{jk} b_k\,,
\qquad
r_{jk} = \delta_{jk} \bm{1} + i \epsilon_{jkl} \sigma_l\,,
\label{eq:bracket}
\end{align}
so that $[\bm a \bm b]$ is a matrix in spin space. It will also be useful to introduce vectors of spin matrices,
\begin{equation}
[r \bm v]_j \equiv r_{jk} v_k\,,
\qquad
[\bm v r ]_k \equiv v_j r_{jk}\,.
\label{eq:bracketr}
\end{equation}
The key property of these bracket operations is their linearity in the vector arguments. Further useful results are
\begin{equation}
[\bm a \bm b]^\dagger = [\bm b \bm a]\,,\quad
[r \bm v]^\dagger = [\bm v r]\,,\quad
r_{jk}^\dagger = r_{kj}\,,
\label{eq:bracketproperties}
\end{equation}
where, in the final result, the hermitian conjugation acts only on the spin indices.

From \Cref{eq:Kdf3I5form,eq:Y5}, the flavor structure of the $I=5/2$ $\cK_0$ term is
\begin{equation}
[\cK_{\rm df,3}^{(0),I=5/2}] =
\begin{pmatrix}
\frac12 \cK_{\df,3}^{(0),NN} & \sqrt{\frac12} \cK_{\df,3}^{(0),N\pi} \\
\sqrt{\frac12} \cK_{\df,3}^{(0),\pi N} & \cK_{\df,3}^{(0),\pi\pi}
\end{pmatrix}\,,
\end{equation}
where the blocks are matrices in the $\bm p \ell m m_s$ space, with superscripts denoting the flavor of the spectators. The blocks are given by
\begin{align}
[\cK_{\df,3}^{(0),NN}]_{\bm p'\ell' m' m_s';\bm p\ell m m_s} &=
\YRp312{\bm p'\ell' m'} \circ \cK_{\rm df,3}^{(0)}(\{\bm p\},\{\bm k\})_{m'_s,m_s} \circ
\YLp312{\bm p\ell m}\,,
\\
[\cK_{\df,3}^{(0),N\pi}]_{\bm p'\ell' m' m_s';\bm p\ell m m_s} &=
\YRp312{\bm p'\ell' m'} \circ \cK_{\rm df,3}^{(0)}(\{\bm p\},\{\bm k\})_{m'_s,m_s} \circ \YLp132{\bm p\ell m}\,,
\\
[\cK_{\df,3}^{(0),\pi N}]_{\bm p'\ell' m' m_s';\bm p\ell m m_s} &=
\YRp132{\bm p'\ell' m'} \circ \cK_{\rm df,3}^{(0)}(\{\bm p\},\{\bm k\})_{m'_s,m_s} \circ \YLp312{\bm p\ell m}\,,
\\
[\cK_{\df,3}^{(0),\pi \pi}]_{\bm p'\ell' m' m_s';\bm p\ell m m_s} &=
\YRp132{\bm p'\ell' m'} \circ \cK_{\rm df,3}^{(0)}(\{\bm p\},\{\bm k\})_{m'_s,m_s} \circ \YLp132{\bm p\ell m}\,,
\end{align}
where the $\boldsymbol \cY$ operators, defined in \Cref{eq:YRdef}, project onto definite pair angular momenta.

We discuss the four flavor blocks in turn.


\subsection{\texorpdfstring{$NN$}{Nucleon-nucleon} block}

This case is very straightforward, since the $\cK_0$ term does not depend on the pair momenta if both spectators are nucleons. Thus only the $\ell'=\ell=0$ entries are nonzero, and are given by
\begin{equation}
[\cK_{\df,3}^{(0),NN}]_{\bm p' 0 0 ; \bm p 0 0} = \sqrt{(\omega(\bm p') + M_N) (\omega(\bm p) + M_N)}
\left\{ \bm 1 - \frac{[\bm p' \bm p]}{(\omega(\bm p')+M_N)(\omega(\bm p)+M_N)} \right\}\,.
\label{eq:K0NN}
\end{equation}
We have abbreviated the notation by keeping the $m'_s$ and $m_s$ indices implicit---they are carried by the identity matrix $\bm 1$ and the quantity $[\bm p' \bm p]$.


\subsection{\texorpdfstring{$N\pi$}{Nucleon-pion} block}

If the final spectator is the nucleon, while the initial spectator is a pion, then $\bm p_3 = \bm p'$ and (because we take the nucleon to be the primary member of the pair) $\bm k_3 = \bm a$. So the starting point for our calculation is the quantity
\begin{multline}
\cK_{\rm df,3}^{(0)}(\{\bm p\},\{\bm k\})_{m'_s,m_s} =
\\
\sqrt{(\omega(\bm p') + M_N) (\omega(\bm a) + M_N)}
\left[
\delta_{m'_s m_s} - \frac{[\bm p' \bm a]_{m'_s m_s}}{(\omega(\bm p')+M_N)(\omega(\bm a)+M_N)} \right]\,.
\label{eq:KdfNpn}
\end{multline}

To proceed, we need to express $\bm a$ and $\omega(\bm a)$ in terms of $\bm a^\star$, the initial-state-pair relative momentum in its CMF. This is done using standard boost results,
\begin{align}
\begin{split}
\bm a &= \bm a^\star + (\gamma_{N\pi}(\bm p) -1) (\hat{\bm \beta}_{N\pi}(\bm p) \cdot \bm a^\star) \hat{\bm \beta}_{N\pi}(\bm p)
- \omega(\bm a^\star)\; \gamma_{N\pi}(\bm p) \; \boldsymbol \beta_{N\pi}(\bm p) \,,
\\
\omega(\bm a) &= \gamma_{N\pi}(\bm p) \left( \omega(\bm a^\star) - \boldsymbol \beta_{N\pi}(\bm p) \cdot \bm a^\star\right)\,.
\end{split}
\label{eq:boosta}
\end{align}

The next step is to decompose into spherical harmonics relative to the direction $\hat {\bm a}^\star$. The angular dependence in \Cref{eq:KdfNpn} arises both from the $[\bm p' \bm a]$ term and from the kinematic factors depending on $\omega (\bm a)$. The former can be written as
\begin{equation}
[\bm p' \bm a] =
[\bm p' \bm a^\star] + (\gamma_{N\pi}(\bm p)-1) \left [ \bm p' \widehat{\bm \beta}_{N\pi}(\bm p) \right ] \, \widehat{\bm \beta}_{N\pi}(\bm p) \cdot \bm a^\star
- \omega_N(\bm a^\star) \gamma_{N\pi}(\bm p) \Big [\bm p' \bm \beta_{N\pi}(\bm p) \Big ]\,,
\end{equation}
and thus contains both $\ell=0$ and $1$ components. To pull these out we use
\begin{equation}
\bm v \cdot \bm a^\star = \frac13 \sum_m \cY_{1m}(\bm v) \cY^*_{1m}(\bm a^\star)\,,
\end{equation}
and the generalization to vectors of spin matrices,
\begin{equation}
[\bm v \bm a^\star] = \frac13 \sum_m \cY_{1m}([\bm v r]) \cY^*_{1m}(\bm a^\star)\,.
\end{equation}
Here $[\bm v r]$ is defined in \Cref{eq:bracketr}, while the $\cY_{\ell m}$ are the harmonic polynomials defined in \Cref{eq:harmonicpolynomials}.

Turning to the kinematic factors in \Cref{eq:KdfNpn}, we observe that, since they are nonlinear in $\omega(\bm a)$, which itself is non-linear in $\bm a^\star$, they contribute to all values of $\ell$. To make this explicit, we write
\begin{align}
\omega(\bm a) + M_N &= A_a + B_a c\,, \\
A_a &= \gamma_{N\pi}(\bm p)\; \omega_N(\bm a^\star) + M_N \,, \\
B_a &= \gamma_{N\pi}(\bm p)\; \beta_{N\pi}(\bm p)\; a^\star\,,
\end{align}
where $c = \widehat{\bm \beta}_{N\pi}(\bm p) \cdot \hat {\bm a}^\star$, $a^\star = |\bm a^\star|$, and $\beta_{N\pi}(\bm p) = |\boldsymbol \beta_{N\pi}(\bm p)|$.

We then expand in terms of Legendre polynomials,
\begin{equation}
f(c) = \sum_{\ell} f_\ell P_\ell(c)\,,\qquad
f_\ell = \frac{2\ell + 1}2 \int_{-1}^1 dc\; P_\ell(c) f(c)\,,
\end{equation}
and use
\begin{equation}
P_\ell(c) = \frac{1}{2\ell+1} \sum_m \cY_{\ell m}(\widehat{\bm \beta}_{N\pi}(\bm p)) \cY^*_{\ell m}(\hat {\bm a}^\star) \,,
\label{eq:PintoY}
\end{equation}
(where we stress that the arguments of the harmonic polynomials are unit vectors) to pull out the angular dependence.

The expansion coefficients that we need are for the functions
\begin{equation}
f^{(1)} = \sqrt{\omega(\bm a)+M_N}\,,\quad
f^{(2)} = \frac1{\sqrt{\omega(\bm a)+M_N}}\,.
\end{equation}
Since we ultimately restrict $\ell \le \ell_{\rm max}=1$, only a small number of coefficients are needed. These are
\begin{align}
f^{(1)}_0 &= \frac{S_-^{(3)}}{3B_a}\,,
\qquad
f^{(1)}_1 = \frac1{5B_a^2}
\left[
3 B_a S_+^{(3)} - 2 A_a S_-^{(3)}
\right]\,,
\end{align}
where $S_\pm^{(n)} = (A_a + B_a)^{n/2} \pm (A_a -B_a)^{n/2}$,
and
\begin{align}
f^{(2)}_0 &= \frac{S_-^{(1)}}{B_a}\,,
\\
f^{(2)}_1 &=
\frac1{B_a^2}
\left[
B_a \, S_+^{(1)} - 2 A_a \, S_-^{(1)}
\right] \,,
\\
f^{(2)}_2 &=
\frac1{B_a^3}
\left[
(4 A_a^2 - B_a^2) \, S_-^{(1)} - 2 A_a \, B_a \, S_+^{(1)}
\right]\,.
\end{align}
A useful result that we need below is that $f^{(i)}_j = \mathcal O [(B_a)^j]$ for small $B_a$, and in general is an even/odd function of $B_a$ for $j$ even/odd.

We next combine the two sources of angular dependence into an overall dependence using Clebsch-Gordon coefficients. The combinations we require are $0\times 0 \to 0$, $0\times 1 \to 1$, $1\times 1 \to 0$, and $1\times 2 \to 1$. The first two are trivial, while for the latter two we use
\begin{equation}
\bigg \{ \cY_{1m}^*(\hat{\bm a}^\star) P_1(c) \bigg \}_{\ell=0} = \frac{1}{3} \cY_{1m}(\hat{\bm \beta}_{N \pi}(\bm p))
\end{equation}
and
\begin{align}
\bigg\{ \cY^*_{1 m'}(\hat {\bm a}^\star) P_2(c)\bigg\}_{\ell = 1} &=
\sum_m
{\bm S}_{m'm} \cY^*_{1m}(\hat {\bm a}^\star) \,,
\\
{\bm S}_{m'm} &=
\frac15 \left\{ \cY^*_{1m'}(\widehat{\bm \beta}_{N\pi}(\bm p)) \cY_{1m}(\widehat{\bm \beta}_{N\pi}(\bm p)) - \delta_{m' m} \right\} \,,
\label{eq:Sdef}
\end{align}
respectively. The combination $1\times 1 \to 1$ vanishes due to symmetry.

Using these results we find the $\ell'=\ell=0$ component to be
\begin{align}
\frac{\left[\cK_{\df,3}^{(0),N\pi}\right]_{\bm p' 0 0 ; \bm p 0 0}}{\sqrt{\omega(\bm p')+M_N}} &=
\bm 1 f^{(1)}_0
+ \frac{\Big [\bm p' \widehat{\bm \beta}_{N\pi}(\bm p) \Big]}{\omega(\bm p')+M_N}
\left\{f^{(2)}_0 \omega(\bm a^\star)\beta_{N\pi}(\bm p) \gamma_{N\pi}(\bm p) - \frac13 f^{(2)}_1 a^\star \gamma_{N\pi}(\bm p) \right\}\,,
\label{eq:K0Np00}
\end{align}
while that for $\ell'=0$, $\ell=1$ is
\begin{align}
\begin{split}
\frac{ \left[\cK_{\df,3}^{(0),N\pi}\right]_{\bm p' 0 0 ; \bm p 1 m} }{a^\star \sqrt{\omega(\bm p')+M_N}} & = \bm 1
\frac{\cY_{1m}(\widehat{\bm \beta}_{N\pi}(\bm p))}{3} \frac{f^{(1)}_1}{a^\star}
\\ & \hspace{-70pt}
- \frac{1}{\omega(\bm p')+M_N} \bigg [ \frac{\cY_{1m}([\bm p' r])}{3}
\left\{ f^{(2)}_0 - \frac{1}{5} f^{(2)}_2 \right\} + \frac{\cY_{1m}(\widehat{\bm \beta}_{N\pi}(\bm p))}{3}
[\bm p' \widehat{\bm \beta}_{N\pi}(\bm p)]
\\
& \hspace{-60pt} \times
\bigg \{ f^{(2)}_0 (\gamma_{N\pi}(\bm p) - 1) - \frac{f^{(2)}_1}{a^\star} \beta_{N\pi}(\bm p) \gamma_{N\pi}(\bm p) \; \omega(\bm a^\star)
+ \frac{1}{5} f^{(2)}_2 (2\gamma_{N\pi}(\bm p) + 1) \bigg \}
\bigg ]\,.
\end{split}
\label{eq:K0Np01}
\end{align}
As noted above, we drop contributions with $\ell'=0$, $\ell > 1$, while those with $\ell'>0$ vanish.

In \Cref{eq:K0Np01}, we have chosen to divide by $a^\star$, since this is what we use in practice in our implementation. This leads to factors of $f^{(1)}_1/a^\star$ and $f^{(2)}_1/a^\star$ on the right-hand side. These ratios are given by
\begin{equation}
\frac{f^{(i)}_1}{a^\star} = \frac{f^{(i)}_1}{B_a} \gamma_{N\pi}(\bm p) \beta_{N\pi}(\bm p)\,,
\end{equation}
and are real functions of $B_a^2$ that are nonsingular at the origin. In particular, when the initial-state pair goes below threshold, these ratios smoothly extend to the resulting negative values of $B_a^2$.


\subsection{\texorpdfstring{$\pi N$}{Pion-nucleon} block}

The result for this block can be obtained by hermitian conjugation from that for the $N\pi$ block. For completeness we quote the results. For $\ell'=\ell=0$ we have
\begin{multline}
\frac{\left[\cK_{\df,3}^{(0),\pi N}\right]_{\bm p' 0 0 ; \bm p 0 0}}{\sqrt{\omega(\bm p)+M_N}} =
\bm 1 f'^{(1)}_0
\\
+ \frac{[\widehat{\bm \beta}_{N\pi}(\bm p')\, \bm p]}{\omega(\bm p)+M_N}
\left\{f'^{(2)}_0 \omega(\bm a'^\star)\beta_{N\pi}(\bm p') \gamma_{N\pi}(\bm p') - \frac13 f'^{(2)}_1 a'^\star \gamma_{N\pi}(\bm p') \right\}\,,
\end{multline}
while for $\ell'=1$, $\ell=0$ the result is
\begin{align}
\begin{split}
\frac{ \left[\cK_{\df,3}^{(0),\pi N}\right]_{\bm p' 1 m' ; \bm p 0 0} }{a'^\star \sqrt{\omega(\bm p)+M_N}} & =
\bm 1
\frac{\cY^*_{1m'}(\widehat{\bm \beta}_{N\pi}(\bm p'))}{3} \frac{f'^{(1)}_1}{a'^\star}
\\ & \hspace{-80pt}
- \frac{1}{\omega(\bm p)+M_N} \bigg [ \frac{\cY^*_{1m'}([r \bm p])}{3}
\left\{ f'^{(2)}_0 - \frac{1}{5} f'^{(2)}_2 \right\} + \frac{\cY^*_{1m'}(\widehat{\bm \beta}_{N\pi}(\bm p'))}{3}
[\widehat{\bm \beta}_{N\pi}(\bm p')\, \bm p]
\\
& \hspace{-80pt} \times
\bigg \{ f'^{(2)}_0 (\gamma_{N\pi}(\bm p') - 1) - \frac{f'^{(2)}_1}{a'^\star} \beta_{N\pi}(\bm p') \gamma_{N\pi}(\bm p') \; \omega(\bm a'^\star)
+ \frac{1}{5} f'^{(2)}_2 (2\gamma_{N\pi}(\bm p') + 1) \bigg \}
\bigg ]\,.
\end{split}
\end{align}
We drop contributions with $\ell'>1$, $\ell =0$, while those with $\ell>0$ vanish.


\subsection{\texorpdfstring{$\pi\pi$}{Pion-pion} block}

In this block, both the final and initial spectators are pions, so that $\bm p_3=\bm a'$ and $\bm k_3=\bm a$. Thus the starting point is
\begin{multline}
\cK_{\rm df,3}^{(0)}(\{\bm p\},\{\bm k\})_{m'_s,m_s} =
\\
\sqrt{(\omega(\bm a') + M_N) (\omega(\bm a) + M_N)}
\left[
\delta_{m'_s m_s} - \frac{[\bm a' \bm a]_{m'_s m_s}}{(\omega(\bm a')+M_N)(\omega(\bm a)+M_N)} \right]\,.
\label{eq:Kdfppn}
\end{multline}
This is the most complicated case to decompose, since there is angular dependence from both final and initial states. Using the boosts in \Cref{eq:boosta}, as well as the corresponding results for the final state, we find
\begin{align}
[\bm a' \bm a] &= C_0 - \bm a'^\star \cdot \bm V - \bm a^\star \cdot \bm V' + \bm a'^\star \cdot \bm T \cdot \bm a^\star\,,
\end{align}
where
\begin{align}
C_0 &= \omega(\bm a'^\star) \omega(\bm a^\star) \gamma_{N\pi}(\bm p') \gamma_{N\pi}(\bm p) [\boldsymbol \beta_{N\pi}(\bm p') \boldsymbol \beta_{N\pi}(\bm p)]\,,
\\
\bm V &= \omega(\bm a^\star)\gamma_{N\pi}(\bm p) \left\{ [ r \boldsymbol \beta_{N\pi}(\bm p) ]
+ (\gamma_{N\pi}(\bm p')\!-\!1) [\widehat{\bm \beta}_{N\pi}(\bm p') \boldsymbol \beta_{N\pi}(\bm p)] \widehat{\bm \beta}_{N\pi}(\bm p') \right\} \,,
\\
\bm V' &= \omega(\bm a'^\star)\gamma_{N\pi}(\bm p') \left\{ [\boldsymbol \beta_{N\pi}(\bm p') r]
+ (\gamma_{N\pi}(\bm p)\!-\!1) [ \boldsymbol \beta_{N\pi}(\bm p') \widehat{\bm \beta}_{N\pi}(\bm p) ] \widehat{\bm \beta}_{N\pi}(\bm p) \right\} \,,
\\
{\bm T}_{jk} &=r_{jk}
+ (\gamma_{N\pi}(\bm p)\!-\!1) [r \widehat{\bm \beta}_{N\pi}(\bm p)]_j \widehat{\bm \beta}_{N\pi}(\bm p)_k
+ (\gamma_{N\pi}(\bm p')\!-\!1) \widehat{\bm \beta}_{N\pi}(\bm p')_j [\widehat{\bm \beta}_{N\pi}(\bm p') r]_k
\\
&\quad
+ (\gamma_{N\pi}(\bm p)\!-\!1)(\gamma_{N\pi}(\bm p')\!-\!1) [\widehat{\bm \beta}_{N\pi}(\bm p') \widehat{\bm \beta}_{N\pi}(\bm p)] \widehat{\bm \beta}_{N\pi}(\bm p')_j \widehat{\bm \beta}_{N\pi}(\bm p)_k \,.
\end{align}
To decompose this into pair angular momenta we need the tools given above, along with the relation between Cartesian vectors and spherical harmonics,
\begin{equation}
\bm a_j = \sqrt{\frac13} \bm R_{jm} \cY^*_{1m}(\bm a)\,,\qquad
\bm R = \begin{pmatrix} -1/\sqrt2 & 0 & 1/\sqrt2 \\ - i/\sqrt2 & 0 & - i/\sqrt2 \\ 0 & 1 & 0 \end{pmatrix}\,,
\end{equation}
where Cartesian indices are ordered $\{x,y,z\}$, while spherical coordinates are ordered $\{1, 0, -1\}$. Thus the tensor $\bm T$ expressed in spherical coordinates is
\begin{equation}
\bm T^S_{m' m} = \bm R^\dagger_{m' i} \bm T_{ij} \bm R_{jm}\,.
\end{equation}

After much tedious algebra we obtain the following results. For $\ell'=\ell=0$ we have
\begin{multline}
\left[\cK_{\df,3}^{(0),\pi\pi}\right]_{\bm p' 0 0 ; \bm p 0 0} =
f'^{(1)}_0 f^{(1)}_0 \bm 1 - f'^{(2)}_0 f^{(2)}_0 C_0
\\
- \frac13 \left\{ f'^{(2)}_0 \frac{f^{(2)}_1}{a^\star} (a^\star)^2 \left( \bm V' \cdot \hat{\bm \beta}_{N\pi}(\bm p) \right)
+ (a'^\star)^2 \frac{f'^{(2)}_1}{a'^\star} f^{(2)}_0 \left( \bm V \cdot \hat{\bm \beta}_{N\pi}(\bm p') \right) \right\}
\\
- \frac{(a'^\star)^2 (a^\star)^2}{9} \frac{f'^{(2)}_1}{a'^\star} \frac{f^{(2)}_1}{a^\star} \left[ \hat{\bm \beta}_{N\pi}(\bm p') \cdot \bm T \cdot \hat{\bm \beta}_{N\pi}(\bm p) \right]
\,,
\end{multline}
where we have made explicit that this is an even function of $a'^{\star}$ and $a^{\star}$.

The result for $\ell'=0$, $\ell=1$, divided by $a^\star$, is given by
\begin{align}
\begin{split}
\frac{\left[\cK_{\df,3}^{(0),\pi\pi}\right]_{\bm p' 0 0 ; \bm p 1 m}}{a^\star} & = \frac{\cY_{1m}(\widehat{\bm \beta}_{N\pi}(\bm p))}{3}
\Bigg\{ \bm 1 f'^{(1)}_0 \frac{f^{(1)}_1}{a^\star}
- f'^{(2)}_0 \frac{f^{(2)}_1}{a^\star} C_0
\Bigg\}
\\
& -
\frac{\cY_{1m}(\widehat{\bm \beta}_{N\pi}(\bm p))}{9}
\Bigg\{
\frac{f'^{(2)}_1}{a'^\star} \frac{f^{(2)}_1}{a^\star} (a'^\star)^2 \left( \bm V \cdot \widehat{\bm \beta}_{N\pi}(\bm p') \right)
\Bigg\}
\\
& - \frac{\cY_{1m}(\bm V')}{3} f'^{(2)}_0 \left\{ f^{(2)}_0 - \frac{1}{5} f^{(2)}_2 \right\}
\\
& - \cY_{1m}(\widehat{\bm \beta}_{N\pi}(\bm p)) \frac{1}{5} f'^{(2)}_0 f^{(2)}_2 \left( \bm V' \cdot \widehat{\bm \beta}_{N\pi}(\bm p) \right)
\\
& - \frac{(a'^\star)^2}{3} \frac{f'^{(2)}_1}{a'^\star} \frac{1}{\sqrt{3}} \left[ \widehat{\bm \beta}_{N\pi}(\bm p') \cdot \bm T \, \cdot \bm R \cdot ( f^{(2)}_0 + f^{(2)}_2 {\bm S} ) \right]_{m}
\,,
\end{split}
\end{align}
while the hermitian conjugate quantity is
\begin{align}
\begin{split}
\frac{\left[\cK_{\df,3}^{(0),\pi\pi}\right]_{\bm p' 1 m' ; \bm p 0 0}}{a'^\star} & =
\frac{\cY^*_{1m'}(\widehat{\bm \beta}_{N\pi}(\bm p'))}{3}
\Bigg\{ \bm 1
\frac{f'^{(1)}_1}{a'^\star} f^{(1)}_0
- \frac{f'^{(2)}_1}{a'^\star} f^{(2)}_0 C_0
\Bigg\}
\\
&
-
\frac{\cY^*_{1m'}(\widehat{\bm \beta}_{N\pi}(\bm p'))}{9}
\Bigg\{
\frac{f'^{(2)}_1}{a'^\star} \frac{f^{(2)}_1}{a^\star} (a^\star)^2 \left( \bm V' \cdot \widehat{\bm \beta}_{N\pi}(\bm p) \right)
\Bigg\}
\\
&
- \frac{\cY^*_{1m'}(\bm V)}{3} \left\{ f'^{(2)}_0 - \frac{1}{5} f'^{(2)}_2 \right\} f^{(2)}_0
\\
&
- \cY^*_{1m'}(\widehat{\bm \beta}_{N\pi}(\bm p')) \frac{1}{5} f'^{(2)}_2 f^{(2)}_0 \left( \widehat{\bm \beta}_{N\pi}(\bm p') \cdot \bm V \right)
\\
&
- \frac{(a^\star)^2}{3} \frac{1}{\sqrt{3}} \left[ (f'^{(2)}_0 + f'^{(2)}_2 {\bm S'}) \cdot \bm R^\dagger \cdot \bm T \cdot \widehat{\bm \beta}_{N\pi}(\bm p) \right]_{m'} \frac{f^{(2)}_1}{a^\star}
\,.
\end{split}
\end{align}
Here
\begin{align}
\bm S'_{m'm} &=
\frac15 \left\{ \cY^*_{1m'}(\widehat{\bm \beta}_{N\pi}(\bm p')) \cY_{1m}(\widehat{\bm \beta}_{N\pi}(\bm p')) - \delta_{m' m} \right\} \,,
\end{align}
is the analogous quantity to $\bm S$, defined in~\cref{eq:Sdef}, but with $\bm p \to \bm p'$.

Finally, the $\ell'=\ell=1$ term (divided by $a'^* a^\star$) is given by
\begin{align}
\begin{split}
\frac{\left[\cK_{\df,3}^{(0),\pi\pi}\right]_{\bm p' 1 m' ; \bm p 1 m}}{a'^\star a^\star} & =
\frac{\cY^*_{1m'}(\widehat{\bm \beta}_{N\pi}(\bm p')) \cY_{1m}(\widehat{\bm \beta}_{N\pi}(\bm p))}{9}
\Bigg\{ \bm 1
\frac{f'^{(1)}_1}{a'^\star} \frac{f^{(1)}_1}{a^\star}
- \frac{f'^{(2)}_1}{a'^\star} \frac{f^{(2)}_1}{a^\star} C_0
\Bigg\}
\\
&
- \frac{\cY^*_{1m'}(\widehat{\bm \beta}_{N\pi}(\bm p')) \cY_{1m}(\bm V')}{9}
\frac{f'^{(2)}_1}{a'^\star} \left( f^{(2)}_0 - \frac{1}{5} f^{(2)}_2 \right)
\\
&
- \frac{\cY^*_{1m'}(\bm V) \cY_{1m}(\widehat{\bm \beta}_{N\pi}(\bm p))}{9}
\left( f'^{(2)}_0 - \frac{1}{5} f'^{(2)}_2 \right) \frac{f^{(2)}_1}{a^\star}
\\
&
- \frac{\cY^*_{1m'}(\widehat{\bm \beta}_{N\pi}(\bm p')) \cY_{1m}(\widehat{\bm \beta}_{N\pi}(\bm p))}{15}
\frac{f'^{(2)}_1}{a'^\star} f^{(2)}_2 \left( \bm V' \cdot \widehat{\bm \beta}_{N\pi}(\bm p) \right)
\\
&
- \frac{\cY^*_{1m'}(\widehat{\bm \beta}_{N\pi}(\bm p')) \cY_{1m}(\widehat{\bm \beta}_{N\pi}(\bm p))}{15}
f'^{(2)}_2 \frac{f^{(2)}_1}{a^\star} \left( \widehat{\bm \beta}_{N\pi}(\bm p') \cdot \bm V \right)
\\
&
- \frac{1}{3} \left[
\left( f'^{(2)}_0 + f'^{(2)}_2 {\bm S}' \right) \cdot \bm T^S \cdot \left( f^{(2)}_0 + f^{(2)}_2 {\bm S} \right)
\right]_{m' m}
\,.
\end{split}
\end{align}
In this case entries other than the $\ell' = \ell = 1$ block are also nonzero, but are dropped in our truncation.


\subsection{Irrep projections}
\label{app:irreps}

We have implemented the above decompositions numerically using two independent codes. One question of interest is which irreps are contained in the $\cK_0$ term, for only levels in these irreps will be shifted by the presence of the three-particle interaction. We recall that the relevant (little) group is the subgroup of transformations of a cube that leave the total momentum, $\bm P = \bm n_P (2\pi/L)$, invariant. (We consider here a cubic spatial volume of size $L$.) Since the nucleon has spin $1/2$, the transformations include the rotation of this spin degree of freedom, and the irreps are actually those of the doubled little groups. These are presented in ref.~\cite{Morningstar:2013bda} for most little groups, and the extension to the remaining cases is given in ref.~\cite{Blanton:2021llb}.

We find for all choices of $\bm n_P$ that the $\cK_0$ term has eight nonzero eigenvalues once the total energy $E$ lies far enough above threshold. The breakdown into irreps is shown in \Cref{tab:K0irreps}. As can be seen, several irreps do not appear. In particular, in the rest frame, the $H_u$ irrep is not present. This is significant as this is the irrep in which a $\Delta + \pi$ state at rest lives. Thus a higher-order term in $\Kdf$ is needed to shift the energy of the lowest-lying $\Delta+\pi$ state in the rest frame. However, the $G_{1g}$ irrep does appear, and this contains $\Delta+\pi$ states in a relative $p$-wave. In the main text (see \Cref{sec:num}) we show illustrative results for this irrep.


\begin{table}[htp]
\begin{center}
\begin{tabular}{ccc}
$\bm n_P$ & irreps appearing & absent irreps
\\
\hline
$(0,0,0)$ & $G_{1g}(2) + 3 G_{1u}(2)$ & $G_{2g}(2), H_g(4), G_{2u}(2), H_u(4)$
\\
$(0,0,a)$ & $4G_1(2)$ & $G_2(2) $
\\
$(a,a,0)$ & $4G(2)$ & $-$
\\
$(a,a,a)$ & $4G(2)$ & $F_1(1),F_2(1)$
\\
$(a,b,0)$ & $4F_1(1)+4 F_2(1)$ & $-$
\\
$(a,a,b)$ & $4F_1(1)+4 F_2(1)$ & $-$
\\
$(a,b,c)$ & $8F(1)$ & $-$
\end{tabular}
\end{center}
\caption{Decomposition of eigenvalues of the $\cK_0^{5/2,SS}$ term into irreps at asymptotic energies. Frames are denoted using nonzero integers $a$, $b$ and $c$, which are all different but otherwise arbitrary. The numbers in parentheses following the irreps gives their dimensions. In the $(a,b,0)$ and $(a,a,b)$ rows, the four eigenvalues in the $F_1$ irrep are the same as the four in the $F_2$ irrep.}
\label{tab:K0irreps}
\end{table}

\bibliographystyle{JHEP}
\bibliography{ref}

\end{document}